%% file: main.tex
\documentclass[a4paper,11pt]{article}
\usepackage{jcappub}
\graphicspath{{figs/}} % directory of images
\usepackage{xspace}
\usepackage{cleveref}

\newcommand{\xmax}{\ensuremath{X_{\mathrm{max}}}\xspace}%
\newcommand{\rcut}{\ensuremath{\log_{10}(R_{\mathrm{cut}}/\mathrm{V})}\xspace}%
\newcommand{\edet}{\ensuremath{\log_{10}(E_{\mathrm{det}}/\mathrm{eV})}\xspace}%

\title{Constraining models for the origin of ultra-high-energy cosmic rays with a novel combined analysis of arrival directions, spectrum, and composition data measured at the Pierre Auger Observatory}

\input{jcappub_authorlist}

\emailAdd{auger\_spokespersons@fnal.gov}

\abstract{The combined fit of the measured energy spectrum and shower maximum depth distributions of ultra-high-energy cosmic rays is known to constrain the parameters of astrophysical models with homogeneous source distributions. Studies of the distribution of the cosmic-ray arrival directions show a better agreement with models in which a fraction of the flux is non-isotropic and associated with the nearby radio galaxy Centaurus A or with catalogs such as that of starburst galaxies.
Here, we present a novel combination of both analyses by a simultaneous fit of arrival directions, energy spectrum, and composition data measured at the Pierre Auger Observatory. The model takes into account a rigidity-dependent magnetic field blurring and an energy-dependent evolution of the catalog contribution shaped by interactions during propagation.

We find that a model containing a flux contribution from the starburst galaxy catalog of around 20\% at 40\,EeV with a magnetic field blurring of around $20^\circ$ for a rigidity of 10~EV provides a fair simultaneous description of all three observables. 
The starburst galaxy model is favored with a significance of $4.5\sigma$ (considering experimental systematic effects) compared to a reference model with only homogeneously distributed background sources. By investigating a scenario with Centaurus A as a single source in combination with the homogeneous background, we confirm that this region of the sky provides the dominant contribution to the observed anisotropy signal. Models containing a catalog of jetted active galactic nuclei whose flux scales with the $\gamma$-ray emission are, however, disfavored as they cannot adequately describe the measured arrival directions.

}

\begin{document}

\maketitle
\flushbottom

\section{Introduction}
The origin and acceleration mechanisms of ultra-high-energy cosmic rays (UHECRs) still remain a mystery today. Several challenges have to be faced in the search for UHECR sources, among others the stochastic nature of the interactions during propagation, the not directly measurable cosmic-ray composition at the highest energies, and deflections by largely uncertain cosmic magnetic fields.
However, significant advances have been made both in theoretical modeling and in the amount and quality of measured data in recent years. With the Pierre Auger Observatory~\cite{the_pierre_auger_collaboration_a_aab_et_al_pierre_2015} an exposure of more than $100{,}000$ km$^2$ sr yr~\cite{the_pierre_auger_collaboration_p_abreu_arrival_2022} has been accumulated, allowing us to determine the characteristics of arriving UHECRs with unprecedented precision. Three properties that could provide insights into the cosmic-ray sources, and that are investigated in this work, are the cosmic-ray energy spectrum, the distribution of shower maxima tracing the mass composition, and the arrival directions.

For instance, the arrival directions have already revealed a dipolar distribution above $8~\mathrm{EeV}$ with a significance exceeding $5\sigma$. It is directed away from the Galactic center, which serves as proof of an extragalactic origin of UHECRs~\cite{the_pierre_auger_collaboration_a_aab_et_al_observation_2017, the_pierre_auger_collaboration_a_aab_et_al_large-scale_2018}. Recently, an indication of a correlation of the UHECR arrival directions above $\sim40~\mathrm{EeV}$ with the directions of powerful extragalactic source candidates was found~\cite{the_pierre_auger_collaboration_p_abreu_update_2010, the_pierre_auger_collaboration_a_aab_et_al_indication_2018, di_matteo_uhecr_2021}. The catalogs that have been tested by the Pierre Auger Collaboration are two different selections of jetted- and non-jetted active galactic nuclei, a selection of starburst galaxies, a broader catalog tracing the large-scale mass distribution, and the nearby radio galaxy Centaurus A. The largest statistical significance of $4.0\sigma$ (1-sided) is achieved for the selection of starburst galaxies~\cite{the_pierre_auger_collaboration_p_abreu_arrival_2022}.
But, mainly because all catalogs contain strong sources located in the same direction around Centaurus A, currently no catalog can truly be favored above the others~\cite{the_pierre_auger_collaboration_p_abreu_arrival_2022}. 
%Also, the arrival directions alone may not be able to provide sufficient information to decide if the catalog objects are really the sources of UHECRs.
%, or if they, for example, just serve as a last deflection center~\cite{kotera_optical_2008, bell_echoes_2022}.

The combination of the arrival directions with the other two observables, the energy spectrum and the distribution of shower maxima, may help to solve this puzzle, as they contain additional information about the nature and propagation of UHECRs. The energy spectrum and shower maximum depth distributions have been used in previous analyses of the Pierre Auger Collaboration~\cite{the_pierre_auger_collaboration_a_aab_combined_2017, e_guido_for_the_pierre_auger_collaboration_combined_2021, the_pierre_auger_collaboration_a_abdul_halim_constraining_2023} in order to constrain generic source models of UHECRs. In those works, the model comprises only homogeneous distributions of identical extragalactic sources, possibly adding a Galactic component when lower energies are included, and the arrival directions are not considered as an observable. Through the comparison of the modeled energy spectrum and shower maximum depth distributions to the measured ones, parameters describing the injected spectrum and composition at the sources have been determined. The previous works will be used as a basis for this study. 

Different kinds of combined analyses of all three observables have been conducted before by other authors, for example Refs.~\cite{eichmann_ultra-high-energy_2018, ding_imprint_2021, vanvliet_extragalactic_2022, allard_what_2022, eichmann_explaining_2022}. 
Due to the large parameter space involved for sophisticated astrophysical models which also describe the arrival directions, these works focus on specific aspects and choose to simplify other ingredients of the model or fit procedure.
Often, a two-step approach is chosen~\cite{eichmann_ultra-high-energy_2018, allard_what_2022}, in which the model parameters are adapted first, and then a comparison of the predicted arrival directions to the measured ones is conducted. In other studies, simplifications regarding the modeling of the source emission or propagation have been made, in order to focus, for example, on the determination of parameters of the extragalactic magnetic field or of the emission parameters of individual sources~\cite{eichmann_explaining_2022, vanvliet_extragalactic_2022, ding_imprint_2021}. 

In this work, all three observables are combined in a simultaneous fit of the energy spectrum, shower maximum distributions, and arrival directions measured at the Pierre Auger Observatory, taking into account propagation effects and determining the parameters describing the source setup and emission. Additionally, the uncertainties and correlations associated with these parameters are also determined during the fit, using a Bayesian approach. A preliminary proof of concept of the method using realistic simulations was presented before in Refs.~\cite{t_bister_for_the_pierre_auger_collaboration_combined_2021, t_bister_for_the_pierre_auger_collaboration_sensitivity_2023}.

The astrophysical model is built in a straightforward way, based on the medium-scale anisotropies seen at energies well above the ankle and correlating with extragalactic objects~\cite{the_pierre_auger_collaboration_p_abreu_arrival_2022, the_pierre_auger_collaboration_a_aab_et_al_indication_2018}. For that reason, this work focuses on the highest energies, taking into account arrival direction data above 16~EeV, and energy spectrum and shower depth distributions above 10~EeV.
The astrophysical model, following the findings of~\cite{the_pierre_auger_collaboration_p_abreu_arrival_2022, the_pierre_auger_collaboration_a_aab_et_al_indication_2018}, not only contains homogeneously distributed sources, but also a variable contribution from the catalogs of either starburst galaxies or jetted active galactic nuclei that correlate with the UHECR arrival directions. Here, the contribution of each source to the total flux is modeled depending on its distance and the injected spectrum and composition common to all sources. Additionally, a rigidity and possibly distance-dependent magnetic field blurring is included. As a comparison, a model containing only Centaurus A as a single source in addition to the homogeneously distributed background sources, is also investigated.

This work is structured as follows: in \cref{sec:model}, the astrophysical model is described, going from the emission at the sources in \cref{sec:sources} over the propagation in \cref{sec:propagation} to the calculation of the energy-dependent contributions of the source candidates in \cref{sec:3dsetup}. Then, the modeling of the three observables on Earth is described in \cref{sec:simulated_observables}. In \cref{sec:fitting_procedure}, the used inference methods, the likelihood function, and the measured data sets are described. The fit results are discussed in \cref{sec:results}, and the influence of the most important experimental systematic uncertainties is evaluated in \cref{sec:results_syst}. Finally, a discussion of the results and a conclusion is provided in \cref{sec:conclusion}.

%\cite{eichmann_ultra-high-energy_2018} model individual radio galaxies, fit to En and lnA, then compare multipole moments afterwards
%\cite{eichmann_explaining_2022} again model individual radio galaxies, compare to dipole and energy spectrum, consider JF12, fit emission parameters gamma & rcut + EGMF & acceleration parameters, simplification of elemental fractions to mean
%\cite{ding_imprint_2021} model based on LSS+JF12, simplified propagation, fit to lnA, sigma(lnA), arrival directions, do not fit source emission
%\cite{van_vliet_extragalactic_2021} emission following best-fit without catalog sources, consider specific source candidates & structured EGMF, scan over source density & magnetic field which best reproduce smearing & signal fraction from Auger
%\cite{allard_what_2021} investigate dipole with source densities and catalogs e.g. 2MRS, simulate 300 maps in each case (3 different injected spectra,EGMF strength, GMF models) and apply large-scale searches, check consistency with spectrum & composition

\section{Astrophysical models for the origin of UHECRs} \label{sec:model}
The astrophysical model in this work is based on the one already successfully employed before~\cite{the_pierre_auger_collaboration_a_aab_combined_2017} in which only homogeneous background sources were used. That setup, without a contribution of catalog sources, will also be investigated in this work as a baseline and will hence be called the \textit{reference model}. 
%In the full setup, when both homogeneous background and catalog sources are included, the background accounts for further away, unidentified sources of the same kind as the catalog sources.

\subsection{Source candidates and acceleration} \label{sec:sources}
In the following, the setup of the full astrophysical model will be explained in detail, starting with the source distribution including catalog and background sources.

\paragraph{Catalogs of starburst galaxies and active galactic nuclei:} \label{sec:candidates}
The arrival directions of UHECRs measured with the Pierre Auger Observatory exhibit intermediate-scale correlations with candidate sources from different catalogs~\cite{the_pierre_auger_collaboration_a_aab_et_al_indication_2018}.
As described above, the astrophysical model in this work is built up as a sum of homogeneous background sources and prominent foreground catalog sources whose contributions are studied individually. As the costly calculation of individual fluxes for all source candidates prohibits the use of catalogs with a large number of sources, the shortest two updated ones from the latest analysis~\cite{the_pierre_auger_collaboration_p_abreu_arrival_2022, j_biteau_for_the_pierre_auger_collaboration_ultra-high-energy_2021} are selected for this analysis. Catalogs with a large number of source candidates, e.g. tracing the extragalactic matter distribution, can rather be studied by calculating the contributions of different distance shells~\cite{ding_imprint_2021} than that of individual sources (which is however outside the scope of this work).
The first catalog, which reaches the highest significance of $4.0\sigma$ (1-sided) for a correlation with the UHECR arrival directions, is a selection of 44 starburst galaxies (SBGs). 
The second catalog is a selection of 26 radio-loud, jetted active galactic nuclei selected by their $\gamma$-ray flux ($\gamma$-AGNs) with a significance of $3.2\sigma$.

For the SBGs, the $\gamma$-ray flux measured with Fermi-LAT~\cite{m_ajello_et_al_the_fermi-lat_collaboration_3fhl_2017} is used as a proxy for the expected UHECR flux. In practice, the flux weights are derived from the radio flux in the 1.4\,GHz band as that scales linearly with the $\gamma$-ray emission~\cite{m_ackermann_et_al_the_fermi_lat_collaboration_gev_2012}. Only SBGs with a flux above 0.3\,Jy, located between 1\,Mpc and 130\,Mpc, are selected. The selection is based on the catalog from~\cite{lunardini_are_2019} with a few changes: the Large and Small Magellanic Clouds are excluded and the Circinus galaxy is added as in~\cite{j_biteau_for_the_pierre_auger_collaboration_ultra-high-energy_2021}. The distances to the catalog sources are derived by crossmatching the selection with the HyperLEDA database\footnote{\url{http://leda.univ-lyon1.fr/}}~\cite{j_biteau_for_the_pierre_auger_collaboration_ultra-high-energy_2021}.
The sources in the jetted $\gamma$-AGN catalog are selected from the 3FHL catalog~\cite{m_ajello_et_al_the_fermi-lat_collaboration_3fhl_2017}. The selection contains only sources with a $\gamma$-ray flux measured with Fermi-LAT $>3.3\times 10^{-11}\,\text{cm}^{-2}\,\text{s}^{-1}$ between 10\,GeV and 1\,TeV, which is also used as a UHECR flux proxy. This leads to a selection of 26 jetted $\gamma$-AGNs between 1\,Mpc and 150\,Mpc, where the distances are again obtained from HyperLEDA.

The most important source candidates from the SBG catalog are NGC 4945, NGC 253, M83, and M82 (outside of the Observatory's exposure), which are all nearby at around $3-5$\,Mpc. For the $\gamma$-AGN catalog, the dominating source candidate is the powerful blazar Markarian 421 at a large distance of $\sim134$\,Mpc. Also, the closest radio galaxy Centaurus A with a distance of $\sim3.5$\,Mpc is part of the $\gamma$-AGN selection. Centaurus A has been discussed as a UHECR source in previous publications~\cite{romero_centaurus_1996, matthews_fornax_2018}, and an overdensity of events around its position has been observed and studied throughout the years~\cite{the_pierre_auger_collaboration_j_abraham_correlation_2007, the_pierre_auger_collaboration_p_abreu_arrival_2022, the_pierre_auger_collaboration_p_abreu_update_2010, the_pierre_auger_collaboration_a_aab_et_al_searches_2015}. 
For that reason, Centaurus A will also be tested as a single source contributing on top of the homogeneous background.
Note that the direction and distance of the strongest source in the SBG selection, NGC 4945, are almost identical to those of Centaurus A, so that a differentiation of the two sources is not possible within the size of the expected magnetic field blurring~\cite{the_pierre_auger_collaboration_p_abreu_arrival_2022}.

\paragraph{Background sources and source evolution:}  \label{sec:hommodel}
In addition to the catalog sources, the model contains a distribution of homogeneous background sources. These should account for unidentified, mostly distant sources of the same kind as the catalog sources. A minimum distance of 3\,Mpc is chosen for the background distribution, and a maximum distance corresponding to redshift $z\approx2.2$ (see above).
% In the astrophysical model, a constant luminosity of the sources is assumed as in Ref.~\cite{the_pierre_auger_collaboration_a_aab_combined_2017}. The source density however can vary over time. 
The evolution of the emissivity of the background sources as a function of redshift $z$ is modeled as
\begin{equation} \label{eq:redshift}
    \psi(z) \propto (1+z)^{m}.
\end{equation}
For starforming regions such as SBGs the evolution follows the star formation rate, which corresponds to $m=3.4$ up to $z=1$~\cite{hopkins_normalisation_2006, yuksel_enhanced_2007}.
Under the assumption that the flux of UHECRs from redshifts beyond $z\sim1$ is negligibly small~\cite{alves_batista_cosmogenic_2019, the_pierre_auger_collaboration_a_aab_combined_2017}, we choose to simplify the evolution by keeping $m=3.4$ constant over the whole range of redshifts of the background up to $z=2.2$. 
For AGNs, the evolution depends on the luminosity of the considered class. 
% As justified in~\cite{alves_batista_cosmogenic_2019}, low- and medium-luminosity AGNs are not expected to be able to accelerate UHECRs, and very-high-luminosity AGNs are rare. So, the evolution of the $\gamma$-AGN model can be reasonably described with the value $m=5.0$ corresponding to the high-luminosity source class~\cite{hasinger_luminosity-dependent_2005}. 
These assumptions for the source evolution are in line with other studies such as Refs.~\cite{the_pierre_auger_collaboration_a_abdul_halim_constraining_2023, heinze_cosmogenic_2016, gelmini_gamma-ray_2012, ahlers_neutrino_2009}.

\paragraph{Emission characteristics:} \label{sec:emission}
We assume that both catalog and background sources accelerate CRs according to a rigidity-dependent process, so that the high-energy cutoff gives rise to a Peters cycle~\cite{peters_primary_1961} leading to an increase in the average mass with energy. Analogously to Ref.~\cite{the_pierre_auger_collaboration_a_aab_combined_2017}, five representative stable elements are studied: hydrogen ($^1$H), helium ($^4$He), nitrogen ($^{14}$N), silicon ($^{28}$Si), and iron ($^{56}$Fe). %These are chosen as the logarithm of their mass numbers log$A_\mathrm{inj}$ is approximately uniformly distributed. The corresponding charge numbers are $Z_\mathrm{A}\in(1, 2, 7, 14, 26)$.
The generation rate at the sources $Q_\mathrm{inj}$ (defined as the number of nuclei ejected per unit of energy, volume and time) as a function of the injected energy $E_\mathrm{inj}$ and the mass number $A_\mathrm{inj}$ of each representative element is modeled as a power-law with a broken exponential cutoff function which sets in when the maximum energy $E_\mathrm{cut}=Z_A R_\mathrm{cut}$ is reached
\begin{equation}
\label{eq:emission}
    Q_\mathrm{inj}(E_\mathrm{inj}, A_\mathrm{inj}) = Q_0 \ a_A \ \Big(\frac{E_\mathrm{inj}}{10^{18}~\mathrm{eV}} \Big)^{-\gamma} \ ~ \begin{cases}
     1 & ; \ Z_A R_\mathrm{cut} > E_\mathrm{inj} \\
    \exp \big( 1-\frac{E_\mathrm{inj}}{Z_A R_\mathrm{cut}} \big) & ; \, Z_A R_\mathrm{cut} \leq E_\mathrm{inj}
    \end{cases}.
\end{equation}
Here, $\gamma$ is the spectral index, $R_\mathrm{cut}$ is the maximum rigidity of the source and $Q_0$ is a normalization for the UHECR generation rate.
$Z_A$ is the atomic number of the injected species with mass number $A_\mathrm{inj}$ and $a_A$ is the fraction of particles of that species. Due to the way \cref{eq:emission} is defined, the charge fractions $a_A$ denote the fractions defined at energies below where the cutoff function sets in. It is often more intuitive (and leads to a more stable fit) to utilize \textit{integral fractions} $I_A$ which denote the total emitted energy fractions of the respective element above a given threshold. They can be calculated as\footnote{Note that this definition denotes integral fractions calculated above a common rigidity threshold and not a common energy threshold as used in~\cite{the_pierre_auger_collaboration_a_abdul_halim_constraining_2023}. Only for hard spectral indices $\gamma<0$ do both definitions result in approximately the same results.}
\begin{equation} \label{eq:int_fracs}
    I_A = \frac{ a_A \ Z_A^{2-\gamma}}{\sum_A a_A \ Z_A^{2-\gamma}}.
\end{equation}

It is important to note that the emission spectrum as described in \cref{eq:emission} does not necessarily represent the accelerated one since propagation and interaction effects inside the source environment may alter it (see e.g. Refs.~\cite{allard_interactions_2009, globus_uhecr_2015, unger_origin_2015, fang_linking_2018, rodrigues_active_2021, condorelli_testing_2023, muzio_probing_2022}). So, if one determines the free parameters of the source emission ($Q_0$, $\gamma$, $R_\mathrm{cut}$, $I_\mathrm{H}$, $I_\mathrm{He}$, $I_\mathrm{N}$, $I_\mathrm{Si}$, $I_\mathrm{Fe}$) by a fit to the observed data, these describe the state of the spectrum and composition \textit{after} leaving the source environment.

\subsection{Propagation through intergalactic space} \label{sec:propagation}
For the propagation of UHECRs from the sources to Earth, the software \texttt{CRPropa3}~\cite{alves_batista_crpropa_2016} is used. With \texttt{CRPropa3}, in principle, propagation through 4-dimensional time-space can be performed. 
But, because of the stochastic nature of interactions and possible deflections in the extragalactic magnetic field (EGMF), 4-dimensional simulations can take a long time, prohibiting the determination of multiple free model parameters. 
%Since our astrophysical model has several free parameters, performing a full four-dimensional simulation for each possible parameter combination is not feasible. 
% Concretely, the EGMF can influence the energy spectrum and arrival directions measured on Earth.
The origin, strength, and structure of the EGMF are largely unknown~\cite{beck_galactic_2008, durrer_cosmological_2013} and models can differ substantially~\cite{alves_batista_diffusion_2014}. 
Following the propagation theorem~\cite{aloisio_diffusive_2004}, a suppression of the energy spectrum at lower energies is expected due to diffusion effects~\cite{mollerach_magnetic_2013, mollerach_ultrahigh_2019, lemoine_extra-galactic_2005}, depending on the field strength and coherence length of the EGMF. 
In this work, however, we only use the energy spectrum at the highest energies $>10^{19}$\,eV in the likelihood function (see \cref{sec:likelihood}). For that reason, we neglect the suppression effect\footnote{For an investigation of the effect of a structured EGMF on the energy spectrum and shower depth distributions $>10^{18.7}$\,eV see~\cite{d_wittkowski_for_the_pierre_auger_collaboration_reconstructed_2018}. Further studies of the EGMF effect are planned for the future, see also~\cite{j_m_gonzalez_for_the_pierre_auger_collaboration_combined_2023}.}. Regarding the arrival directions, the modeling of the effect of possible deflections by cosmic magnetic fields will be described in \cref{sec:simulated_observables:AD}.
% According to the propagation theorem~\cite{aloisio_diffusive_2004}, a suppression of the flux due to diffusion effects of UHECRs in the EGMF becomes significant if the distances between sources are larger than typical length scales like the Larmor radius $r_L\approx 1.1\(E/\mathrm{EeV})\(\mathrm{nG} / B)\(1/Z)\\mathrm{Mpc}$~\cite{mollerach_magnetic_2013, mollerach_ultrahigh_2019}.
% In this work, only energies above $\sim10$\,EeV are taken into account, so even for iron with $Z=26$ the EGMF suppression can be neglected for magnetic fields below $B / \mathrm{nG} = 1.1\ 10 / 26 \ 1/10 = 0.04$ for typical source distances of $\lesssim 10$\,Mpc. For lighter nuclei, which are rather expected at lower energies due to the injection following a Peters cycle in combination with the hard spectral indices seen in previous analyses~\cite{the_pierre_auger_collaboration_a_aab_combined_2017, e_guido_for_the_pierre_auger_collaboration_combined_2021, the_pierre_auger_collaboration_a_abdul_halim_constraining_2023}, the magnetic field strength $B$ could be even larger without an effect on the energy spectrum above 10\,EeV. Given that the EGMF strength is expected to be below $\mathcal{O}(\mathrm{nG})$ according to Faraday rotation measurements~\cite{pshirkov_new_2016}, we neglect the suppression effect of the EGMF for this work.

Because we neglect the effect of the EGMF on the propagation, we can use 1-dimensional simulations instead of full 4-dimensional ones. Thus, a \textit{propagation database} of 1-dimensional \texttt{CRPropa3} simulations is generated. The database is then reweighted to account for the three-dimensional spatial setup as will be described in \cref{sec:3dsetup}. 
%For other works including specific four-dimensional simulations in simplified astrophysical models we refer to~\cite{d_wittkowski_for_the_pierre_auger_collaboration_reconstructed_2018, van_vliet_extragalactic_2021, vanvliet_extragalactic_2022}.
Interactions with the CMB and EBL are included in the form of photo-pion production, electron pair production, and photodisintegration. To describe the photodisintegration, the \texttt{TALYS} model~\cite{koning_talys_2005, koning_modern_2012} is used. For describing the EBL, the \texttt{Gilmore} model~\cite{gilmore_semi-analytic_2012} is taken. The influence of the interaction model on the results has been examined already in e.g. Refs.~\cite{the_pierre_auger_collaboration_a_aab_combined_2017, e_guido_for_the_pierre_auger_collaboration_combined_2021, the_pierre_auger_collaboration_a_abdul_halim_constraining_2023}. Additionally, the propagation considers nuclear decay, cosmological evolution of sources and background radiation fields, and adiabatic energy losses of the cosmic rays during propagation. For the latter two, the standard cosmological parameters of \texttt{CRPropa3} assuming a flat universe ($H_0=67.3$\,km/s/Mpc, $\Omega_m=0.315$, $\Omega_\Lambda=0.685$)~\cite{alves_batista_crpropa_2016} are used.

Simulations are performed for multiple distances, injected energies, and mass numbers, using $10^4$ injected particles for each bin. The injected energy is binned in 150 logarithmic bins of width $\log_{10}(E_\mathrm{inj}/\mathrm{eV})=0.02$ between $10^{18.0}$ eV and $10^{21.0}$ eV. The light-travel distances $d$ are binned in 118 logarithmic bins between 3\,Mpc and 3342\,Mpc (which approximately corresponds to redshift $z\approx2.2$). Simulations are done for the five representative elements described in \cref{sec:emission}, so that in total this amounts to $5\times118\times150=88{,}500$ simulations of $10^4$ injected particles each.
Upon detection, the arriving UHECRs are grouped into detected mass number bins $A_\mathrm{det} \in (1,\ 2\text{--}4,\ 5\text{--}22,\ 23\text{--}38,\ $\textgreater39) and detected energy bins $E_\mathrm{det} \in (10^{18.0}, 10^{18.02}..., 10^{20.98}, 10^{21.0})$\,eV. In the following, the propagation database will be denoted as $P(E^h_\mathrm{inj}, A^i_\mathrm{inj}, E^j_\mathrm{det}, A^k_\mathrm{det}, d^l)$, where $h, i, j, k, l$ run over the bins described above.
The detected energies are rebinned to larger bins of width $\edet=0.1$ to match the measurements and to be in line with previous analyses~\cite{the_pierre_auger_collaboration_a_aab_combined_2017, the_pierre_auger_collaboration_a_abdul_halim_constraining_2023}. These larger energy bins will then be denoted as $E^e_\mathrm{det} \in (10^{18.0}, 10^{18.1}, ..., 10^{20.4})$\,eV with the running index $e$. The initial finer energy binning is necessary to include systematic uncertainties on the energy spectrum, which will be discussed in \cref{sec:systematic}.

\subsection{Modeling the contributions from individual sources} \label{sec:3dsetup}
The propagation database can now be used to map the injected flux from \cref{eq:emission} to Earth. For that, we calculate the histogram $\mu(E^e_\mathrm{det}, A^k_\mathrm{det}, d^l)$ describing the number of arriving UHECRs in each energy $E^e_\mathrm{det}$ and mass bin $A^k_\mathrm{det}$ from different distances $d^l$. It can be calculated via matrix multiplication as
\begin{equation}
    \frac{\Delta \mu}{\Delta d^l}(E^e_\mathrm{det}, A^k_\mathrm{det}, d^l)= \frac{1}{4 \pi} \sum_h \sum_i Q_\mathrm{inj}(E^h_\mathrm{inj}, A^i_\mathrm{inj}) \ P(E^h_\mathrm{inj}, A^i_\mathrm{inj}, E^e_\mathrm{det}, A^k_\mathrm{det}, d^l) \Delta E_\mathrm{inj}^h.
\end{equation}
% Here, $w_\mathrm{inj}(E^h_\mathrm{inj}, A^i_\mathrm{inj}) \propto J_\mathrm{inj}(E_\mathrm{inj}, A_\mathrm{inj}) \ \Big(\frac{E^h_\mathrm{inj}}{10^{18}~\mathrm{eV}} \Big)$ with the last factor accounting for the logarithmic binning of the injected energies. The proportionality constant is chosen such that $p$ represents a correctly normalized probability as will be described below.

\paragraph{Calculation of the homogeneous background flux:}
From this, the background arrival histogram $\mu_\mathrm{back}$ for the homogeneous background sources can be calculated as
\begin{equation}
    \mu_\mathrm{back}(E^e_\mathrm{det}, A^k_\mathrm{det}) = \sum_l  \frac{\Delta \mu}{\Delta d^l}(E^e_\mathrm{det}, A^k_\mathrm{det}, d^l) \ (1 + z(d^l))^{m-1}  \Delta d^l.
\end{equation}
% The multiplication with $d^l$ again accounts for the logarithmic binning of the distances in the propagation database. 
The function $z(d^l)$ describes the redshift as a function of the distance of the respective bin. The factor $(1 + z(d^l))^{-1}$ accounts for the conversion from light-travel distance $d$ to comoving distance. The additional factor $(1 + z(d^l))^{m}$ takes care of the source evolution as described in \cref{sec:hommodel}.

\paragraph{Calculation of the flux from individual catalog sources:}
For the catalog sources, the distances $d(C^s):=d^s$ and relative flux weights $w_\mathrm{flux}(C^s)$ of the respective candidates $C^s$ from the catalogs described in \cref{sec:candidates} are used. Since the propagation database $P$ contains explicit distance bins $d^l$ and not specifically the source distances $d^s$, a linear interpolation is taken into account to calculate the fluxes for the distances $d^s$ (here denoted by the Kronecker $\delta(d^s-d^l)$):
% \begin{equation}
% \label{eq:weighting_factor}
%      w_\mathrm{interpolate}(d^s, d^l) = \frac{d^l_+ - d^s}{d^l_+ - d^l_-} \ \delta_{d^s, d^l_-} + \frac{d^s - d^l_-}{d^l_+ - d^l_-} \ \delta_{d^s, d^l_+}
% \end{equation}
% Here, $d^l_+$ represents the next higher distance that has been simulated for the propagation database and $d^l_-$ the next lower one seen from the wanted distance $d^s$ of the respective candidate source $C^s$. $\delta$ denotes the Kronecker Delta. 
% Both the interpolation weight and the flux weight factors can now be used to calculate the expected arrivals for every candidate source via
% \begin{equation}
%     \mu_\mathrm{sig}(E^e_\mathrm{det}, A^k_\mathrm{det}, C^s) = \sum_l \mu(E^e_\mathrm{det}, A^k_\mathrm{det}, d^l) \ w_\mathrm{interpolate}(d^s, d^l) \ w_\mathrm{flux}(C^s).
% \end{equation}
\begin{equation}
    \mu_\mathrm{sig}(E^e_\mathrm{det}, A^k_\mathrm{det}, C^s) = \sum_l \frac{\Delta \mu}{\Delta d^l}(E^e_\mathrm{det}, A^k_\mathrm{det}, d^l) \ \delta(d^s-d^l) \ w_\mathrm{flux}(C^s) \Delta d^l
    %\Big( (d_+^l-d^s)\delta_{d^s,d_-^l} + (d_-^l-d^s)\delta_{d^s,d_+^l} \Big) / \Delta d^l.
\end{equation}

% \begin{equation}
%     \mu_\mathrm{sig}(E^e_\mathrm{det}, A^k_\mathrm{det}, C^s) = \frac{1}{4 \pi} \sum_h \sum_i Q_\mathrm{inj}(E^h_\mathrm{inj}, A^i_\mathrm{inj}, C^s) \ \bar{P}(E^h_\mathrm{inj}, A^i_\mathrm{inj}, E^e_\mathrm{det}, A^k_\mathrm{det}, d^s) \delta(d^s-d^l) \Delta E_\mathrm{inj}^h 
% \end{equation}

% where $Q_\mathrm{sig}(E^h_\mathrm{inj}, A^i_\mathrm{inj}, C^s)$ is in units of particles per energy and time:
% \begin{equation}
%     \frac{Q_\mathrm{sig}(E^h_\mathrm{inj}, A^i_\mathrm{inj}, C^s)}{4 \pi (d^s)^2} = Q_\mathrm{inj}(E^h_\mathrm{inj}, A^i_\mathrm{inj}) \ w_\mathrm{flux}(C^s).
% \end{equation}
% Here, $d^l_{+/-} \in d^l$ is the next higher / lower distance to $d^s$ contained in the propagation database, and $\delta$ is the Kronecker delta.
The histogram $\mu_\mathrm{sig}(E^e_\mathrm{det}, A^k_\mathrm{det}, C^s)$ describes the \textit{contribution} of each source to the flux arriving at Earth for each mass number bin in each detected energy bin. Overall, this contribution depends on the flux weight $w_\mathrm{flux}(C^s)$. The energy dependency of the contribution, however, is determined by the source distance in combination with propagation effects.

If the exposure on Earth in the direction of each source $C^s$ would be the same, a simple summation over all candidate sources would lead to the detected energies and masses for CRs from all catalog sources combined. For a ground-based observatory, however, the differing exposure for the source candidates has to be taken into account when computing the arrival histogram. Here, the 3-dimensional setup becomes important, even if one is only interested in calculating energies and charges and not the arrival directions of the incoming particles. The flux contribution of each catalog source due to the non-uniform exposure of the Pierre Auger Observatory depends not only on the source direction, but also on the modeling of the arrival directions and the consideration of magnetic field deflections.

For a fast calculation of the exposure influence, a weight matrix $w_\mathrm{exp}(E^e_\mathrm{det}, A^k_\mathrm{det}, C^s)$ is defined and then multiplied with the arrival histogram $\mu_\mathrm{sig}$ to get the exposure-weighted histogram $\hat{\mu}_\mathrm{sig}$.
% \begin{equation} ~\label{eq:p_arrival_directions}
%     \hat{p}_\mathrm{sig}(E^e_\mathrm{det}, A^k_\mathrm{det}, C^s) = p_\mathrm{sig}(E^e_\mathrm{det}, A^k_\mathrm{det}, C^s) \ w_\mathrm{exposure}(E^e_\mathrm{det}, A^k_\mathrm{det}, C^s)
% \end{equation}
The weights $w_\mathrm{exp}$ for each catalog source, mass number, and energy bin are calculated considering the rigidity-dependent arrival directions modeling which will be described below in \cref{sec:simulated_observables:AD}.
The sum over all catalog sources now gives the arrival histogram for the signal part:
\begin{equation}
    \hat{\mu}_\mathrm{sig}(E^e_\mathrm{det}, A^k_\mathrm{det}) = \sum_s \hat{\mu}_\mathrm{sig}(E^e_\mathrm{det}, A^k_\mathrm{det}, C^s) = \sum_s \mu_\mathrm{sig}(E^e_\mathrm{det}, A^k_\mathrm{det}, C^s) \ w_\mathrm{exp}(E^e_\mathrm{det}, A^k_\mathrm{det}, C^s).
\end{equation}

\paragraph{Definition of a signal fraction:}
% The two histograms $\hat{\mu}_\mathrm{sig}(E^e_\mathrm{det}, A^k_\mathrm{det})$ and $\mu_\mathrm{back}(E^e_\mathrm{det} A^k_\mathrm{det})$ denote the arrival probabilities from either the catalog sources (\textit{signal}) or the background. 
%Note that the weights $w_\mathrm{inj}$ entering the probabilities of the signal and the background are normalized such that
%$\sum_{e,k} p_\mathrm{sig / back}(E^e_\mathrm{det}, A^k_\mathrm{det})=1$.

The two histograms for the catalog sources (\textit{signal}) and the background have to be combined to get the total arrival histogram. For that, another free model parameter $f_0$ is introduced, the \textit{signal fraction} that weights the two parts. To have a signal fraction that is easily comparable to the arrival directions correlation analysis~\cite{the_pierre_auger_collaboration_a_aab_et_al_indication_2018}, $f_0$ is defined as the signal fraction in the detected energy bin $\edet=19.5-19.6$ ($\approx40$\,EeV), corresponding to the running index $e=15$. Accordingly, the two arrival histograms are weighted in the following way:
\begin{equation}
\label{eq:ptot}
    \mu_\mathrm{tot} (E^e_\mathrm{det}, A^k_\mathrm{det}) = \underbrace{f_0 \ \frac{\hat{\mu}_\mathrm{sig}(E^e_\mathrm{det}, A^k_\mathrm{det})}{\sum_k \hat{\mu}_\mathrm{sig}(E^{15}_\mathrm{det}, A^k_\mathrm{det})}}_{s_\mathrm{sig}} + \underbrace{(1-f_0) \ \frac{\mu_\mathrm{back}(E^e_\mathrm{det}, A^k_\mathrm{det})}{\sum_k \mu_\mathrm{back}(E^{15}_\mathrm{det}, A^k_\mathrm{det})}}_{s_\mathrm{back}}:=\mu^{e,k}.
\end{equation}
The catalog contribution as a function of the detected energy bin can be calculated from the two summands of \cref{eq:ptot} as follows:
\begin{equation}
\label{eq:fs}
    f_S(f_0, E^e_\mathrm{det}) = \sum_k \frac{s_\mathrm{sig}}{s_\mathrm{sig} + s_\mathrm{back}}.
\end{equation}
As the catalog sources are on average much closer than the homogeneous background, especially for a strong evolution, $f_S(f_0, E^e_\mathrm{det})$ usually rises with energy.

\subsection{Simulated observables} \label{sec:simulated_observables}
The histogram $\mu_\mathrm{tot}(E^e_\mathrm{det}, A^k_\mathrm{det})$ describes the arriving energies and masses on Earth according to the astrophysical model. In order to find the best-fit free model parameters, the observables predicted from $\mu_\mathrm{tot}(E^e_\mathrm{det}, A^k_\mathrm{det})$ have to be compared to the measured ones on Earth. In our case, these observables are the energy spectrum, shower maxima (\xmax) distributions, and arrival directions as measured at the Pierre Auger Observatory. 

In this section, it will be described how the observables are calculated from the arrival histogram.
For that, the detector effects that the measuring devices induce on the observables have to be considered.
A forward-folding approach will be used to include the detector effects, meaning that the detector response is part of the model so that the comparison between simulations and data is performed at the detector level. Additionally, the systematic uncertainties on the detector effects are taken care of, as will be described in \cref{sec:systematic}.

\subsubsection{Energy spectrum} \label{sec:simulated_observables:E}
% The raw number of events $\mu^e$ expected in each energy bin, before detector effects, can be obtained directly from the arrival histogram at observation $\mu_\mathrm{tot}^{e, k}$ via a summation over $k$.
% \begin{equation}
%     p(E^e_\mathrm{det}) = \sum_k \hat{p}_\mathrm{tot}(E^e_\mathrm{det}, A^k_\mathrm{det})
% \end{equation}
The effects of the measurement with the surface detector (SD) of the Pierre Auger Observatory on the event counts per energy bin can be expressed in terms of an energy resolution, a bias, and an acceptance. Above $10^{18.4}$\,eV, the SD is fully efficient, so the acceptance is 100\% in the energy range used for this analysis. 
% The impact of the resolution can be described by a convolution of the simulated energies with a Gaussian with a standard deviation equal to the SD detector resolution $\sigma_\mathrm{SD}$, and the mean equal to the bias $b_\mathrm{SD}$. This is described in detail in~\cite{the_pierre_auger_collaboration_a_aab_et_al_measurement_2020}. 
We take the response matrix $R^{e, e'}$ from~\cite{the_pierre_auger_collaboration_a_aab_et_al_measurement_2020} which transforms the energy spectrum predicted by the model to the reconstructed one including resolution and bias effects. The spectrum $J(E^e_\mathrm{det})$ is then calculated as
\begin{equation} \label{eq:J_det}
    J(E^e_\mathrm{det}) = \frac{\sum_{e'} R^{e, e'} \ \sum_k \mu^{e',k}}{\mathcal{E}_\mathrm{\theta\leq60^\circ} \ \Delta E^e_\mathrm{det}}
\end{equation}
with the total vertical (zenith angle $\theta\leq60^\circ$) exposure $\mathcal{E}_\mathrm{\theta\leq60^\circ}$, the arrival histogram $\mu^{e,k}$ from \cref{eq:ptot}, and the width of the energy bins $\Delta E^e_\mathrm{det}$. Here, the vertical exposure is taken as we use only vertical events for the energy spectrum data set while for the arrival directions inclined events are also included (see \cref{sec:data_sets}).

\subsubsection{Depth of the shower maximum} \label{sec:simulated_observables:Xmax}
The Gumbel distributions $g(\tilde{X}_\mathrm{max}^x | E^e_\mathrm{det}, A^k_\mathrm{det})$~\cite{de_domenico_reinterpreting_2013} are used to model the expected \xmax distribution for each energy and mass bin as in Ref.~\cite{the_pierre_auger_collaboration_a_aab_combined_2017}. Here, the parameters are taken from the updated generalized version of the Gumbel distributions, as in Ref.~\cite{the_pierre_auger_collaboration_a_abdul_halim_constraining_2023} for the EPOS-LHC~\cite{pierog_epos_2013} hadronic interaction model. The \xmax bins denoted by $\tilde{X}_\mathrm{max}^x$ are given as $\tilde{X}_\mathrm{max}^x \in (550, 570, ..., 1050)\,\text{g/cm}^{-2}$.

The shower maximum depth distributions are influenced by the acceptance $\mathcal{A}$ and the resolution $\mathcal{R}$ of the fluorescence detector (FD) of the Pierre Auger Observatory, with which the longitudinal shower development is observed. The parameterizations of the acceptance and resolution can be found in Ref.~\cite{the_pierre_auger_collaboration_a_aab_et_al_depth_2014}. For both effects, we use the updated parameters as in Ref.~\cite{the_pierre_auger_collaboration_a_abdul_halim_constraining_2023}. 
The influence of the detector effects is folded into the Gumbel distributions in the following way:
\begin{equation} ~\label{eq:gumbel_det}
    g(\xmax^x | E^e_\mathrm{det}, A^k_\mathrm{det}) = \Big(g(\tilde{X}_\mathrm{max}^x | E^e_\mathrm{det}, A^k_\mathrm{det}) \ \mathcal{A}(\tilde{X}_\mathrm{max}^x | E^e_\mathrm{det})\Big) \otimes \mathcal{R}(\xmax^x | \tilde{X}_\mathrm{max}^x, E^e_\mathrm{det}).
\end{equation}

For the comparison of the measured \xmax with the predicted model values, a histogram is produced:
\begin{equation}~\label{eq:p_xmax}
    \mu(E^e_\mathrm{det}, \xmax^x) = c \sum_k \mu(E^e_\mathrm{det}, A^k_\mathrm{det}) \ g(\tilde{X}_\mathrm{max} | E^e_\mathrm{det}, A^k_\mathrm{det}):=\mu^{e,x}.
\end{equation}
Here, $c$ normalizes the right side of the equation over all $\xmax^x$ bins to one.

\subsubsection{Arrival directions} \label{sec:simulated_observables:AD}
In analogy with the energy spectrum and the shower maximum depth distributions, the arrival directions are included as an observable to compare them to the measured data. For that, the arrival directions need to be binned in order to calculate the probability of cosmic rays arriving in different directions for different energies. For the binning, the \texttt{healpy} package~\cite{zonca_healpy_2019} is used which is based on \texttt{HEALPix}\footnote{\url{http://healpix.sf.net}}~\cite{gorski_healpix_2005}. Considering $\texttt{nside}=64$ it divides the sky into $n_\mathrm{pix}=49{,}152$ pixels of the same angular size. The binning approximately corresponds to the detector resolution of $~0.9^\circ$~\cite{the_pierre_auger_collaboration_a_aab_et_al_searches_2015} and can hence be considered to account for this effect.

The modeling of the arrival directions is done separately for the background and the catalog sources. For the background distribution, the arrival directions are expected to be isotropic, so they are modeled following the SD exposure. For that, the exposure $\mathcal{E}_\mathrm{\theta\leq80^\circ}$ up to zenith angle $\theta=80^\circ$ in each \texttt{healpy} pixel $\mathrm{pix}^p$ is calculated and a normalized isotropic arrival map is defined which contains the exposure values of the $49{,}152$ pixels indexed with the running index $p$:
\begin{equation}
    B(\mathrm{pix}^p)=\mathcal{E}_\mathrm{\theta\leq80^\circ}(\mathrm{pix}^p)/\sum_p \mathcal{E}_\mathrm{\theta\leq80^\circ}(\mathrm{pix}^p).
\end{equation}

For the modeling of the arrival directions of the catalog sources, the effect of magnetic fields has to be considered. 
As described above, we neglect the flux suppression induced by diffusion effects in the EGMF. Similarly, the effect of the EGMF on the arrival directions is expected to be small~\cite{hackstein_simulations_2018} as we are only interested in the highest energies in this work. The effect can be parameterized by a beam-widening, based on the assumption that UHECRs at the highest energies travel mostly in the non-resonant scattering regime~\cite{mollerach_ultrahigh_2019}.

Regarding the Galactic magnetic field (GMF), several sophisticated models exist, e.g. Refs.~\cite{jansson_galactic_2012, pshirkov_deriving_2011, sun_radio_2008, jaffe_comparing_2013}. These do, however, differ substantially regarding the predictions of the arrival directions~\cite{erdmann_nuclear_2016, unger_uncertainties_2018}.
Most catalog sources are not located in the direction of the Galactic disk, but rather the Galactic halo, whose coherent component is even less known. Additionally, the effect of the magnetic field of the Local Sheet on the UHECR arrival directions is fairly unknown.
% As this work aims at gaining insights into possible parameters of specific CR source candidates which have been identified in~\cite{the_pierre_auger_collaboration_a_aab_et_al_indication_2018} because of the observed flux overdensities around them. If a GMF model with a strong coherent component was included in the model, the source identification would be broken
As this work aims at investigating scenarios in which source candidates correlating with the UHECR arrival directions are actually the sources of UHECRs, we at this point refrain from including strong coherent deflections which could break the source association\footnote{For alternative studies, taking into account the possibility that the observed overdensities around the catalog sources could actually be generated by different sources whose flux is deflected into the candidate source directions by a stronger coherent Galactic magnetic field model, compare to Refs.~\cite{ding_imprint_2021, eichmann_explaining_2022}}.

Hence, similar to the arrival-directions correlation analysis which originally revealed the overdensities around the catalog sources~\cite{the_pierre_auger_collaboration_a_aab_et_al_indication_2018}, the arrival directions are described using a circular blurring around the source direction (which could be caused by an extragalactic or a mostly turbulent Galactic magnetic field) which is inserted into a \textit{von~Mises--Fisher distribution}~\cite{fisher_dispersion_1953} $F(x | \mu, \kappa)$, the equivalent of a 2-dimensional Gaussian on the sphere. The concentration parameter $\kappa$ describes the width of the distribution. It is linked with the blurring angle $\delta$ via $\kappa=1/\delta^2$. 
In Ref.~\cite{the_pierre_auger_collaboration_a_aab_et_al_indication_2018} this blurring angle is a fixed value for all sources. In this study, however, we include the rigidity dependency that is expected for magnetic field deflections. This only becomes possible in our astrophysical model because the model contains the whole simulation from sources to Earth, and hence a prediction for the rigidities of the arriving particles for each source candidate.

We parameterize the blurring angle $\delta(E^e_\mathrm{det}, A^k_\mathrm{det})$ as
\begin{equation} \label{eq:delta_tot}
    \delta(E^e_\mathrm{det}, A^k_\mathrm{det}) = \frac{\delta_0}{R_\mathrm{det}(E^e_\mathrm{det}, A^k_\mathrm{det}) / 10 \ \mathrm{EV}}
\end{equation}
with the free model parameter $\delta_0$ corresponding to the spread of protons with energies of 10\,EeV (or that of nitrogen with 70\,EeV). 
%Here, $\beta_e$ is intended to describe the beam widening by the extragalactic magnetic field, which scales proportionally to the square-root of the source distance $d$, and 
Here, the detected rigidity $R_\mathrm{det}=E_\mathrm{det}/(e \ Z_\mathrm{det}(A_\mathrm{det}))$ is used, and one should keep in mind that when nuclei photodisintegrate during propagation the rigidity of the leading fragment is essentially unchanged. It is determined from $E_\mathrm{det}$ and $A_\mathrm{det}$ by using the representative atomic numbers $Z_\text{det} = 1$, $2$, $7$, $14$ and~$26$ for the detected mass number bins~$A_\text{det} \in \{1\}$, $\{2, 3, 4\}$, $\{5,...,22\}$, $\{23,...,38\}$ and~$\{39,...,56\}$, respectively.
%Here, the detected mass on Earth is used as we assume the largest deflections happen in the Galactic magnetic field which is small enough so interactions can be neglected for ultra-high energies.

The von~Mises--Fisher distributions are then stored in a histogram 
$F(\mathrm{pix}^p | E^e_\mathrm{det}, A^k_\mathrm{det}, C^s)$ using again the same \texttt{healpy} pixels $\mathrm{pix}^p$ as for the background. The final arrival probability map for the catalog sources is then calculated by matrix multiplication:
\begin{equation} \label{eq:S}
    S(E^e_\mathrm{det}, \mathrm{pix}^p) = c^e \sum_{k, s}  \mu_\mathrm{sig}(E^e_\mathrm{det}, A^k_\mathrm{det}, C^s) \ F(\mathrm{pix}^p | E^e_\mathrm{det}, A^k_\mathrm{det}, C^s) \ \mathcal{E}_\mathrm{\theta\leq80^\circ}(\mathrm{pix}^p)
\end{equation}
with $c^e$ normalizing $\sum_p S(E^e_\mathrm{det}, \mathrm{pix}^p) = 1$ in each energy bin as expected for a probability density. The multiplication in \cref{eq:S} weights each von~Mises--Fisher distribution with the probability of the respective element in the respective energy bin and also takes into account the exposure weights of the different sources. The same von~Mises--Fisher distributions have also been used before to calculate the exposure weighted flux (note that for the spectrum only zenith angles $\theta\leq60^\circ$ are used, see \cref{sec:data_sets}):
\begin{equation}
    w_\mathrm{exp}(E^e_\mathrm{det}, A^k_\mathrm{det}, C^s) = \sum_p F(\mathrm{pix}^p | E^e_\mathrm{det}, A^k_\mathrm{det}, C^s) \ \mathcal{E}_\mathrm{\theta\leq60^\circ}(\mathrm{pix}^p).
\end{equation}

Finally, the probability densities for background and catalog sources are added using the energy-dependent signal fraction function $f_S(f_0, E^e_\mathrm{det})$ from \cref{eq:fs}:
\begin{equation}~\label{eq:pdf}
    \mathrm{pdf}(E^e_\mathrm{det}, \mathrm{pix}^p) = f_S(f_0, E^e_\mathrm{det}) \ S (E^e_\mathrm{det}, \mathrm{pix}^p) + (1-f_S(f_0, E^e_\mathrm{det})) \ B(\mathrm{pix}^p).
\end{equation}
In~\cite{the_pierre_auger_collaboration_a_aab_et_al_indication_2018}, the signal fraction and the magnetic field blurring have a fixed value for all energies above the considered energy threshold, and no energy dependency above the threshold is taken into account. In our astrophysical model, however, the signal fraction function $f_S(f_0, E^e_\mathrm{det})$ is still determined only by one free model parameter $f_0$, but the energy dependency of the signal contribution is generated by propagation effects. The same argument holds for the magnetic field blurring, which is described by one parameter only, $\delta_0$. But, because the rigidity-dependency of the blurring is taken into account, the mean blurring decreases consequently when the rigidity increases with increasing energy. This allows us to accumulate the signal with one likelihood function summing over all energy bins (see \cref{sec:likelihood}), instead of relying on an energy-threshold scan that would have to be penalized for.

\subsection{Consideration of experimental systematic uncertainties} \label{sec:systematic}
Uncertainties on the detector response can induce systematic uncertainties on the observables. The impact of these experimental systematic uncertainties on the analysis results and fitted model parameters can be investigated by introducing nuisance parameters $\nu_{E / \xmax}$ in the astrophysical model, which are here given in units of standard scores.

The energy scale uncertainty can have a significant impact on the fitted model parameters. For the FD, and hence also for the cross-calibrated SD, the resolution is almost energy-independent and around $14\%$~\cite{the_pierre_auger_collaboration_a_aab_et_al_measurement_2020}. A systematic shift of the energy scale, parameterized by the nuisance parameter $\nu_E$, corresponds to a shift of the bin content in the detected energy bins.

The shower maximum depth distributions are influenced by the detector acceptance, resolution, and scale as discussed in \cref{sec:simulated_observables:Xmax}. There are systematic uncertainties on the parameterizations of the acceptance and resolution, but the effect on the observables is very minor as described in Refs.~\cite{the_pierre_auger_collaboration_a_abdul_halim_constraining_2023, the_pierre_auger_collaboration_a_aab_combined_2017}, so they are neglected here. This is mostly due to a cut on the field of view of the FD applied on the measured data, keeping it within a high-acceptance region. The \xmax scale uncertainty is influenced by effects like the calibration and the atmospheric conditions, leading to an energy-dependent uncertainty between  $\pm 6$\,g/cm$^{2}$ and  $\pm 10$\,g/cm$^{2}$~\cite{the_pierre_auger_collaboration_a_aab_et_al_depth_2014}. It was shown in~\cite{the_pierre_auger_collaboration_a_aab_combined_2017, the_pierre_auger_collaboration_a_abdul_halim_constraining_2023} that the hadronic interaction model used in the Gumbel distributions can have a major impact on the fit results, and this can partially be parameterized by a shift of the $\xmax$ scale. The \xmax scale uncertainty is included in the astrophysical model by shifting the \xmax values used for the evaluation of the Gumbel distributions in \cref{eq:gumbel_det} by $\nu_{\xmax}$, taking into account the energy dependence of the scale uncertainty. 

Note that, in contrast with~\cite{the_pierre_auger_collaboration_a_abdul_halim_constraining_2023}, in this work we shift the modeled spectrum and \xmax instead of the data. This is because here our model also contains the forward-folding of the detector effects (instead of the unfolding approach used in~\cite{the_pierre_auger_collaboration_a_abdul_halim_constraining_2023}). However, for better comparability, the same convention as in~\cite{the_pierre_auger_collaboration_a_abdul_halim_constraining_2023} is applied to define the shift direction (a ``positive" versus a ``negative" shift). 
Hence, for example for \xmax, a negative shift of the \xmax scale (implying that the true composition is actually heavier than measured) is represented here by shifting the mean modeled \xmax to larger values.
As the underlying uncertainties concern the data and not the model, in all figures visualizing the effects of experimental systematic uncertainties (\cref{fig:spectra_syst} and \cref{fig:xmax_syst}) we shift the data points instead of the model.
% lighter model composition. Similar results would be reached by shifting the measured \xmax to smaller values, suggesting that the composition is actually heavier than measured.

Experimental systematic uncertainties on the arrival directions are significantly smaller than the size of the magnetic field blurring, and hence they are neglected for this work.

\section{Fitting procedure and data sets} \label{sec:fitting_procedure}
The astrophysical model described in \cref{sec:model} can be used to infer the values of the free model parameters by comparing the predicted observables to the data measured with the Pierre Auger Observatory. The inference methods including the likelihood function will be discussed in \cref{sec:inference}, and the measured data sets will be presented in \cref{sec:data_sets}.

\subsection{Inference of model parameters} \label{sec:inference}
The astrophysical model contains several free parameters ($\gamma$, $R_\mathrm{cut}$, $I_A$, $f_0$, $\delta_0$, $Q_0$) $\in \Theta$ as well as the nuisance parameters $\nu_E$ and $\nu_{\xmax}$.

\subsubsection{Fitting techniques}  \label{sec:fitting_techniques}
The inference of the model parameters is performed with two methods. The first one is a gradient-based minimizer based on \texttt{scipy}~\cite{virtanen_scipy_2020}. It determines the best fit according to the likelihood function $\mathcal{L}(\Theta)$ (maximum-likelihood estimate or MLE).
The second method is a Bayesian Markov-Chain-Monte-Carlo (MCMC) sampler which is used to investigate uncertainties on the fit parameters via posterior distributions $p(\Theta | d, \mathcal{M})$, where $d$ denotes the data and $\mathcal{M}$ the model.
% Bayesian statistics is based on the idea to not only accept or reject a model $\mathcal{M}$ or to find a best-fit model and a confidence for it, but to assign probabilities to the different models. For a parameterized, continuous model, these probabilities become probability distributions of the parameters of interest $\Theta$ dependent on the data $d$ called posterior distributions $p(\theta | d, \mathcal{M})$.
For the estimation of the posterior distributions two ingredients are needed, the likelihood function and the prior distributions $p(\Theta | \mathcal{M})$, according to Bayes theorem:
\begin{equation}
\label{eq:Bayes}
    p(\Theta | d, \mathcal{M}) = \frac{\mathcal{L}(\Theta) \ p(\Theta | \mathcal{M})}{p(d | \mathcal{M})} \propto p(\Theta | \mathcal{M}) \ \mathcal{L}(\Theta).
\end{equation}
Here, $p(d | \mathcal{M}) := b$ is the Bayesian evidence, the probability of the model itself given the data. It is not needed for the estimation of the posterior distributions, but can be used to compare the probabilities of different models~\cite{pullen_bayesian_2014, kass_bayes_1995}. For the sampling of the posterior distributions, we use the Sequential Monte Carlo sampler form \texttt{PyMC}~\cite{salvatier_probabilistic_2016} which can also estimate the Bayesian evidence. Note that the use of Bayesian methods for the identification of UHECR sources was explored before in Refs.~\cite{watson_bayesian_2011, soiaporn_multilevel_2013, capel_impact_2019} with promising outcomes.

\subsubsection{Likelihood function} \label{sec:likelihood}
The total log-likelihood function is given as a sum of the single likelihood functions for each independent observable:
\begin{equation}
\log \mathcal{L} = \log \mathcal{L}_E + \log \mathcal{L}_{\xmax} + \log \mathcal{L}_\mathrm{ADs}
\end{equation}
where ``ADs" is used for the arrival directions.
The likelihood for the energy spectrum $\mathcal{L}_E$ is given as a Poissonian~\cite{the_pierre_auger_collaboration_a_aab_combined_2017}
\begin{equation}
    \log \mathcal{L}_E =  \sum_e \Big( n^e \log(\mu^e) - \log(n^e !) - \mu^e \Big). % := \sum_e l_e(p^e, k^e)
\end{equation}
Here, $n^e$ is the measured number of events per energy bin $e$, and $\mu^e=\sum_{e'} R^{e,e'} \mu^{e'}$ gives the number of events per energy bin $e$ following \cref{eq:J_det}, predicted by the model after detector effects.
The minimum energy for the combined fit of energy spectrum and shower maximum depth distributions above the ankle with a homogeneous background model only was set to $\edet=18.7$ in previous works~\cite{the_pierre_auger_collaboration_a_aab_combined_2017, d_bergman_for_the_telescope_array_collaboration_telescope_2021}. In Ref.~\cite{the_pierre_auger_collaboration_a_abdul_halim_constraining_2023} it was shown that an additional sub-ankle component can still have an influence around $\edet=18.7$, and that the extragalactic component only completely dominates the flux above $\sim10$\,EeV. For that reason, the minimum energy is set to $\edet=19.0$ in this work. Nevertheless, it is enforced that the flux predicted by the astrophysical model between $\edet=18.7$ and $\edet=19.0$ does not exceed the measured one by using a one-sided Poissonian likelihood for these energy bins.

The shower maxima $X_\mathrm{max}$ information is binned into the histogram $\mu^{\tilde{e},x}$ following \cref{eq:p_xmax}. Due to the small statistics of the \xmax measurements with the FD at the highest energies, the bins above $\edet=19.6$ are combined into one, consequently denoted by $\tilde{e}$ instead of $e$.
Since the energy spectrum information is already incorporated by the energy likelihood and should not have an influence twice, the \xmax likelihood is given as a multinomial instead of a Poissonian
\begin{equation}
\label{eq:likelihood_xmax}
    \mathcal{L}_{\xmax} = \prod_{\tilde{e}} n^{\tilde{e}}! \prod_x \ \frac{(\mu^{\tilde{e},x})^{n^{\tilde{e},x}}}{n^{\tilde{e},x}!}.
\end{equation}
For the \xmax distributions, the minimum energy is also set to $\edet=19.0$.

For the arrival directions, a similar likelihood function as in Ref.~\cite{the_pierre_auger_collaboration_a_aab_et_al_indication_2018} is used. The only difference is that the events are binned into the detected energy bins $e$, and hence that the modeled energy-dependent probability maps $\mathrm{pdf}^{e,p}$ from \cref{eq:pdf} are used. Then, the pdf value is read out at the arrival directions of the measured data $n^{e, p}$, also binned in energy and pixels. This leads to the likelihood function
\begin{equation}
\label{eq:likelihood_ad}
    \mathcal{L_\text{ADs}} = \prod_e \prod_p (\mathrm{pdf}^{e, p})^{n^{e, p}}.
\end{equation}

For the arrival directions, a minimum energy of $\edet=19.2$ is used which is higher than that for energy and \xmax in the likelihood. The reason for this is mainly that at $E=10^{19.0}$\,eV the dipole~\cite{the_pierre_auger_collaboration_a_aab_et_al_observation_2017} is very prominent. With our simplified treatment of the foreground and background sources, the dipole is not expected to be described properly by the astrophysical model. Additionally, with a minimum energy of $E=10^{19.2}\,\text{eV}\approx16$\,EeV, this analysis is in line with the other recent arrival directions analyses we performed~\cite{the_pierre_auger_collaboration_p_abreu_arrival_2022, r_de_almeida_for_the_pierre_auger_collaboration_large-scale_2021}, and can use the same data set for better comparison.

As the experimental systematic uncertainties on the energy and \xmax scale, $\nu_E$ and $\nu_{\xmax}$, are expressed in units of standard deviations, the nuisance parameters $\nu$ are expected to follow a Gaussian likelihood with mean $0$ and standard deviation $1$
\begin{equation}
\label{eq:likelihood_syst}
    \mathcal{L}_\mathrm{syst} = \frac{1}{\sqrt{2\pi}} \exp (-\nu^2 / 2).
\end{equation}
When the experimental systematic uncertainties are considered in the fit, $\mathcal{L}_\mathrm{syst}$ has to be included in the total likelihood (\cref{eq:likelihood_xmax}).

For the Poissonian energy and the multinomial \xmax likelihood, a \textit{deviance} $D$ can be defined which characterizes the goodness of fit~\cite{lindsey_parametric_1996}. It is given as twice the negative likelihood ratio between the fitted model and the saturated model $\mathcal{L}^\mathrm{sat}$ which describes the data perfectly.
For the arrival directions, the definition of a deviance is not necessarily useful. This is because a saturated model would not have a physical meaning: setting the arrival probability histogram $\mathrm{pdf}^{e, p}$ equal to the measured arrival directions $v^{e, p}$ would lead to a map of only very sparsely filled pixels, which is not in agreement with the physics of UHECR sources and the assumptions of the astrophysical model. For that reason, in the following, the deviance will be stated only for energy, $\xmax$, and systematics, and the likelihood value will be given for the arrival directions.

\subsubsection{Prior distributions} \label{sec:priors}
For the spectral index $\gamma\in[-4, 3]$ and the logarithmic rigidity cutoff \rcut$\in[18.0, 20.5]$, flat priors are used as in previous work~\cite{the_pierre_auger_collaboration_a_aab_combined_2017}. The lower border for $\gamma$ is extended from $-3$ to $-4$ in this work because harder best-fit values for the spectral index were observed than in Ref.~\cite{the_pierre_auger_collaboration_a_aab_combined_2017}. The signal fraction prior is flat $f_0\in[0, 50\%]$. Before, we also tested the range $f_0\in[0, 100\%]$, but found that models with very large signal fractions cannot recreate the findings of Refs.~\cite{the_pierre_auger_collaboration_a_aab_et_al_indication_2018, the_pierre_auger_collaboration_p_abreu_arrival_2022}, so the maximum allowed signal fraction was accordingly set to 50\%.
The blurring prior is also flat and can have values $\delta_0 \in [0^\circ, 86^\circ$]. The upper border value (1.5 rad) corresponds, for reasonable UHECR compositions (e.g. nitrogen at $\approx40~\mathrm{EeV}$), to an almost isotropic distribution of arrival directions at Earth.

For the elemental fractions, the sum over the five representative fractions has to return unity. This circumstance can be included in the prior distributions by using the unit simplex~\cite{onn_generating_2011}.
Instead of using the elemental fractions $a_A$ defined below the cutoff for the fit as in Ref.~\cite{the_pierre_auger_collaboration_a_aab_combined_2017}, in this work the integral fractions $I_A \in [0, 1]$ (\cref{eq:int_fracs}) will be utilized. In the case of very hard spectral indices even extremely small fractions $a_A$ can still lead to large integrated fractions, and hence large total emissions of the respective element. It was tested that the sensitivity of the fit in that case is limited due to the values of $a_A$ spanning over multiple orders of magnitude. This can introduce an unwanted bias on the posterior distributions.

The generation rate normalization $Q_0$ is a special case as it only contributes to the energy likelihood, so it can be deduced simply by normalizing the number of events predicted by the model to the one in the data (uninformative prior). 

Due to the Gaussian likelihood definition for the systematic uncertainties (\cref{eq:likelihood_syst}), their expected distribution must not be taken into account again in the prior, so a uniform prior between $-4\sigma$ and $4\sigma$ is used to keep the nuisance parameters in a reasonable range.

% The prior distributions, as well as the borders of the parameter space, are summarized in table~\ref{tab:priors}.
% %\renewcommand{\arraystretch}{1.5} % Default value: 1
% \begin{table}[htb]
%   \centering
%     \begin{tabular}{| l | l | l |}
%       \hline
%       \textbf{fit parameter} & \textbf{prior} & \textbf{borders}\\ % <-- 
%       \hline
%       \hline
%       spectral index $\gamma$ & uniform & $-4$ to $3$\\ % <--
%       rigidity cutoff $\rcut$ & uniform & $19.0$ V to $20.5$ V \\ % <--
%       element fractions $I_A$ & unit simplex & 0 to 1\\ %
%       flux normalization $J_0$ & uninformative & \\
%       \hline
%       signal fraction $f_0$ & uniform & 0 to 1 \\ % <--
%       local magnetic field blurring $\delta_0$ & uniform & 0 to 1.5 (0.0$^\circ$ - 85.9$^\circ$)\\ %
%       distance-dependent magnetic field blurring $\beta_e$ & uniform & 0 to 0.2 (0.0$^\circ$ - 11.5$^\circ$)\\ %
%       \hline 
%       syst. energy scale $\nu_E$ & uniform & -4 to 4 \\
%       syst. \xmax scale $\nu_{\xmax}$ & uniform & -4 to 4 \\
%       \hline
%     \end{tabular}
% \caption{Prior distributions}
% \label{tab:priors}
% \end{table}

\subsection{The data sets}
%for energy spectrum, shower maximum distributions, and arrival directions}
\label{sec:data_sets}
The Pierre Auger Observatory~\cite{the_pierre_auger_collaboration_a_aab_et_al_pierre_2015} is located in Malargüe, Argentina. It uses a hybrid detector design.
For the energy spectrum, the measurements of the surface detector taken between 01/2004 and 08/2018 with zenith angles below $60^\circ$ from Ref.~\cite{the_pierre_auger_collaboration_a_aab_et_al_measurement_2020} are used. The exposure of the data set is $60{,}426$ km$^2$~sr~yr. Above $10^{18.7}$\,eV, the data set contains $55{,}730$ events. 
% As described in \cref{sec:likelihood}, the data set and the simulations are compared in energy bins $\log_{10}(E_\mathrm{det})=(18.7, 18.8, 18.9, ...20.4)$, so the measured data are binned similarly. 
No unfolding of the detector effects is necessary for this study because the astrophysical model contains a forward folding of these effects as described in \cref{sec:simulated_observables:E}.

The shower maximum depth distributions are taken from the measurements of the fluorescence detector of the Pierre Auger Observatory~\cite{a_yushkov_for_the_pierre_auger_collaboration_mass_2019}. Above $10^{19.0}$\,eV, the data set contains 1022 events which are binned into a 2-dimensional histogram in energy and \xmax as described in \cref{sec:likelihood}.

For the arrival directions, the data set is based on the one from~\cite{the_pierre_auger_collaboration_p_abreu_arrival_2022}, containing the largest set of events from Phase 1 of the Pierre Auger Observatory. It consists of SD data measured between 01/01/2004 and 31/12/2020. In total, the data set contains $12{,}606$ events above the minimum energy of 16\,EeV, so it also extends to lower energies than the one described in~\cite{the_pierre_auger_collaboration_p_abreu_arrival_2022} ($>32$\,EeV). The data set combines both vertical (zenith angle smaller than $60^\circ$) and inclined events (zenith angle larger than $60^\circ$ and up to $80^\circ$).
The exposure amounts to $95{,}700$~km$^2$~sr~yr for the vertical subset and $26{,}300$~km$^2$ sr yr for the inclined subset.

\section{Fit results}  \label{sec:results}
In the following, the fit results will be presented, starting with the models containing SBGs or Centaurus A in \cref{sec:SBG+CenA}. The results of the $\gamma$-AGN model will be discussed separately in \cref{sec:AGNs}. For comparison, the results of the reference models with only homogeneous background are given in \cref{sec:ref}.
%A comparison of the models in regard to how well they describe the data will be given combined for all tested models in \cref{sec:model_performances}.

\subsection{Starburst galaxy and Centaurus A models} \label{sec:SBG+CenA}
\subsubsection{Fitted model parameters}
\begin{figure}[ht]
\centering
\includegraphics[width=0.9\textwidth]{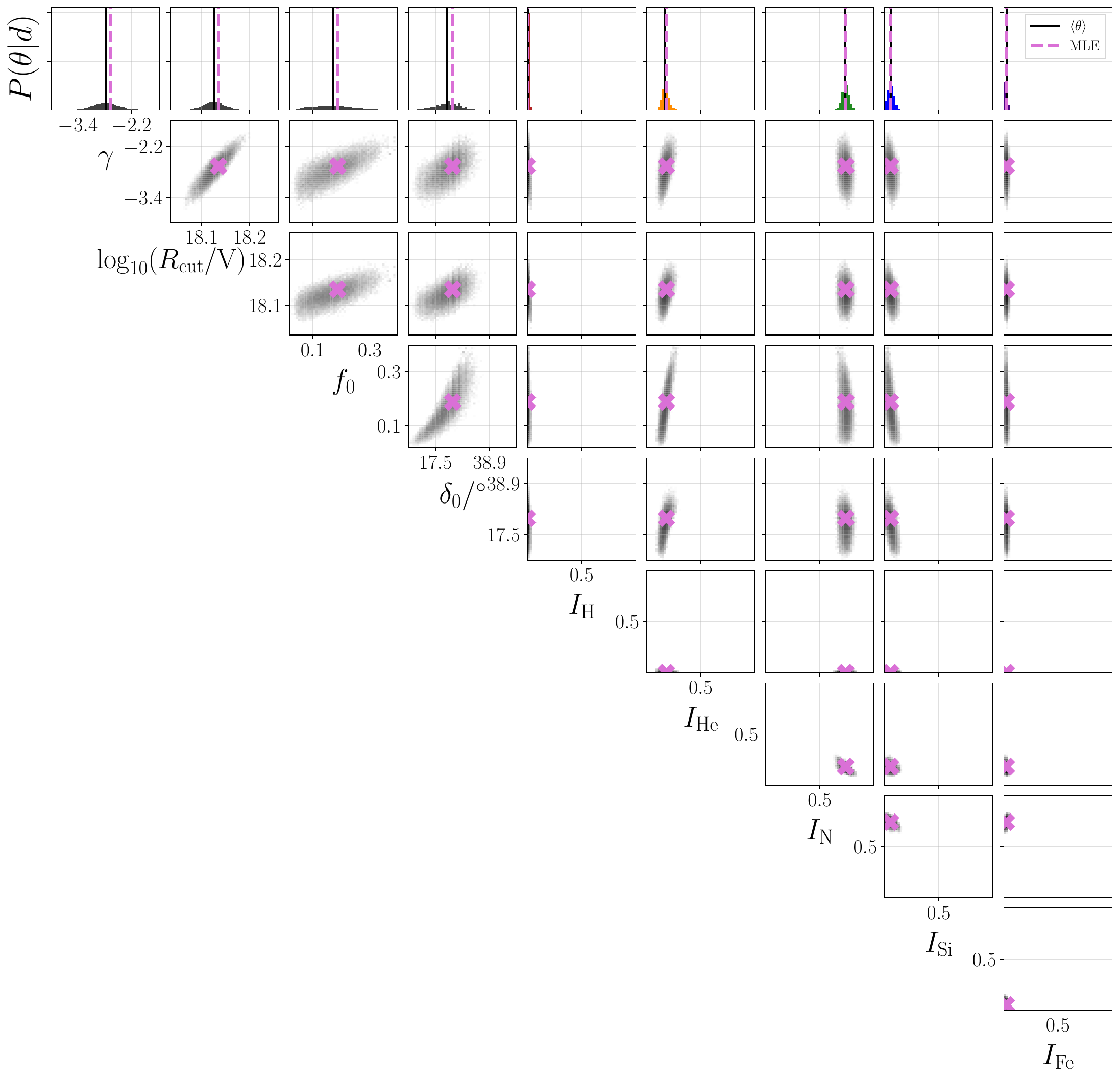}
\caption{Posterior distributions for the SBG model. The pink cross marks the MLE, the black vertical line the posterior mean.}
\label{fig:posterior}
\end{figure}

% \subsubsection{Best fit parameters}
The fit results for the SBG and Centaurus A models are given in \cref{tab:results}. Here, the best fit (MLE) is stated including the respective deviance and likelihood values. Additionally, the posterior mean and the borders of the highest posterior density interval\footnote{Defined as the shortest interval in which $68\%$ of the posterior resides. Note that these borders can be asymmetric with respect to the posterior mean for not fully-Gaussian posterior distribution. So, the values stated in the tables should not be confused with $1\sigma$ uncertainties.} are provided. As an example for posterior distributions, \cref{fig:posterior} depicts them for the SBG model. For all models, a strong correlation between the spectral index $\gamma$ and the maximum rigidity \rcut is visible as in Ref.~\cite{the_pierre_auger_collaboration_a_aab_combined_2017}, as well as a correlation between the signal fraction $f_0$ and the blurring $\delta_0$ as expected from Ref.~\cite{the_pierre_auger_collaboration_a_aab_et_al_indication_2018}. The integral composition fractions are typically almost uncorrelated with the other parameters.

\renewcommand{\arraystretch}{1.2} % Default value: 1
\begin{table}[th]
\centering
\resizebox{\textwidth}{!}{%
\begin{tabular}{l | l l | l l | l l }
& \multicolumn{2}{c|}{\textbf{Cen A, }$\boldsymbol{m=0}$ (flat)} & \multicolumn{2}{c|}{\textbf{Cen A, }$\boldsymbol{m=3.4}$ (SFR)} & \multicolumn{2}{c}{\textbf{SBG, }$\boldsymbol{m=3.4}$ (SFR)}  \\ 
& posterior & MLE & posterior & MLE & posterior & MLE\\
\hline \hline
$\gamma$ & $-1.67_{-0.47}^{+0.48}$ & $-2.21$ & $-3.09_{-0.24}^{+0.23}$ & $-3.05$ & $-2.77_{-0.29}^{+0.27}$ & $-2.67$ \\
$\log_{10} (R_\mathrm{cut}$/V) & \phantom{+}$18.23_{-0.06}^{+0.04}$ & \phantom{+}18.19 & \phantom{+}$18.10_{-0.02}^{+0.02}$ & \phantom{+}18.11 & \phantom{+}$18.13_{-0.02}^{+0.02}$ & \phantom{+}18.13 \\
$f_0$ & \phantom{+}$0.16_{-0.14}^{+0.06}$ & \phantom{+}$0.028$ & \phantom{+}$0.05_{-0.03}^{+0.01}$ & \phantom{+}$0.028$ & \phantom{+}$0.17_{-0.08}^{+0.06}$ & \phantom{+}0.19 \\
$\delta_0 / ^\circ$ & \phantom{+}$56.5_{-12.8}^{+29.4}$ & \phantom{+}16.5 & \phantom{+}$27.6_{-16.3}^{+2.7}$ & \phantom{+}16.8 & \phantom{+}$22.2_{-4.0}^{+5.3}$ & \phantom{+}24.3 \\
$I_\mathrm{H}$ & $5.9_{-1.7}^{+2.5} \times 10^{-2}$ & $7.1\times 10^{-2}$ & $8.3_{-8.3}^{+2.0} \times 10^{-3}$ & $1.6\times 10^{-5}$ & $6.4_{-6.4}^{+1.3} \times 10^{-3}$ & $4.3\times 10^{-5}$ \\
$I_\mathrm{He}$ & $2.3_{-0.5}^{+0.3} \times 10^{-1}$ & $1.9\times 10^{-1}$ & $1.3_{-0.2}^{+0.2} \times 10^{-1}$ & $1.4\times 10^{-1}$ & $1.7_{-0.4}^{+0.3} \times 10^{-1}$ & $1.8\times 10^{-1}$ \\
$I_\mathrm{N}$ & $6.3_{-0.3}^{+0.3} \times 10^{-1}$ & $6.2\times 10^{-1}$ & $7.4_{-0.3}^{+0.3} \times 10^{-1}$ & $7.3\times 10^{-1}$ & $7.4_{-0.3}^{+0.3} \times 10^{-1}$ & $7.4\times 10^{-1}$ \\
$I_\mathrm{Si}$ & $6.5_{-3.3}^{+3.6} \times 10^{-2}$ & $9.9\times 10^{-2}$ & $9.2_{-2.3}^{+3.2} \times 10^{-2}$ & $1.1\times 10^{-1}$ & $5.7_{-3.1}^{+2.5} \times 10^{-2}$ & $5.4\times 10^{-2}$ \\
$I_\mathrm{Fe}$ & $1.6_{-1.0}^{+0.7} \times 10^{-2}$ & $2.0\times 10^{-2}$ & $2.5_{-0.9}^{+0.8} \times 10^{-2}$ & $2.3\times 10^{-2}$ & $2.5_{-0.9}^{+0.8} \times 10^{-2}$ & $2.3\times 10^{-2}$ \\
\hline
$\boldsymbol{\log b }$ & $-264.0 \pm 0.2$ &  & $-272.6 \pm 0.2$ &  & $-266.9  \pm 0.1$ &  \\
$\boldsymbol{D_E} \ (N_J=14)$ &  & \phantom{+}22.3 &  & \phantom{+}28.5 &  & \phantom{+}33.3 \\
$\boldsymbol{D_{X_\mathrm{max}}} \ (N_{{X}_\mathrm{max}}=74)$ &  & \phantom{+}124.9 &  & \phantom{+}130.6 &  & \phantom{+}126.2 \\
$\boldsymbol{D}$ &  & \phantom{+}147.2 &  & \phantom{+}159.1 &  & \phantom{+}159.5 \\
$\boldsymbol{\log}$ $\boldsymbol{\mathcal{L}_\mathrm{ADs}}$ &  & \phantom{+}$10.5$ &  & \phantom{+}$10.4$ &  & \phantom{+}$13.3$ \\
$\boldsymbol{\log}$ $\boldsymbol{\mathcal{L}}$ &  & $-239.1$ &  & $-245.1$ &  & $-242.4$ \\
\end{tabular}}
\caption{Fit results for the Centaurus A and SBG models. The best-fit (MLE) parameters and the corresponding deviance $D$ and log-likelihood values $\log \mathcal{L}$ are stated. Also, the posterior mean and highest posterior density interval from the MCMC sampler are given, as well as the logarithmic Bayesian evidence $\log b$ (see text). Note that in this and all following tables the value for the arrival directions likelihood $\log \mathcal{L}_\mathrm{ADs}^\mathrm{ref}$ for fully isotropic arrival directions (as in the reference models, see \cref{tab:results_ref}) has been set to 0 for better readability of the values of $\log b$, $\log \mathcal{L}$, and $\log \mathcal{L}_\mathrm{ADs}$.}
\label{tab:results}
\end{table}

From the reference models without catalog sources (see \cref{sec:ref} and \cref{fig:TS_all_overview} and compare to Ref.~\cite{the_pierre_auger_collaboration_a_abdul_halim_constraining_2023}) it is clear that a strong source evolution with $m\gtrsim5$ is disfavored by the data. This is due to the great amount of low-energy secondaries from photodisintegration which overshoot the measured spectrum at low energy. They are produced over the large propagation distances when the background sources emit predominantly at large redshifts for a strong source evolution.
The deviance for flat evolution is smaller than that for SFR by around $\Delta D\approx 10$. The maximum rigidity is relatively small with $\rcut \approx18.2$~V. The spectral index is very hard already for flat evolution, but can become as hard as $\gamma\approx-3$ for the SFR case, as also found in~\cite{the_pierre_auger_collaboration_a_abdul_halim_constraining_2023}. The hard injection spectrum allows for a good description of the pronounced features of the energy spectrum, as well as the small \xmax variance. As described above, $\gamma$ is not the spectral index of the spectrum at acceleration, but rather after leaving the source environment. Thus, in-source interactions or magnetic field confinement in the source could lead to hard best-fit values of $\gamma$ even for shock acceleration scenarios for which a softer injection $\propto E^{-2}$ is expected~\cite{unger_origin_2015, condorelli_testing_2023, winchen_energy_2018}. Other explanations could be a low-energy flux suppression by the EGMF~\cite{j_m_gonzalez_for_the_pierre_auger_collaboration_combined_2023, d_wittkowski_for_the_pierre_auger_collaboration_reconstructed_2018}, or the effect of systematic uncertainties (see below).

\subsubsection{Influence of the catalog sources on the observables at Earth}
When adding Centaurus A as a single source on top of the background, the injection spectrum softens slightly compared to the reference model for both tested evolutions (a strong evolution with $m=5.0$ still leads to a poor description of spectrum and composition when Centaurus A is included, so it is not further discussed here, see however \cref{fig:TS_all_overview}). The best-fit signal fraction (MLE) associated with Centaurus A is estimated to be around $f_0\approx3\%$ for both evolutions. But, especially for the flat evolution case, the posterior distributions for signal fraction and blurring are quite broad. This leads to a posterior mean value of $f_0=0.16^{+0.06}_{-0.14}$ in the case of flat evolution, and $f_0=0.05^{+0.01}_{-0.03}$ for SFR evolution.

\begin{figure}[th]
\centering
\includegraphics[width=0.32\textwidth]{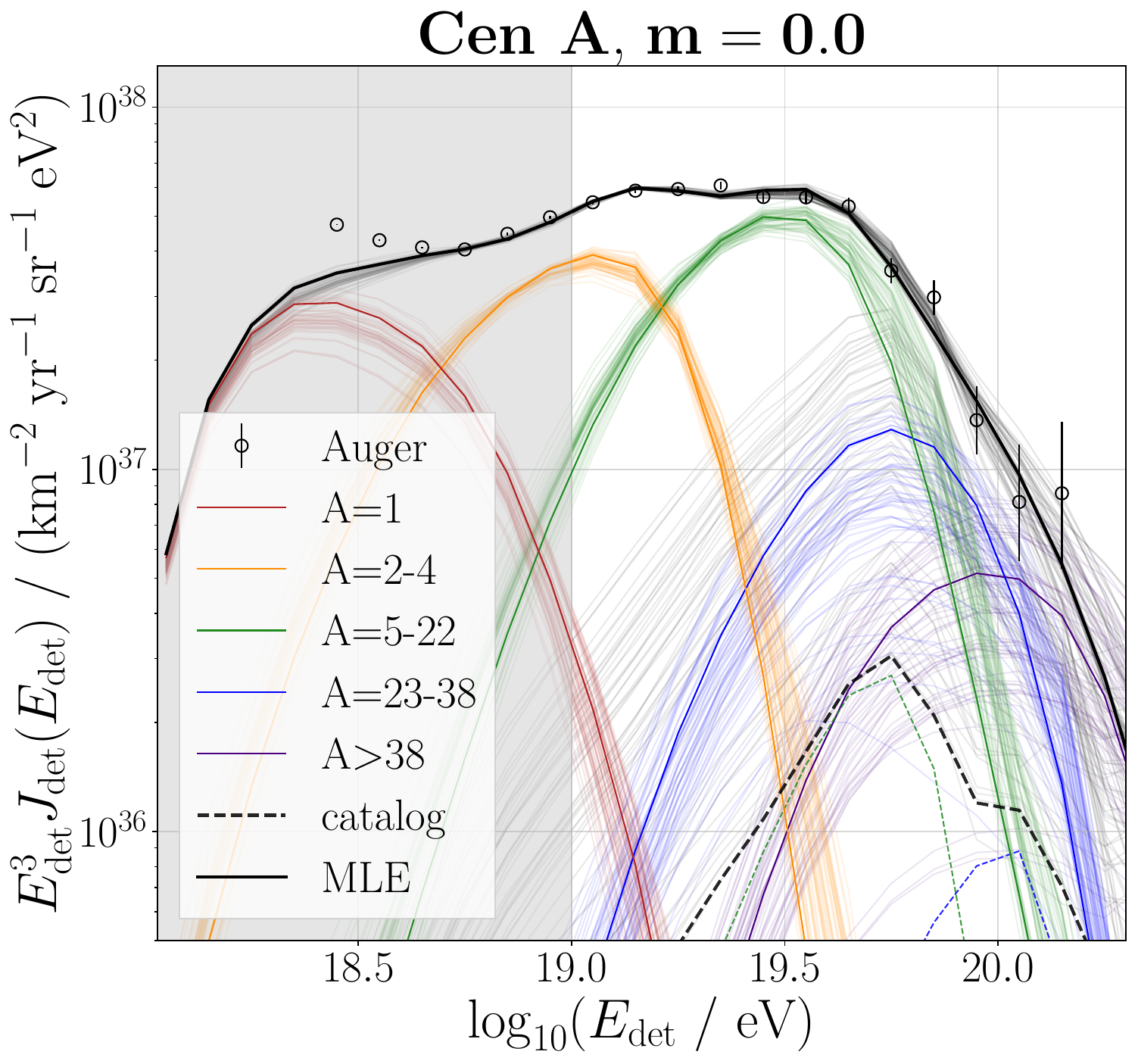}
\includegraphics[width=0.32\textwidth]{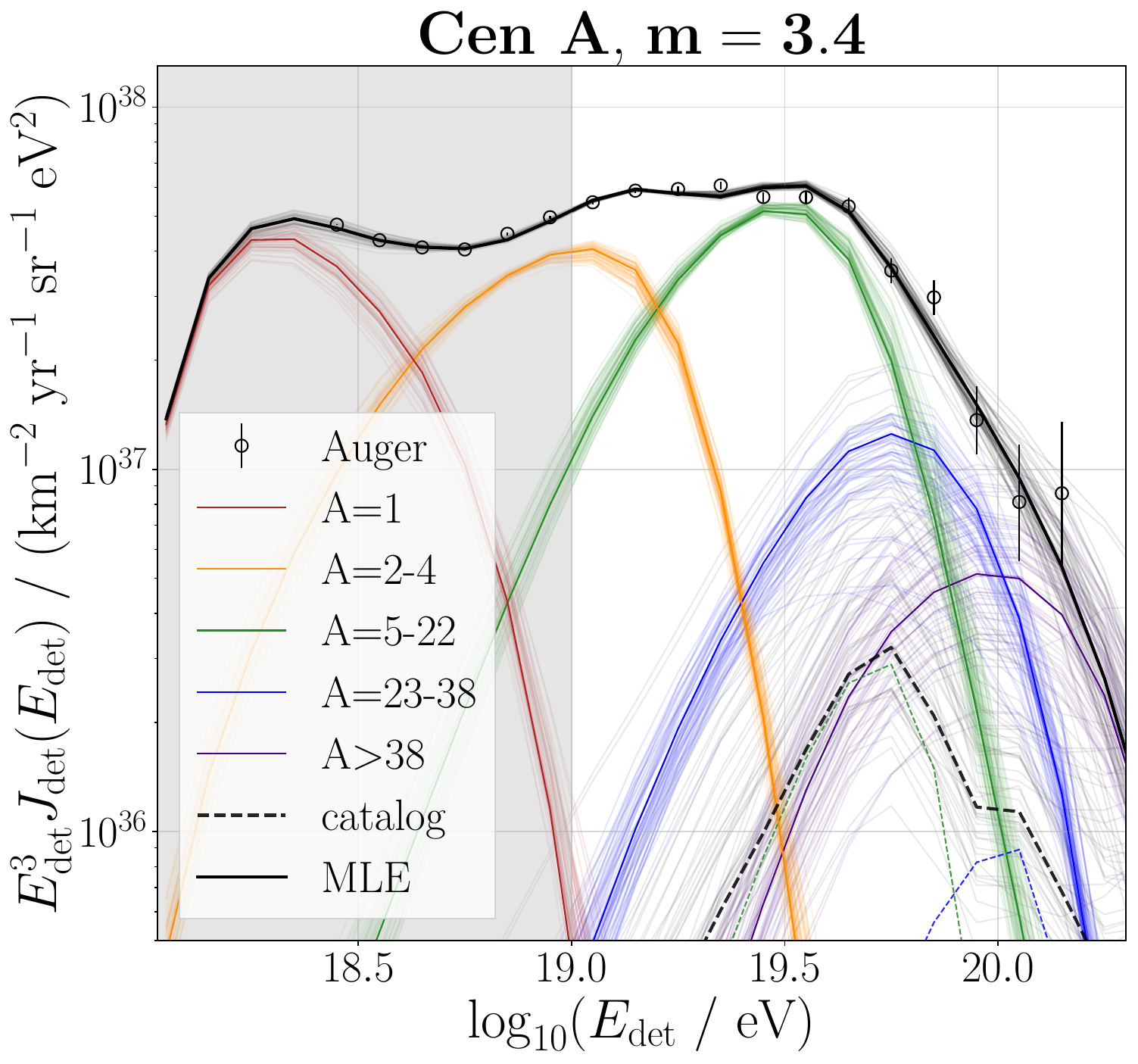}
\includegraphics[width=0.32\textwidth]{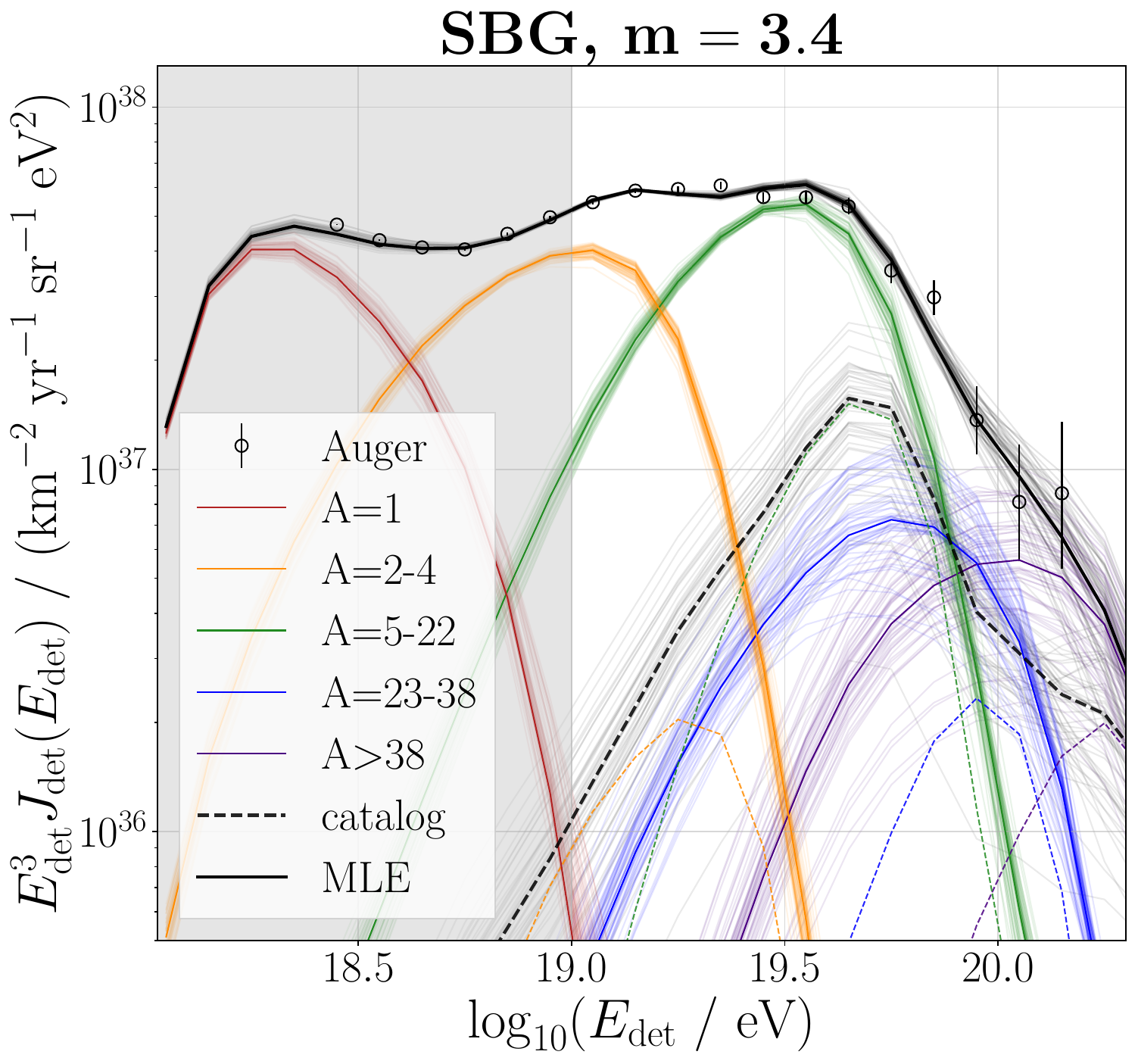}
\includegraphics[width=0.32\textwidth]{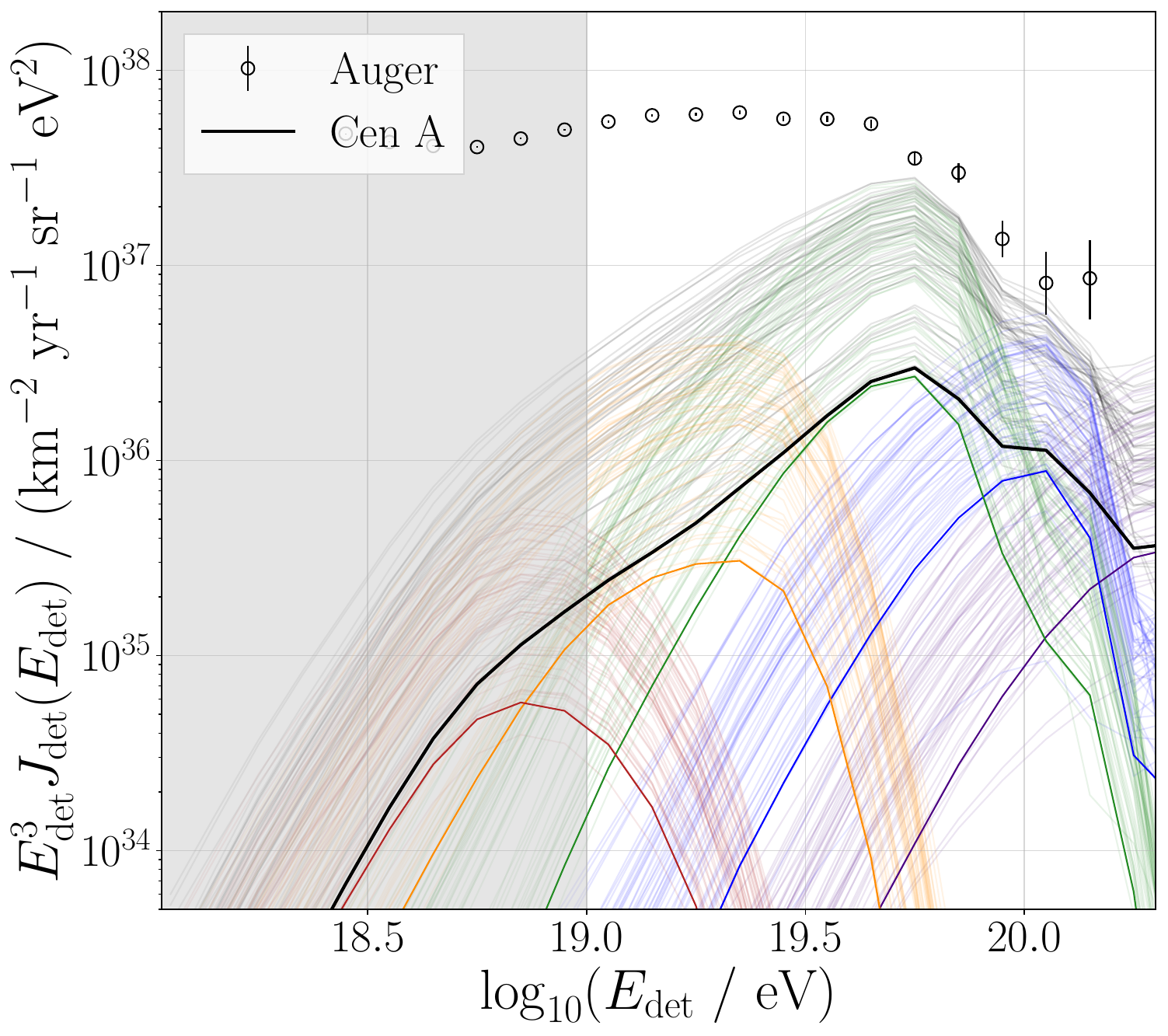}
\includegraphics[width=0.32\textwidth]{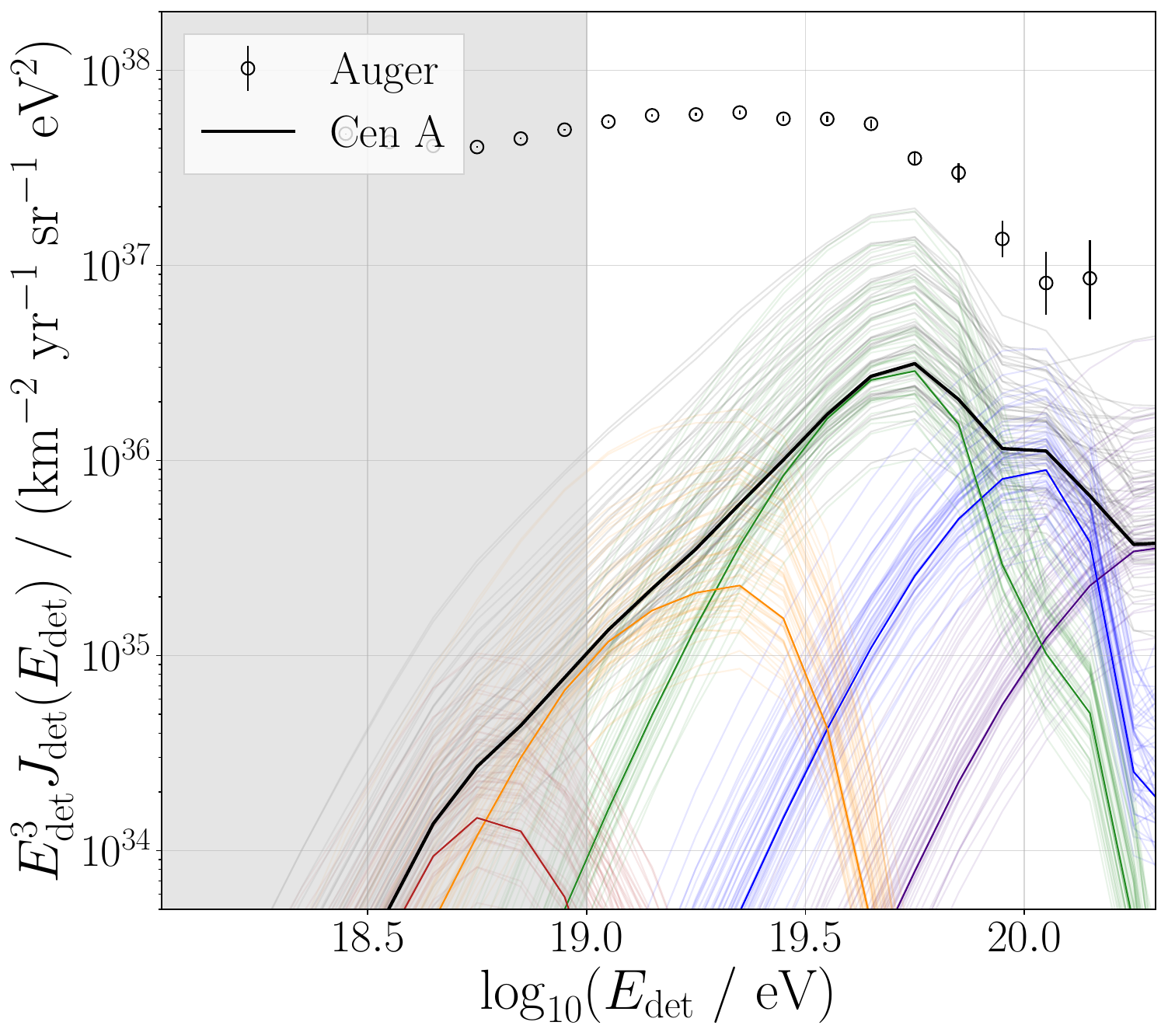}
\includegraphics[width=0.32\textwidth]{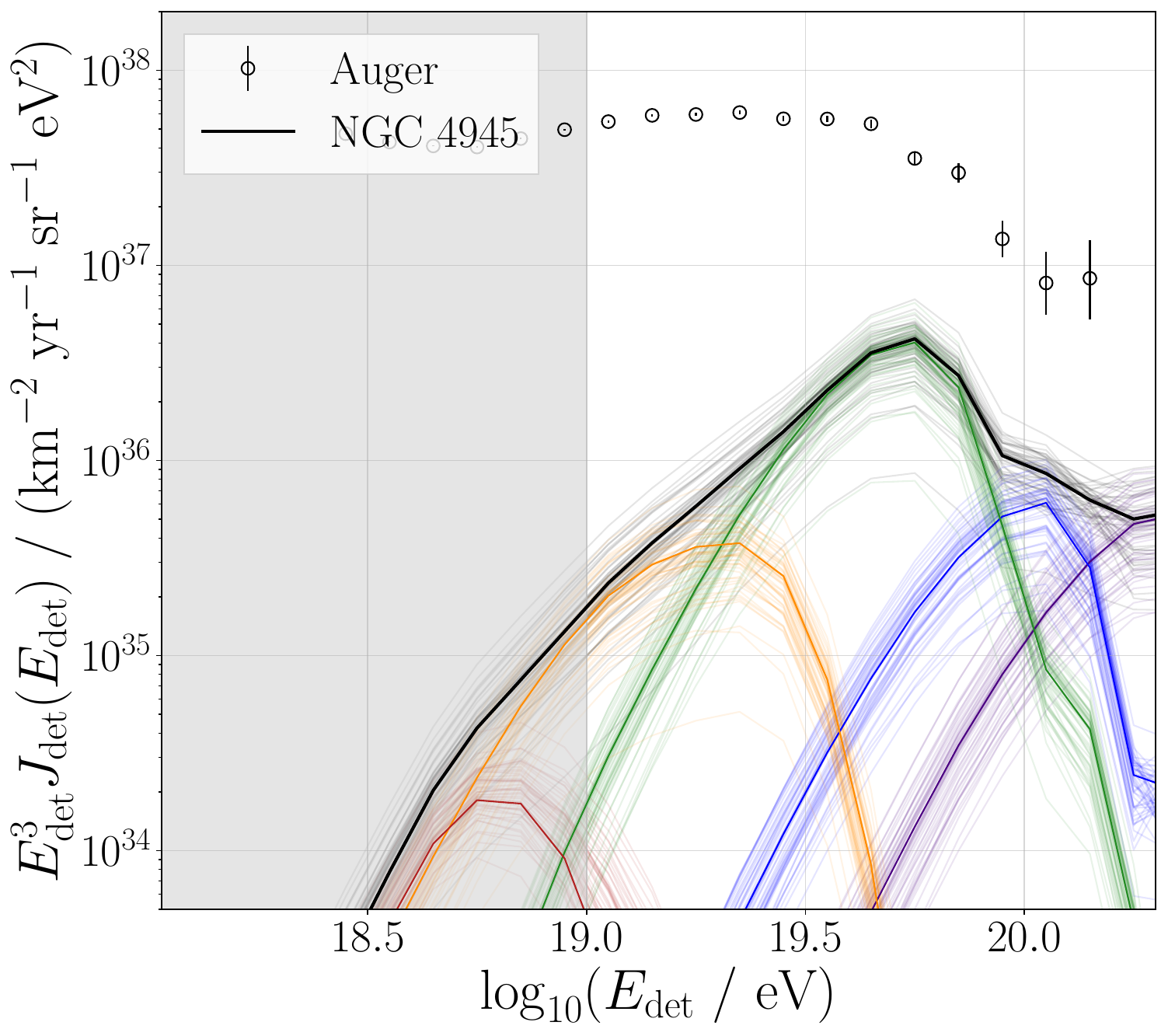}
\caption{Modeled spectra on Earth for the Centaurus A model with $m=0$ (\textit{left}), with $m=3.4$ (\textit{middle}) and the SBG model with $m=3.4$ (\textit{right}). In this and all following figures the thick lines indicate the best-fit and the thin lines are drawn from the posterior distribution demonstrating the uncertainty. The markers represent the measured data of the Pierre Auger Observatory. The grey area symbolizes the energy bins which are not (fully) included in the fit (see \cref{sec:likelihood}). In the \textit{upper} row the total spectrum is depicted with the different element contributions in different colors. The contribution by the source catalog is indicated by the dashed line. The \textit{lower} row shows the individual spectrum of the strongest source in the respective catalog.}
\label{fig:spectra}
\end{figure}

\begin{figure}[ht]
\centering
\includegraphics[width=0.32\textwidth]{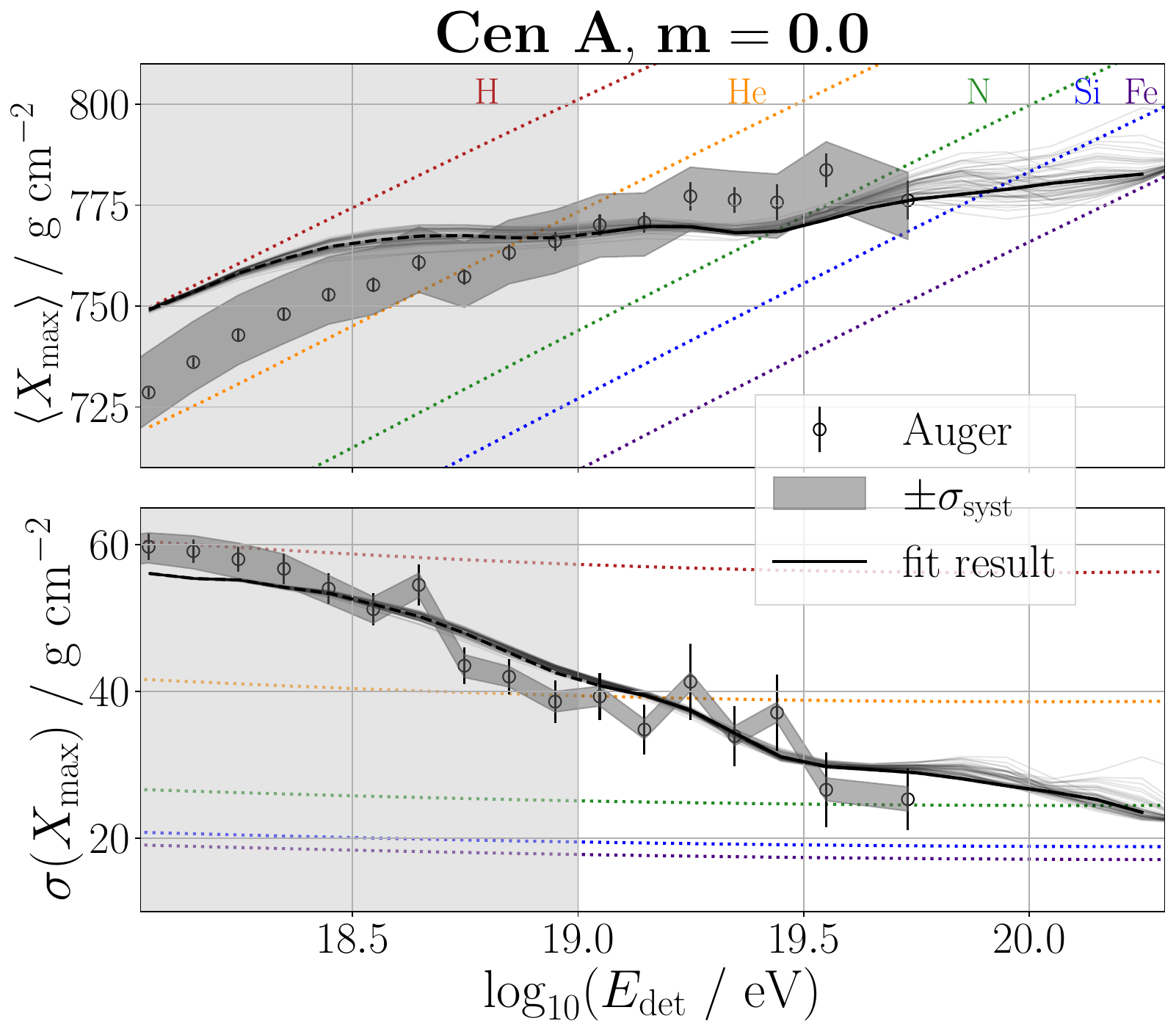}
\includegraphics[width=0.32\textwidth]{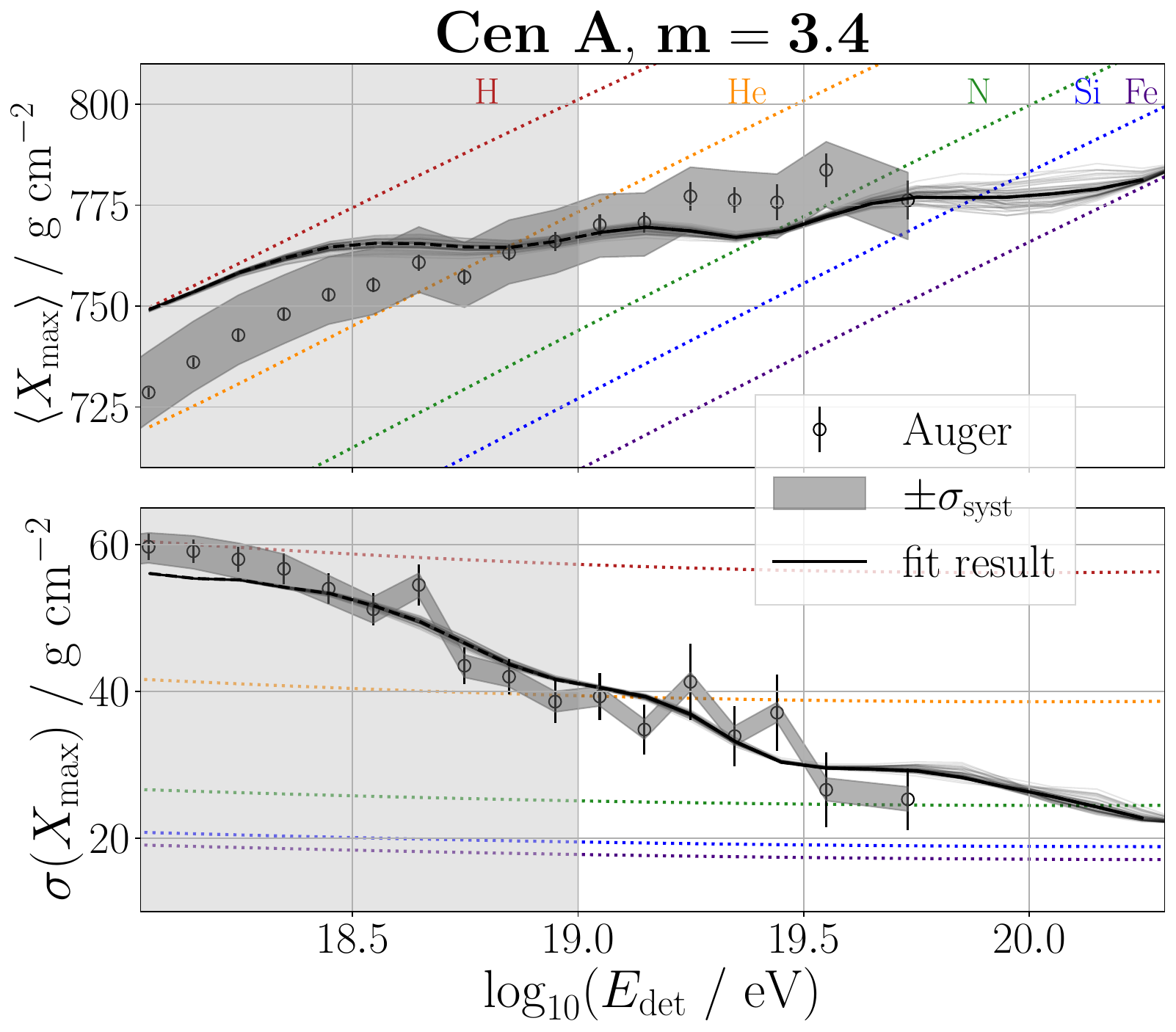}
\includegraphics[width=0.32\textwidth]{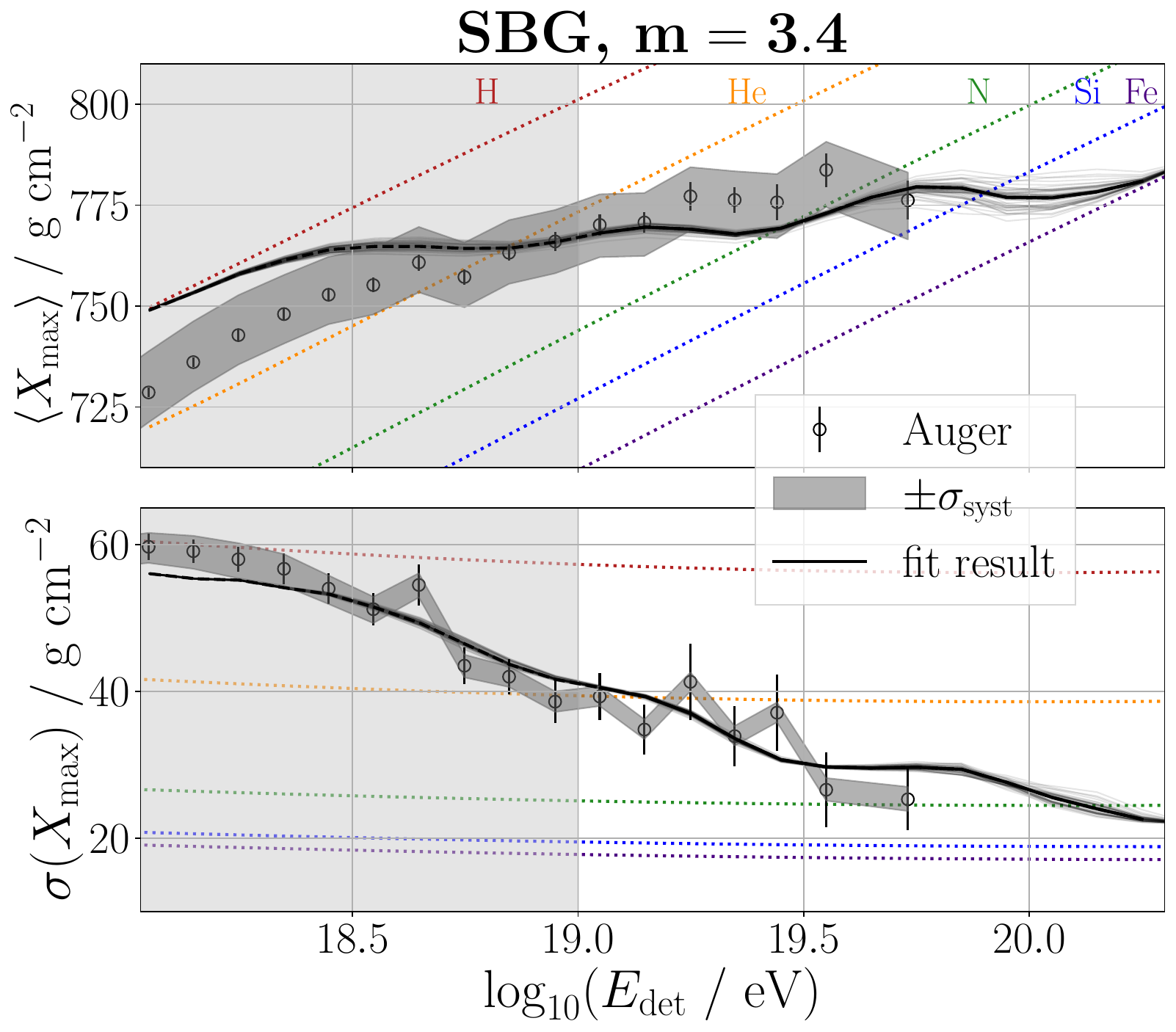}
\caption{First two moments of the \xmax distributions for the Centaurus A model with $m=0$ (\textit{left}), with $m=3.4$ (\textit{middle}) and the SBG model with $m=3.4$ (\textit{right}).}
\label{fig:xmax}
\end{figure}

The modeled spectra on Earth are shown in \cref{fig:spectra}. For both evolutions, the spectra look quite similar so that conclusions can be drawn independently of the model evolution of the background sources. It is visible how the nearby source Centaurus A contributes mainly to the flux at higher energies. The best-fit signal contribution to the total flux rises from $\sim3\%$ at $40$\,EeV to $\sim10\%$ at 100\,EeV. The uncertainty on the signal fraction is large as described above, so contributions up to $\sim50\%$ at 100\,EeV are still within the 68\% highest posterior density interval. 
The composition at the sources is dominated by nitrogen. This leads to a nitrogen flux at Earth up to very high energies, which has a rigidity of around $R\simeq10$\,eV at $70$\,EeV. For this, the model predicts a blurring of around $\delta_0=20^\circ$. The contribution of lighter elements to the flux at the highest energies (both from the nearby sources and from the background) turns out to be very small, and is completely negligible above $\sim30$~ EeV where the helium contribution cuts off. Most of the light elements present below those energies are actually of secondary origin, produced in the photodisintegration of the heavier primaries from the background sources. The first and second moments of the modeled shower maximum depth distributions are depicted in \cref{fig:xmax}. The model describes the measured data fairly well above the minimum energy employed in the fit. It is however visible that the model mean mass is slightly heavier than the measured one, which can be resolved by including the experimental \xmax scale uncertainty as a nuisance parameter in the model, as will be shown in sec.~\ref{sec:results_syst}.

% \begin{figure}[ht]
% \centering
% \includegraphics[width=0.32\textwidth]{img/CenA/m0/SKYMAP_S_woExp_19.3_exposureFalse.pdf}
% \includegraphics[width=0.32\textwidth]{img/CenA/m0/SKYMAP_S_woExp_19.6_exposureFalse.pdf}
% \includegraphics[width=0.32\textwidth]{img/CenA/m0/SKYMAP_S_woExp_19.900000000000002_exposureFalse.pdf}
% \includegraphics[width=0.32\textwidth]{img/CenA/m3/SKYMAP_S_woExp_19.3_exposureFalse.pdf}
% \includegraphics[width=0.32\textwidth]{img/CenA/m3/SKYMAP_S_woExp_19.6_exposureFalse.pdf}
% \includegraphics[width=0.32\textwidth]{img/CenA/m3/SKYMAP_S_woExp_19.900000000000002_exposureFalse.pdf}
% \includegraphics[width=0.32\textwidth]{img/SBG/SKYMAP_S_woExp_19.3_exposureFalse.pdf}
% \includegraphics[width=0.32\textwidth]{img/SBG/SKYMAP_S_woExp_19.6_exposureFalse.pdf}
% \includegraphics[width=0.32\textwidth]{img/SBG/SKYMAP_S_woExp_19.900000000000002_exposureFalse.pdf}
% \caption{Modeled arrival directions map $S$ from the catalog sources, \textit{upper} row for Cen A model with $m=0$, \textit{middle} row for Cen A model with $m=3.4$ and \textit{lower} row for SBG model with $m=3.4$. The energy bins \edet=19.3 (\textit{left}), \edet=19.6 (\textit{middle}) and \edet=19.9 (\textit{right}) are shown exemplarily. }
% \label{fig:}
% \end{figure}

\begin{figure}[ht]
\centering
\includegraphics[width=0.32\textwidth]{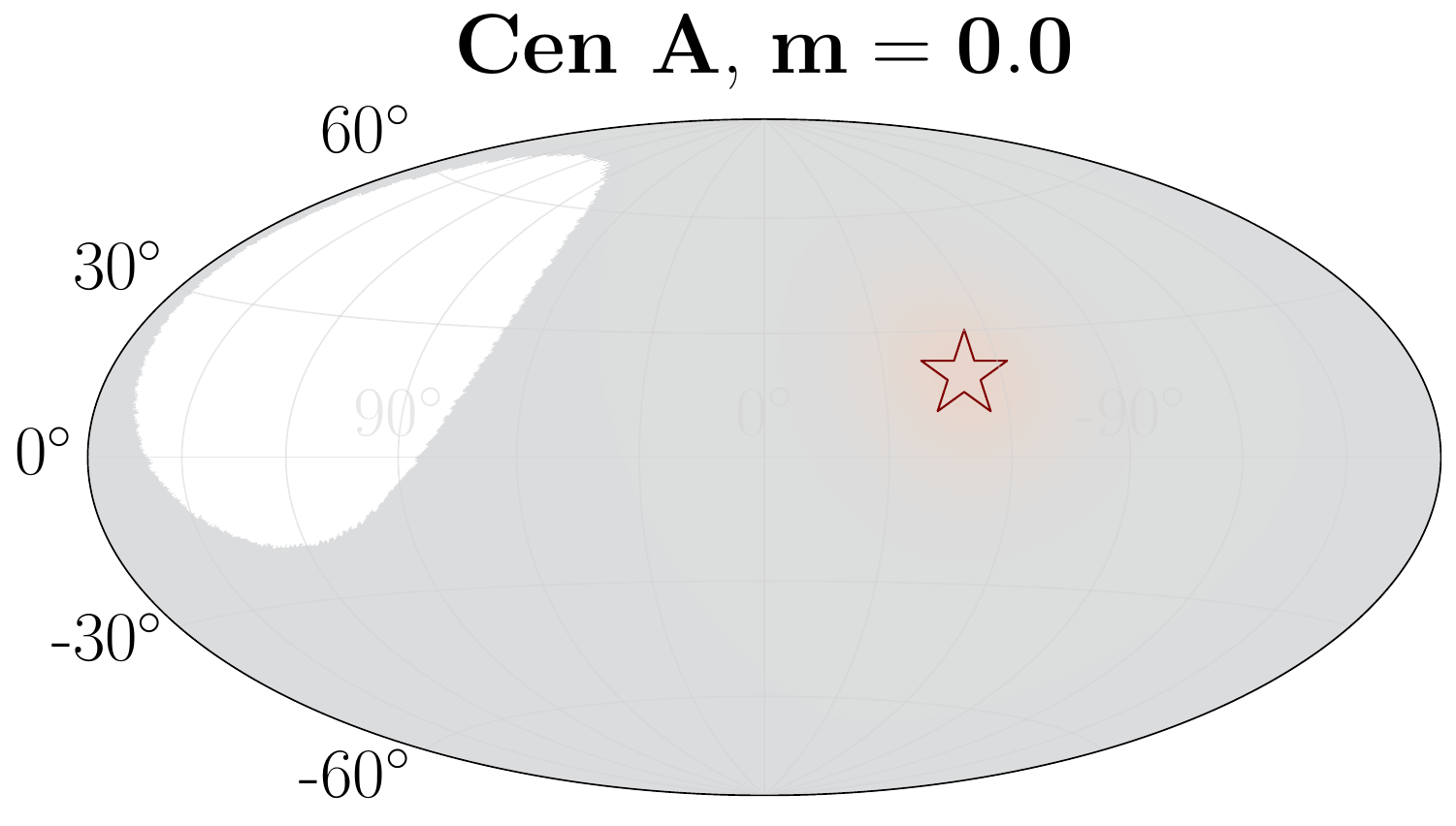}
\includegraphics[width=0.32\textwidth]{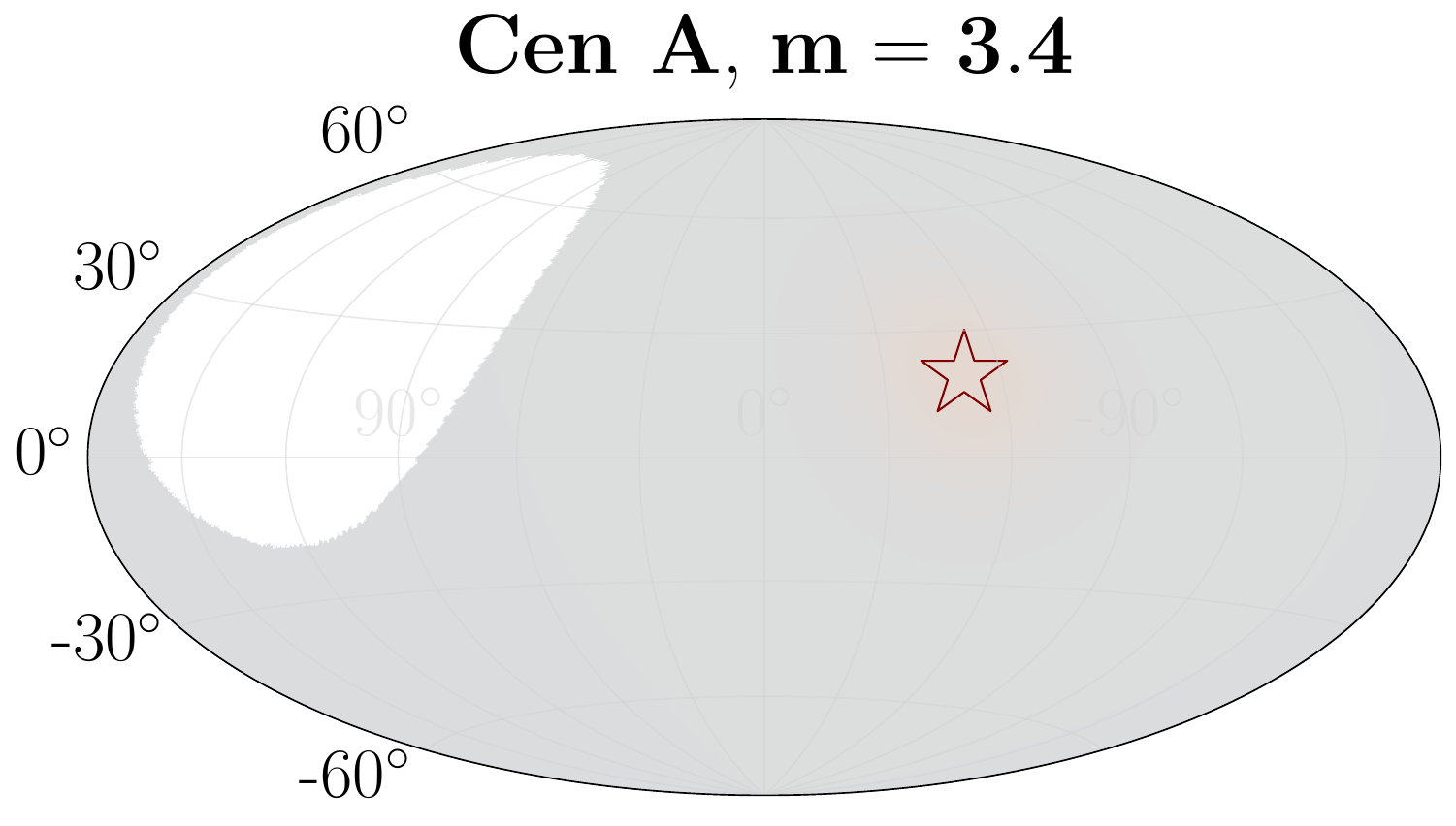}
\includegraphics[width=0.32\textwidth]{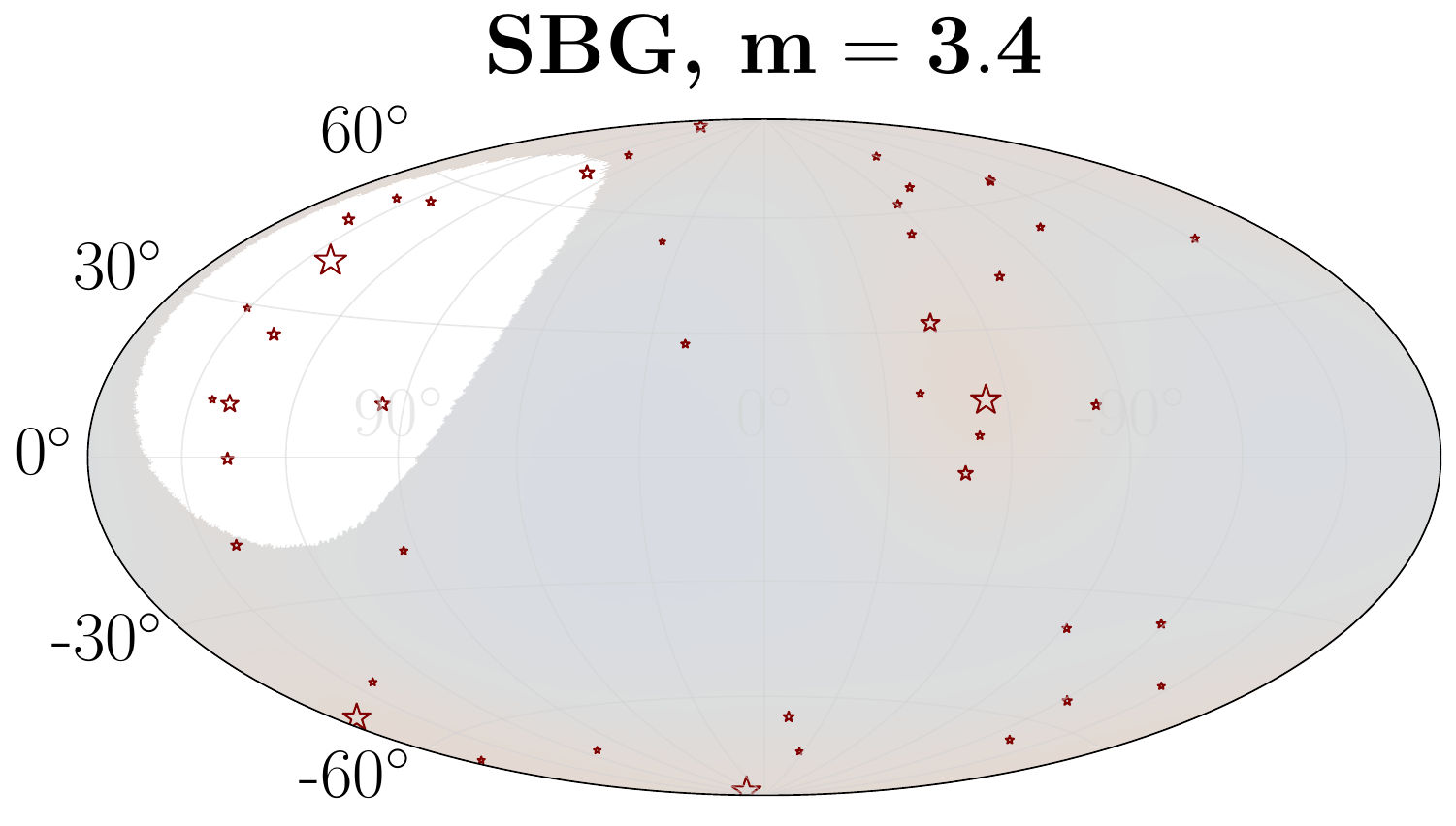}

\includegraphics[width=0.32\textwidth]{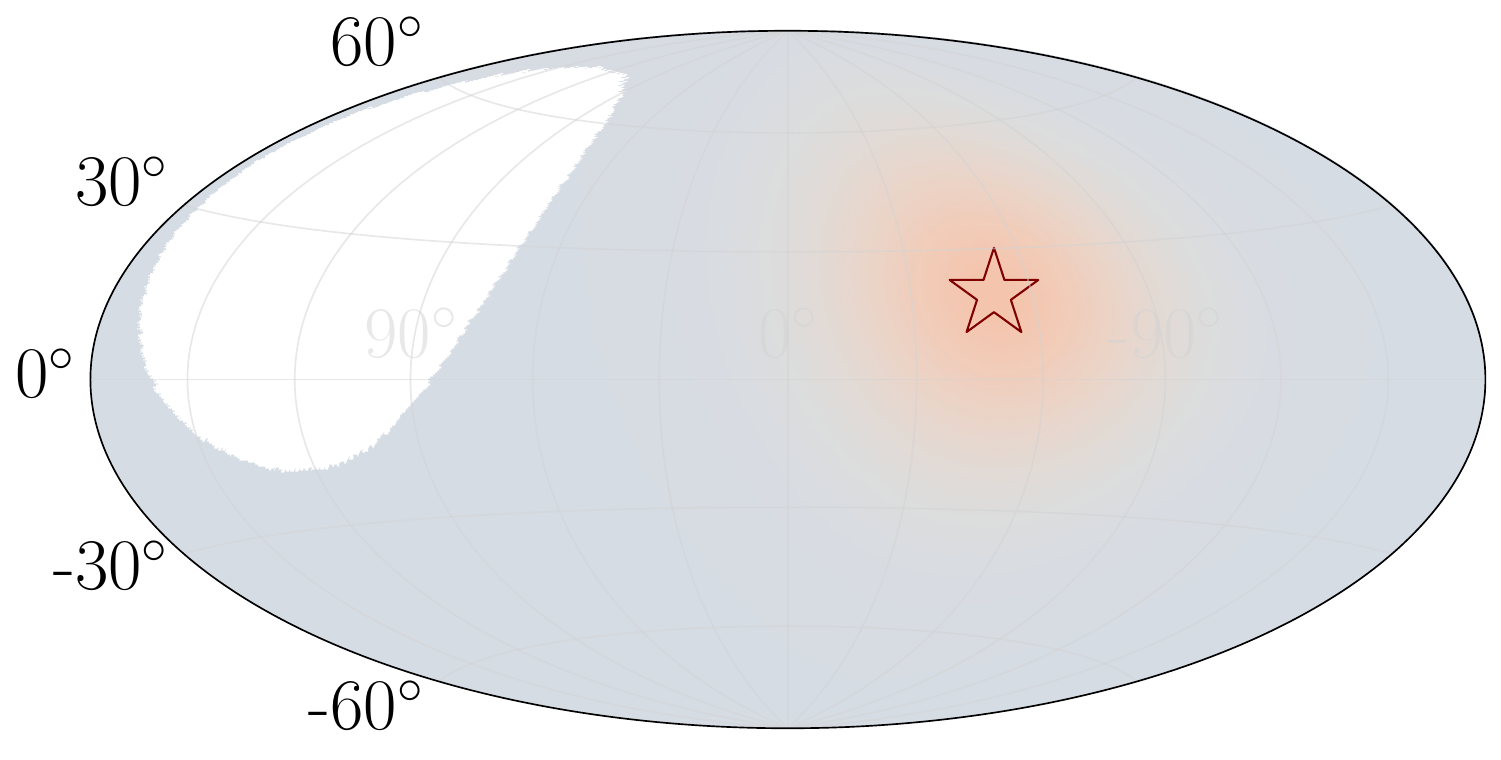}
\includegraphics[width=0.32\textwidth]{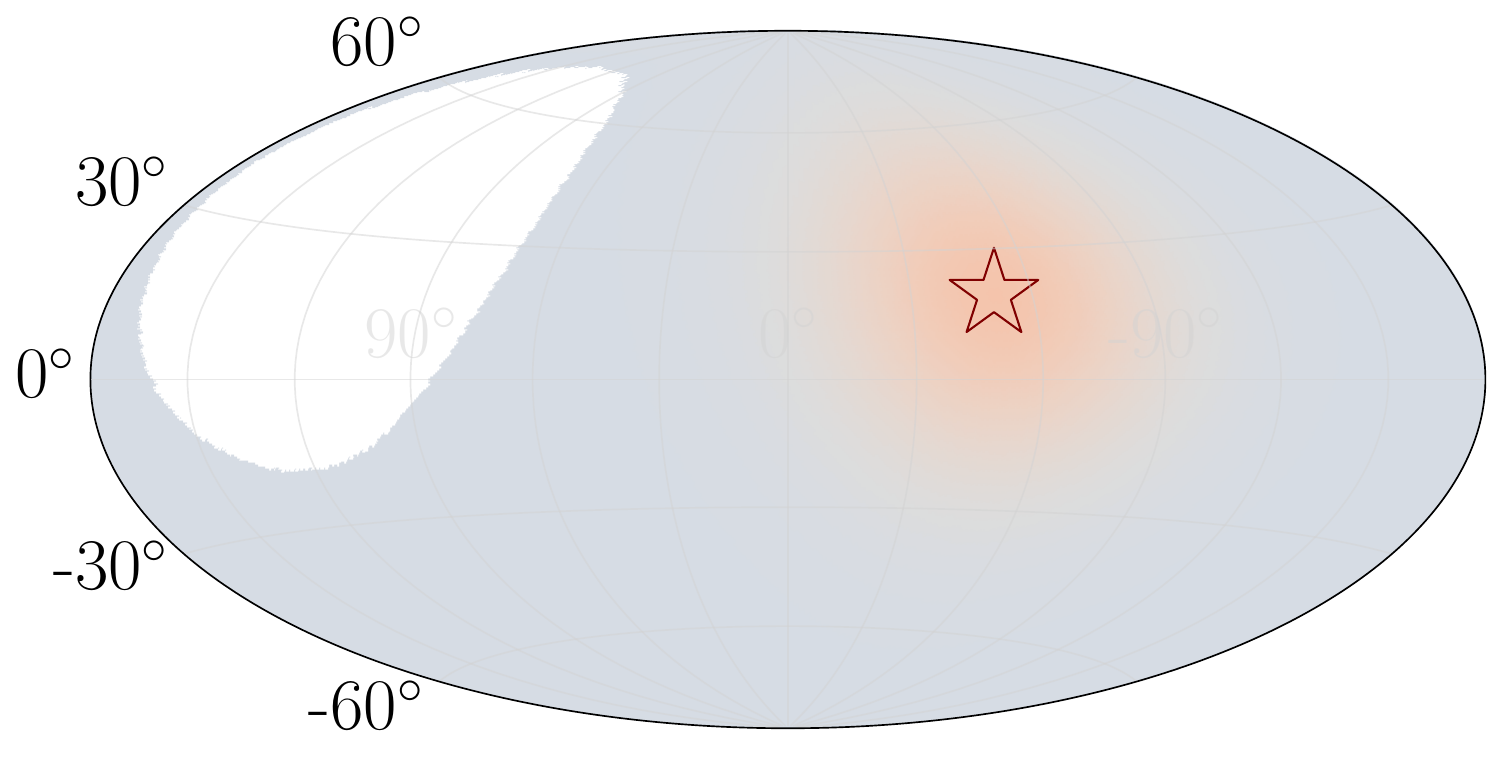}
\includegraphics[width=0.32\textwidth]{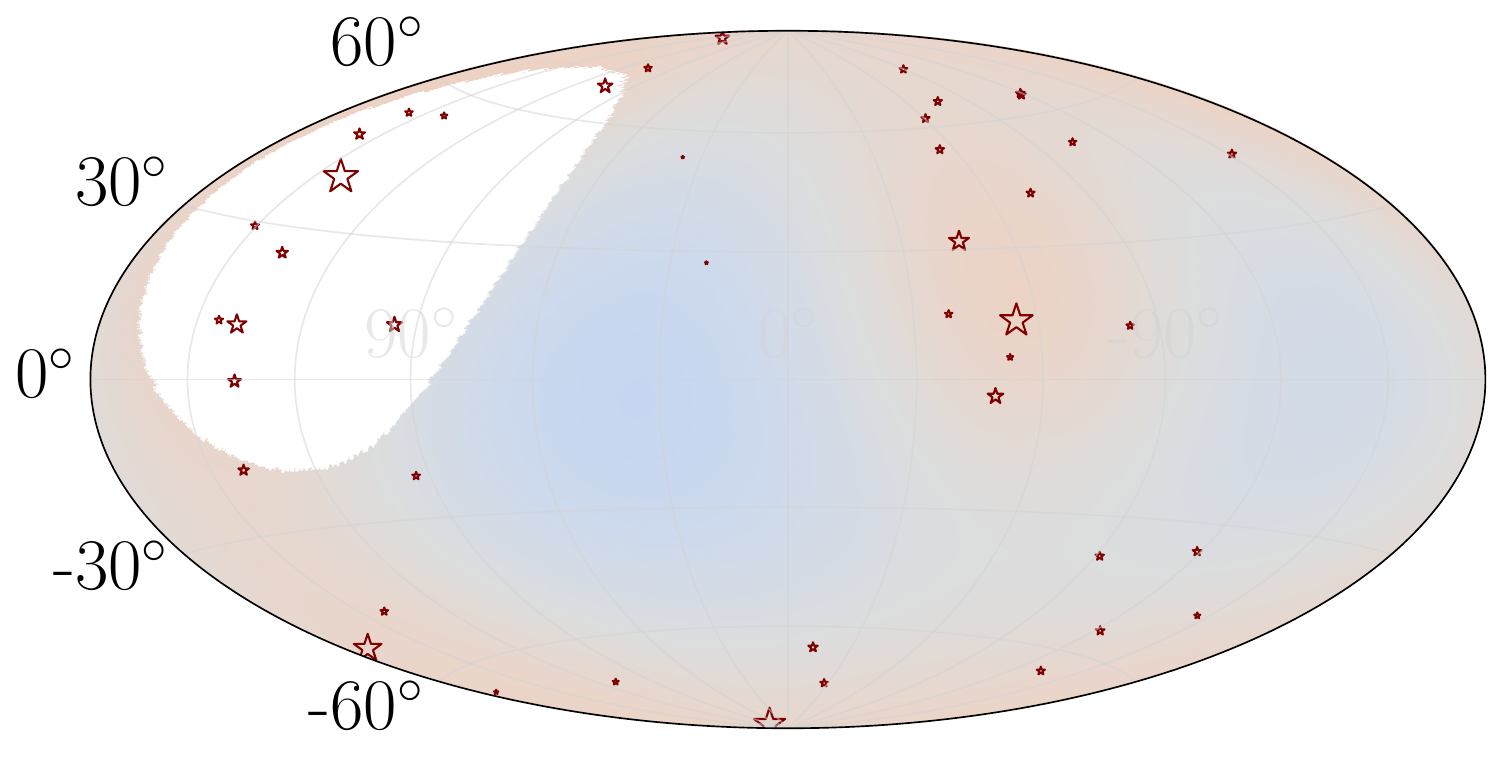}

\includegraphics[width=0.32\textwidth]{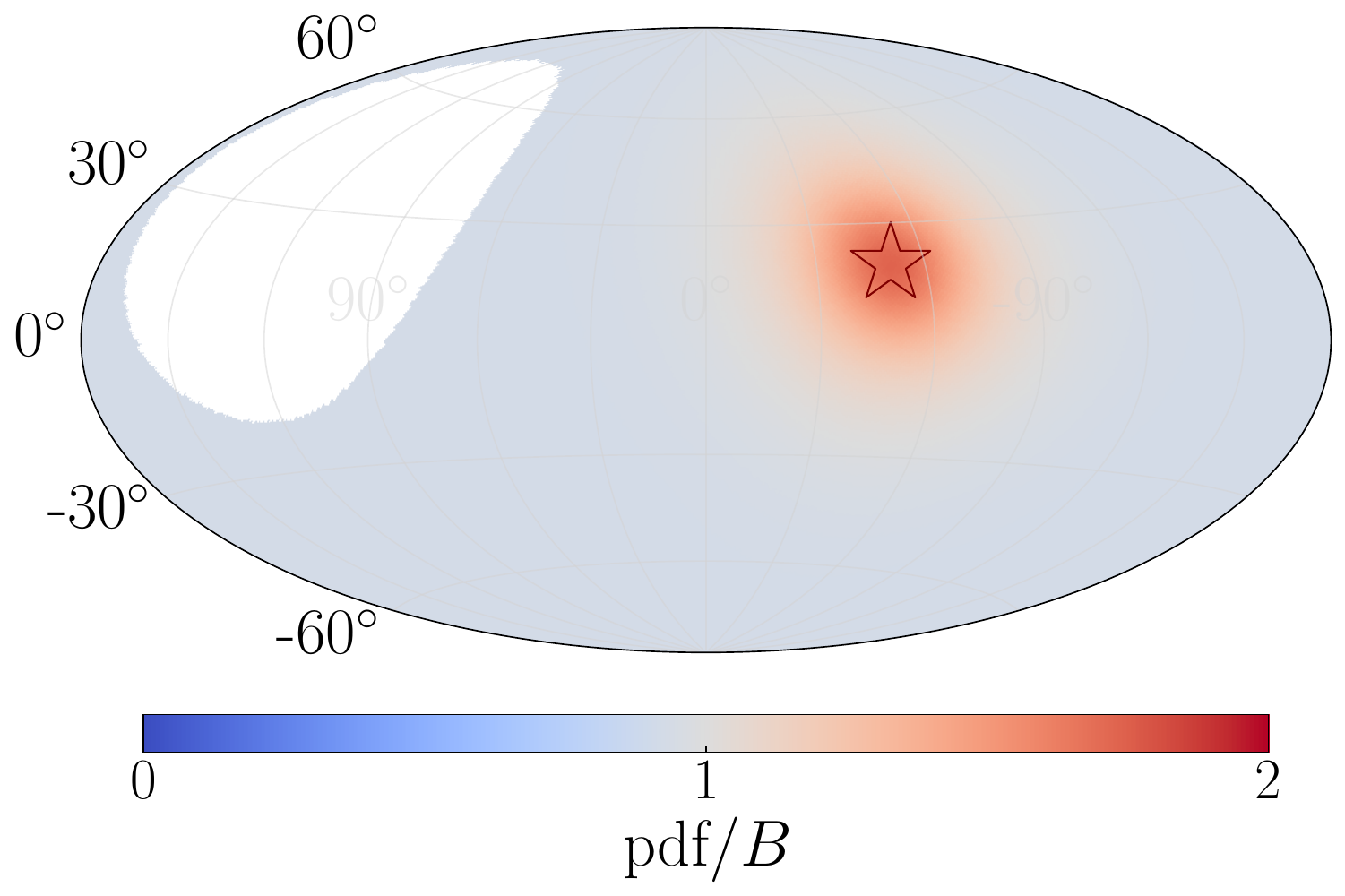}
\includegraphics[width=0.32\textwidth]{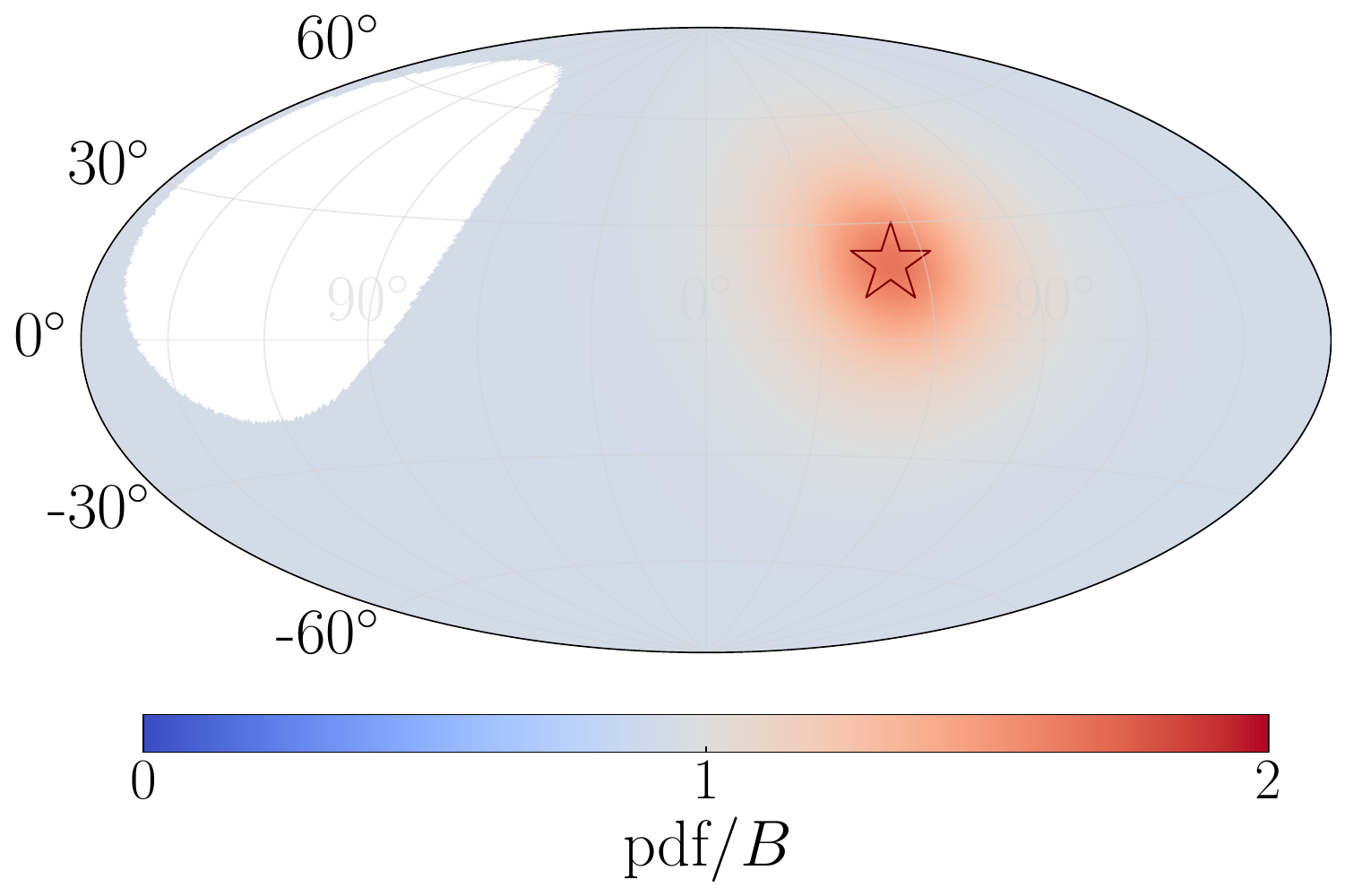}
\includegraphics[width=0.32\textwidth]{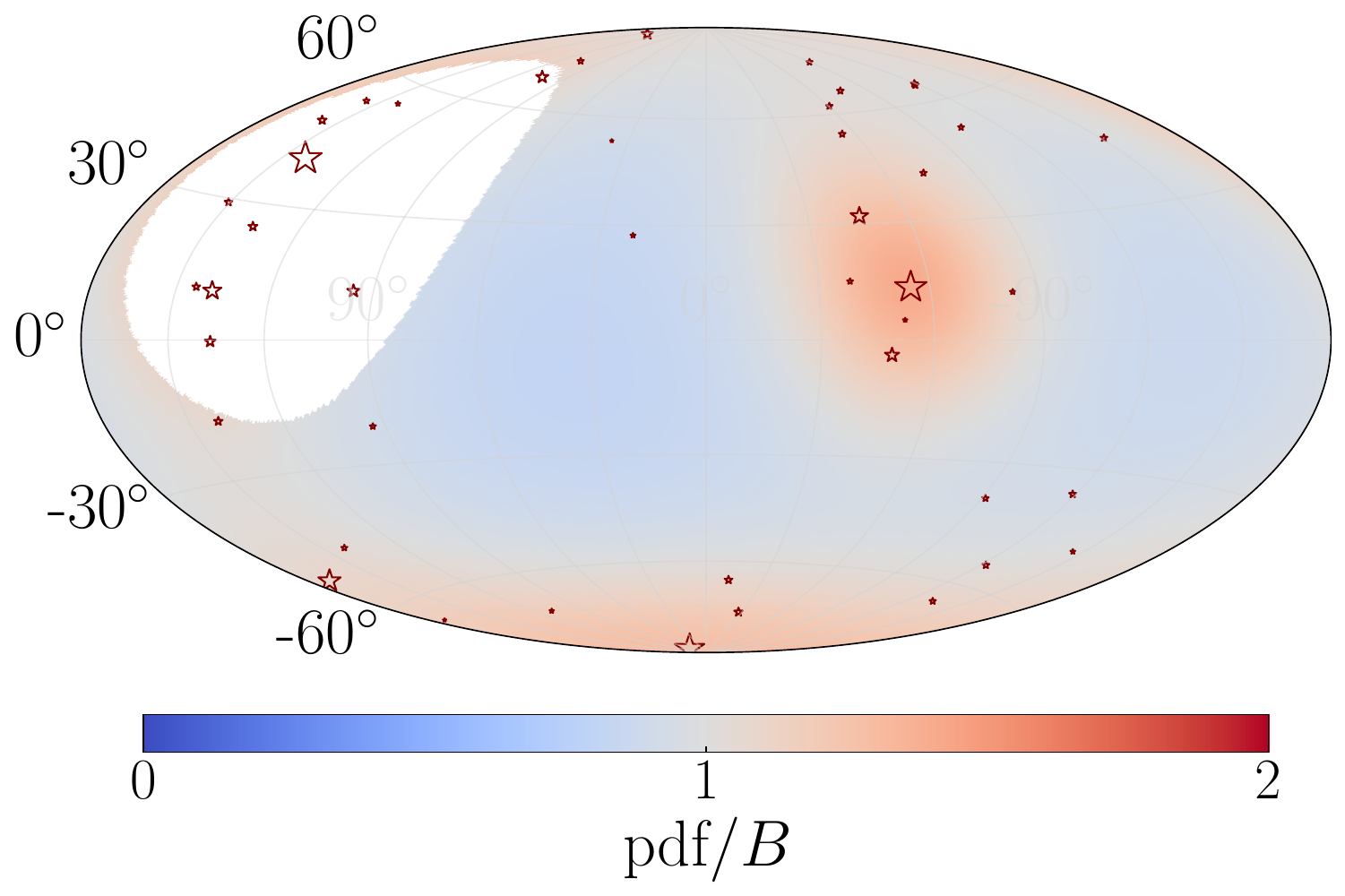}
\caption{Modeled arrival directions pdf (\cref{eq:pdf}), \textit{left} for Centaurus A model with $m=0$, \textit{middle} for Centaurus A model with $m=3.4$ and \textit{right} for SBG model with $m=3.4$. The energy bins \edet=19.3 (\textit{upper row}), \edet=19.6 (\textit{middle row}) and \edet=19.9 (\textit{lower row}) are shown as examples. The catalog contribution and with that the level of anisotropy rises with the energy, while the overall blurring decreases. Additionally, the contribution of individual sources depends on the energy through their distances and flux weights. The stars indicate the directions of the source candidates with the size scaling with the relative flux contribution before the observatory exposure is applied.}
\label{fig:ADs}
\end{figure}

The best-fit arrival directions for three representative energy bins are shown in \cref{fig:ADs}. The growing flux contribution of Centaurus A with the energy, as well as the decreasing blurring, is visible. 

For the SBG model with $m=3.4$, the spectral index has softened visibly compared to the reference model with $m=3.4$. The softening decreases the number of high-energy particles emitted at the background sources. This compensates the increased number of high-energy particles from the catalog sources which can reach Earth easily due to the short propagation distances. 
The signal fraction is around $f_0\approx20\%$. In \cref{fig:spectra}, it is again visible how the catalog contribution becomes relevant at higher energies, where it reaches values of up to $\approx40\%$ at 100\,EeV. Additionally, \cref{fig:spectra} displays the individual spectrum of the strongest source in the SBG catalog, NGC 4945. The modeled spectrum looks rather similar to the one of Centaurus A, only with smaller uncertainties due to the additional constraints from the other candidates in the SBG catalog. The two source candidates NGC 4945 and Centaurus A are located in similar directions and distances of around $3.5$\,Mpc. Hence, the contribution of that sky region is modeled consistently, independent from the number of other subdominant candidate sources in the catalog. This is also visible in the arrival directions in \cref{fig:ADs}, where both the size of the blurring and the overall anisotropy level is similar for both NGC 4945 and Centaurus A.  

\begin{figure}[th]
\centering
\includegraphics[width=0.32\textwidth]{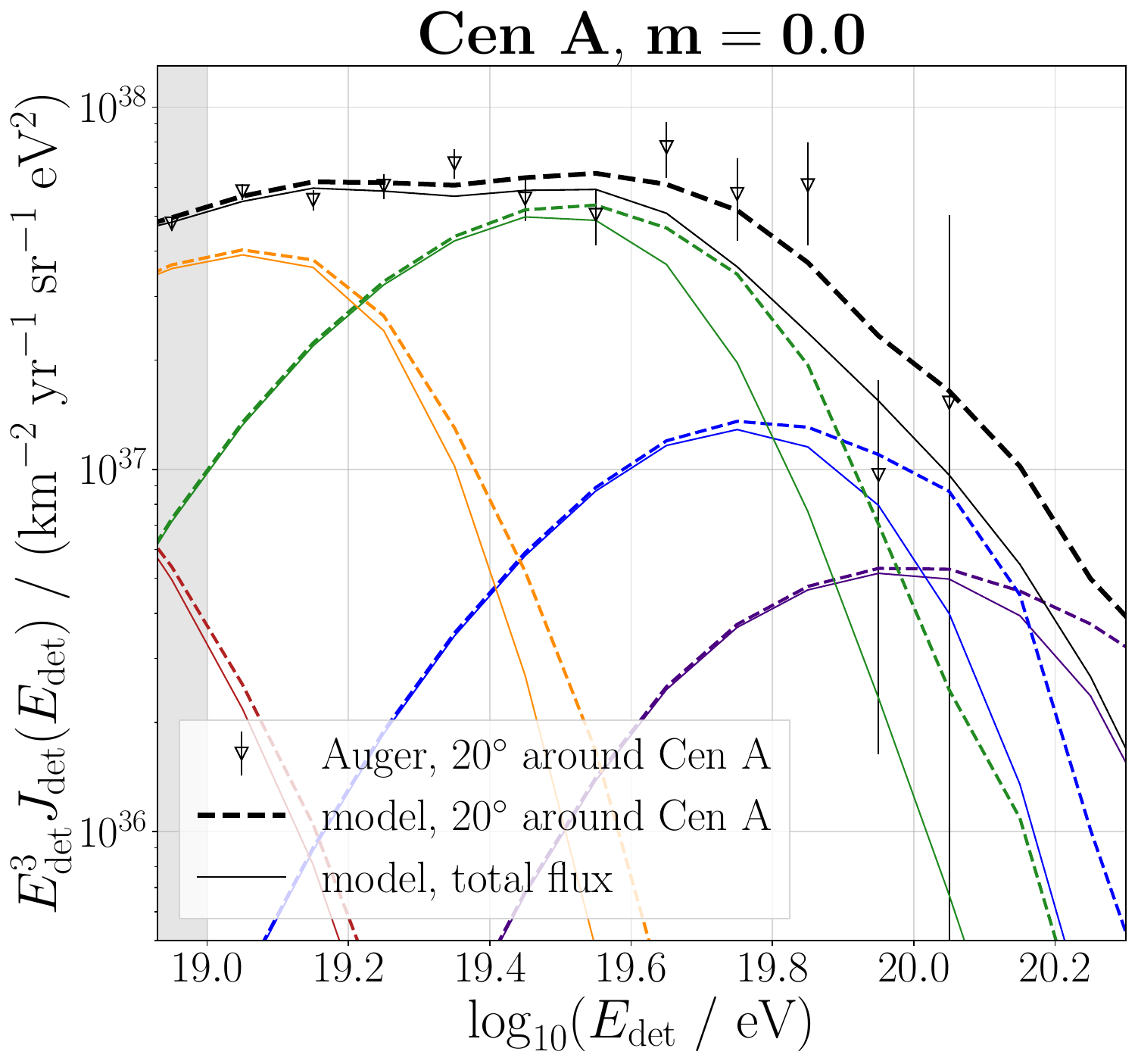}
\includegraphics[width=0.32\textwidth]{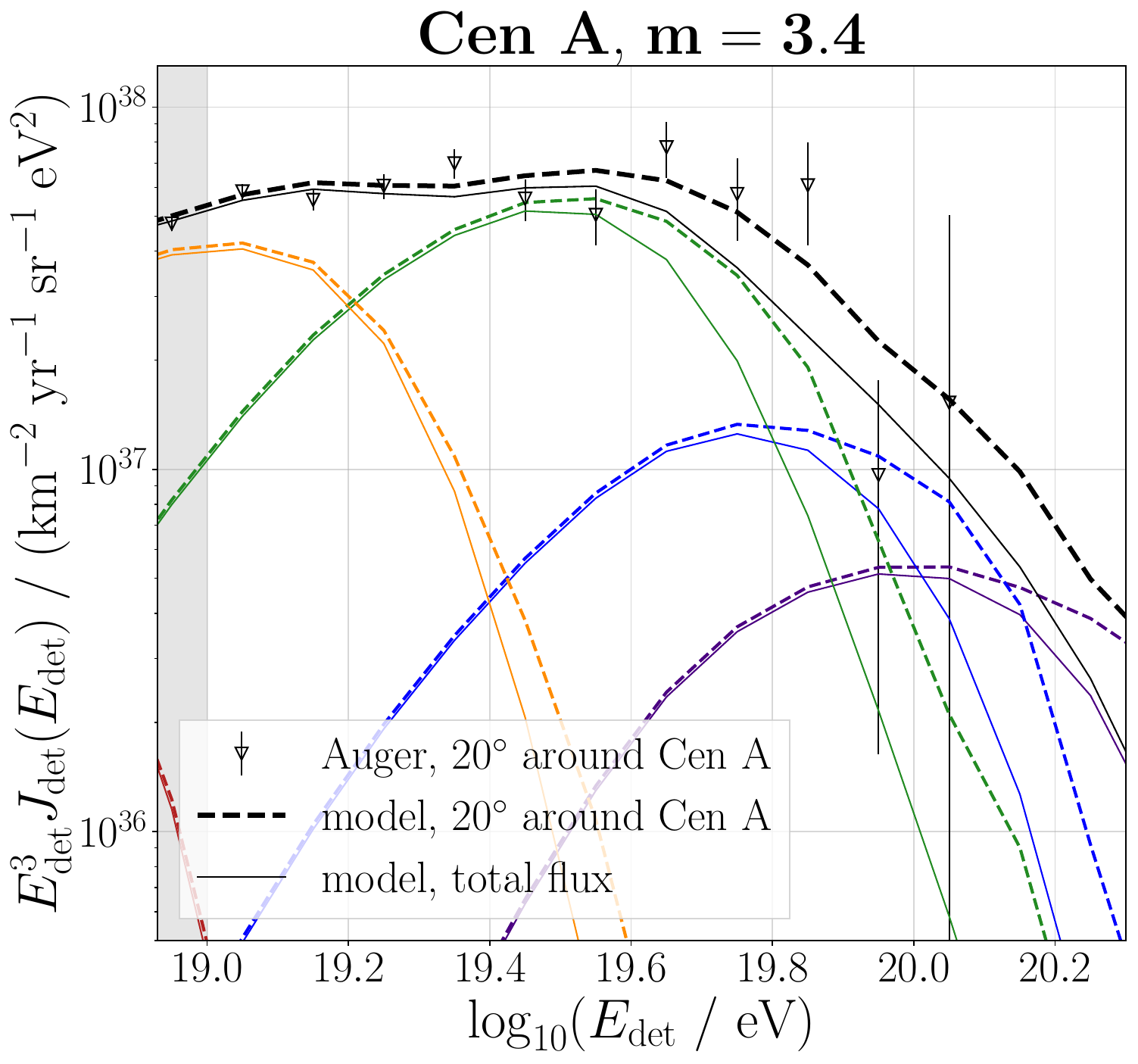}
\includegraphics[width=0.32\textwidth]{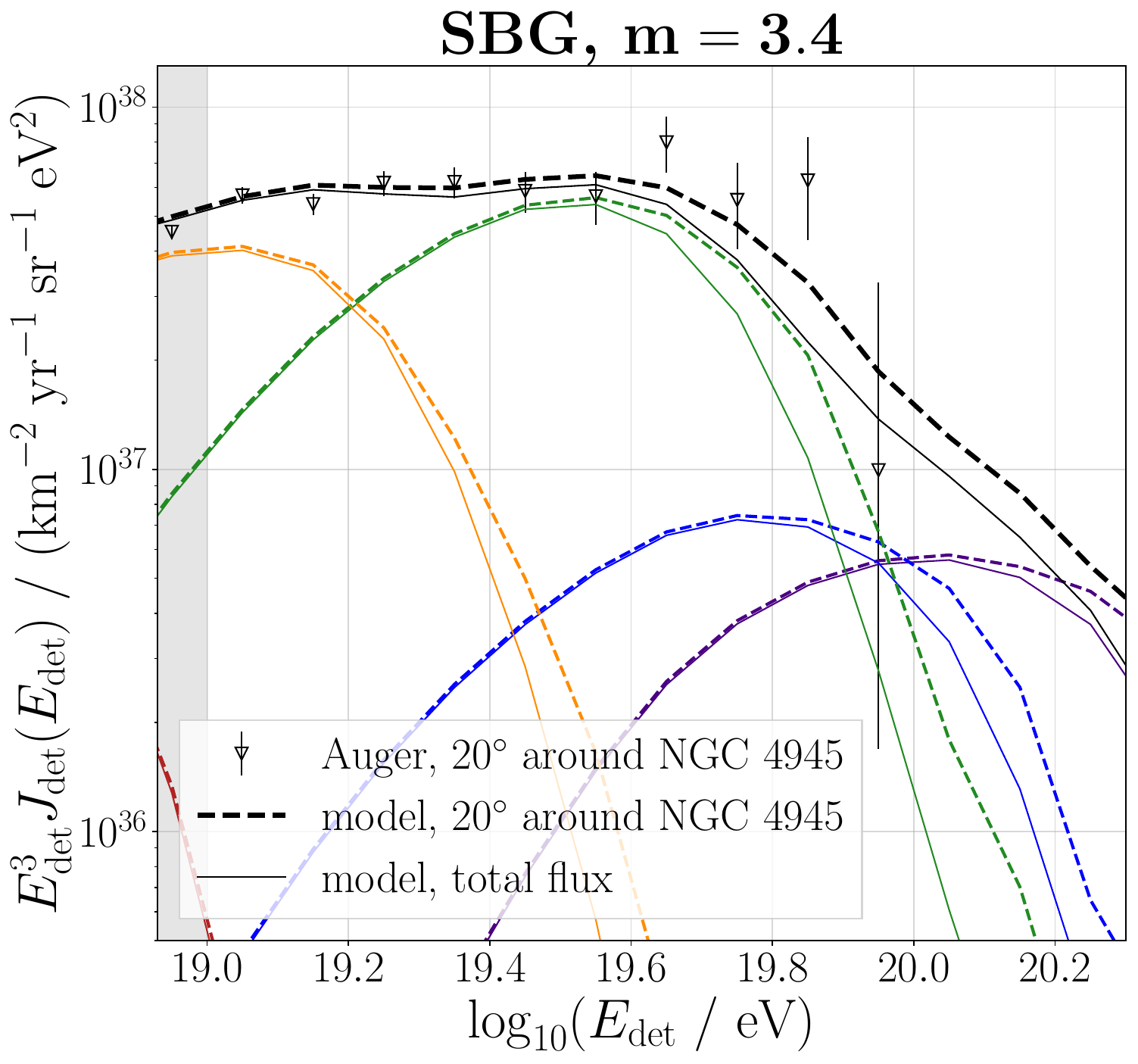}
\caption{Modeled spectra on Earth inside a circular region of angular size $20^\circ$ (radius) around the direction of Centaurus A (\textit{left and middle}) or NGC 4945 (\textit{right}). Displayed are the Centaurus A model with $m=0$ (\textit{left}), with $m=3.4$ (\textit{middle}), and the SBG model with $m=3.4$ (\textit{right}). The colors are the same as in \cref{fig:spectra}. The markers represent the measured data from the Pierre Auger Observatory inside the $20^\circ$ region, the dashed lines the best-fit model in that region, and the solid lines the full best-fit model (same as in \cref{fig:spectra}). Note that the spectra in the $20^\circ$ region have not been fitted explicitly (see \cref{sec:likelihood}), so this serves as a cross check of the model.}
\label{fig:spectra_CenA}
\end{figure}

To verify explicitly that the model describes the overdensity in the region of Centaurus A and NGC 4945, we investigate the spectrum of all events in a circular region with variable angular size centered on the direction of the respective source candidate. The size of the region is a trade-off between wanting to fully contain the contribution of the source candidate, and minimizing the contamination by background and neighboring candidates. The modeled and measured (calculated from the energy spectrum data set described in \cref{sec:data_sets}) spectra for one example selection angle of $20^\circ$ (radius) are displayed in \cref{fig:spectra_CenA}. Here, it is visible how the model predicts an increased flux for the region around the source candidate. In particular, more high-energy events are expected from the close-by candidate. Due to this, the model spectrum from the candidate region (dashed lines) agrees better with the measured flux (markers) than the whole-sky model spectrum (solid lines). For all tested angles between $5^\circ$ and $30^\circ$, the reduced $\chi^2$ for the candidate region model is smaller than that of the whole-sky model by around $10\%$ to $40\%$. With increased statistics in the candidate region, a more detailed assessment of this effect could be performed in the future.

Note that for a source such as Centaurus~A or NGC~4945 at about 4~Mpc distance, one can estimate from the inferred spectral shape and the derived composition that the CR luminosity above 10~EeV should be $L_{>10\,\mathrm{EeV}}\simeq (f_\mathrm{source}/0.05) \times 10^{39}$ erg/s, with $f_\mathrm{source}$ being the overall fractional contribution from the specified source to the flux at 40~EeV (directly corresponding to $f_0$ for the case of the Centaurus A model, and to around $f_0 \times \omega_\mathrm{flux}^\mathrm{NGC 4945}$ for the SBG model). For a scenario like the one of the SBGs, one would expect that the remaining catalog sources should have CR luminosities approximately scaling as the respective adopted flux weights.

\subsubsection{Impact of experimental systematic uncertainties}
\label{sec:results_syst}
\begin{figure}[ht]
\centering
\includegraphics[width=0.32\textwidth]{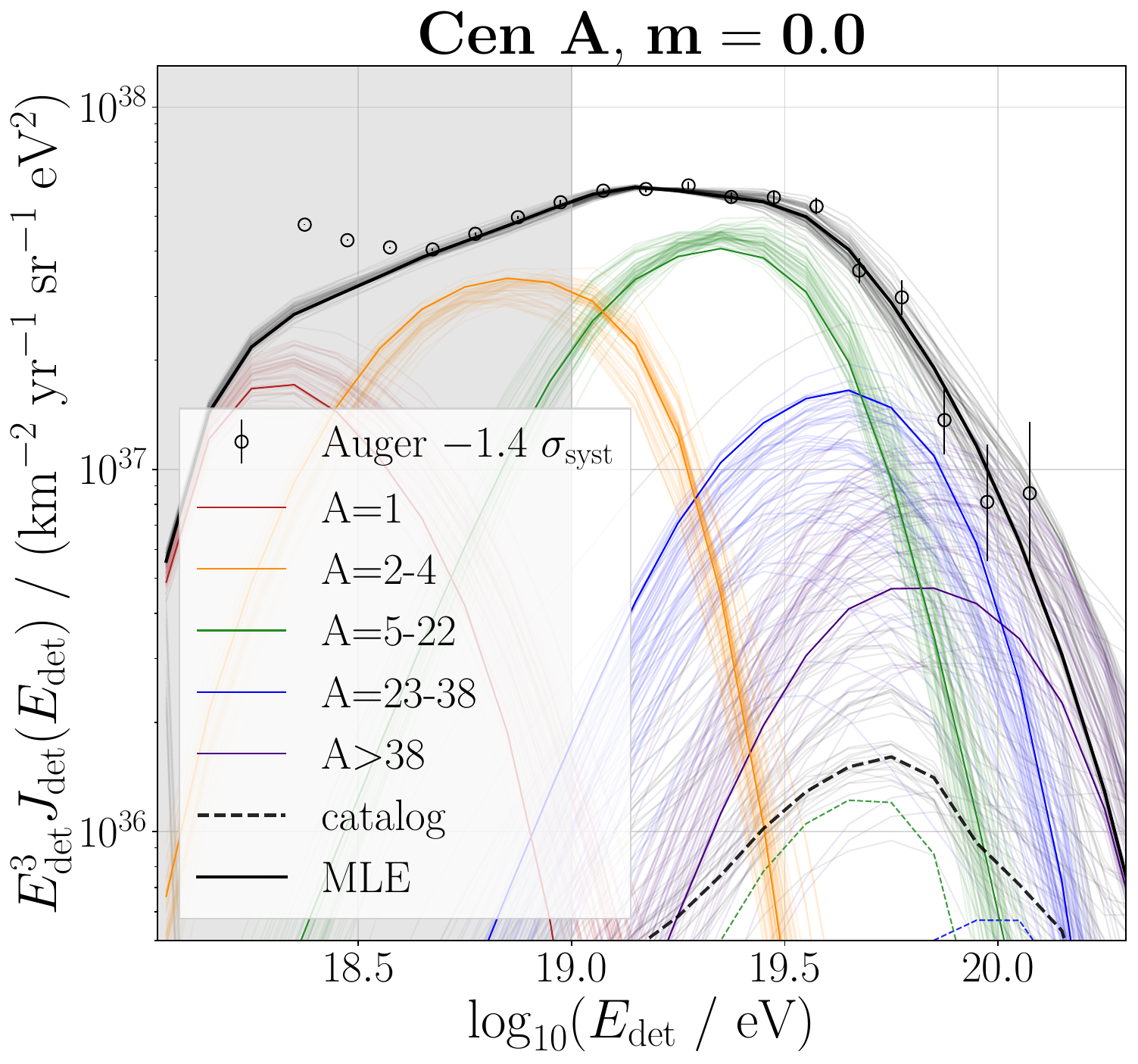}
\includegraphics[width=0.32\textwidth]{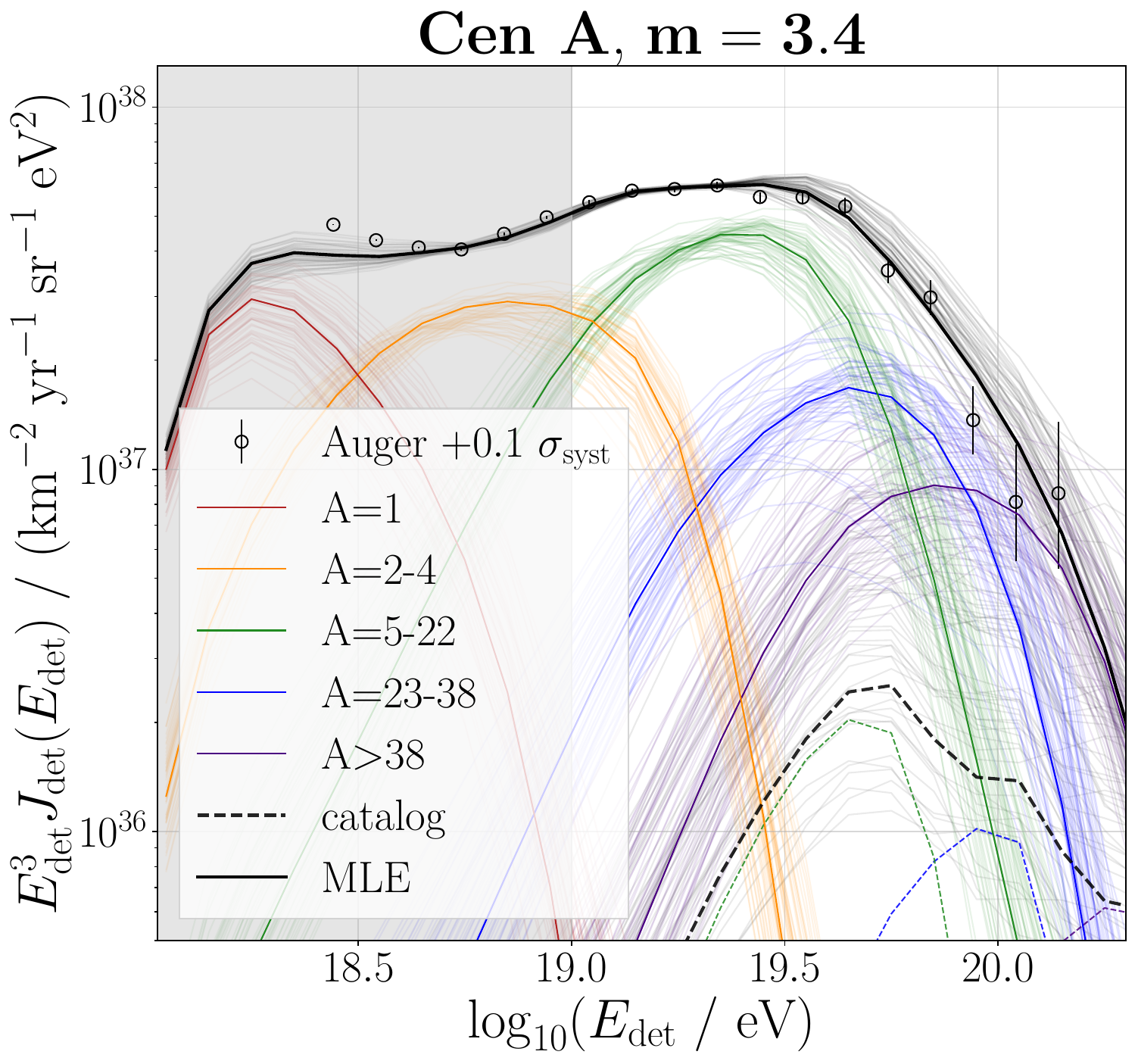}
\includegraphics[width=0.32\textwidth]{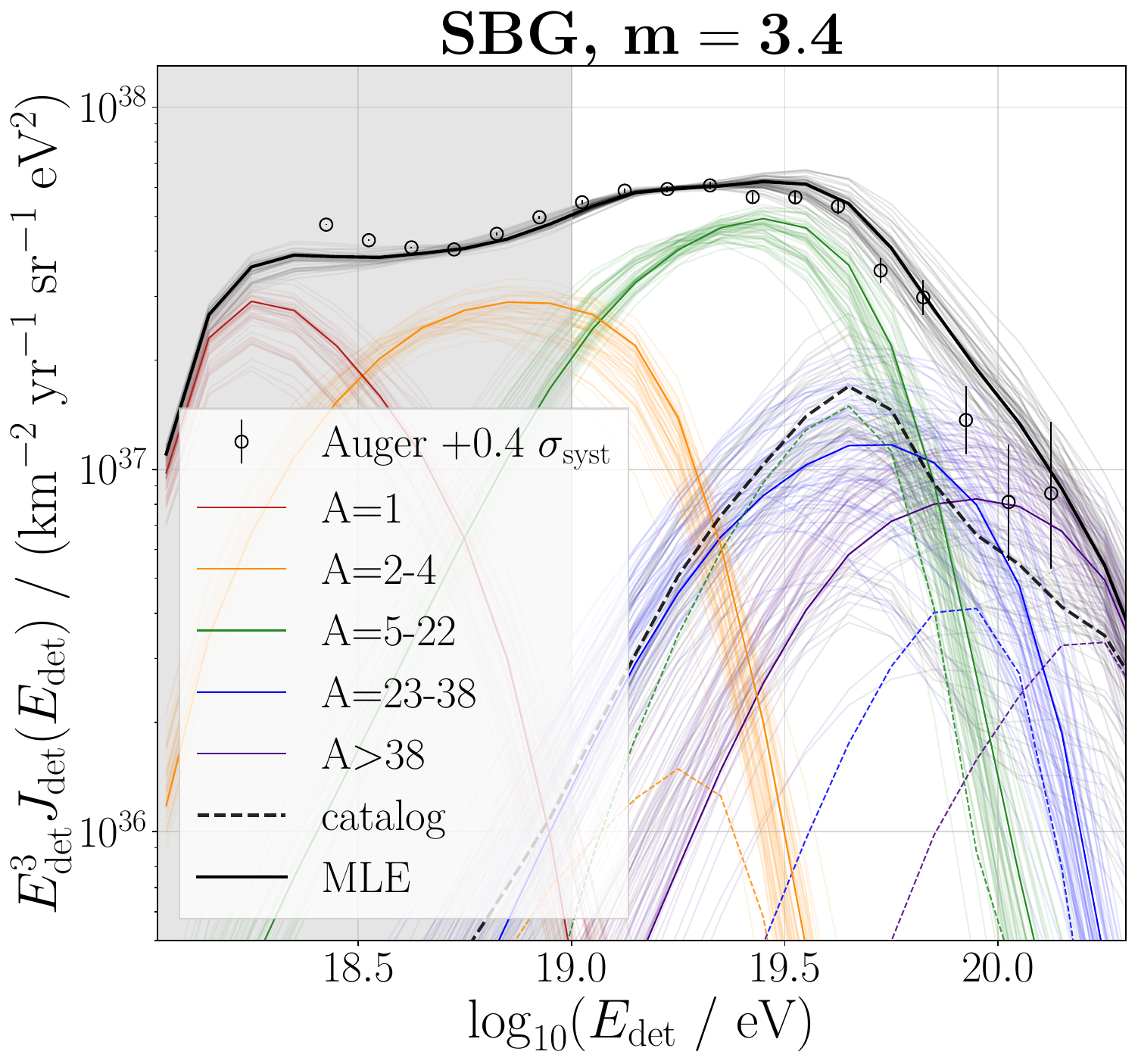}
\caption{Same as \cref{fig:spectra} (\textit{upper row}) but including experimental systematic uncertainties as nuisance parameters.}
\label{fig:spectra_syst}
\end{figure}

The effect of experimental systematic uncertainties on the fit result is tested by including the uncertainty on the energy and \xmax scales as nuisance parameters in the model (see \cref{sec:systematic}). The fit results for that are given in \cref{tab:results_syst}. The preferred shift on the energy scale is estimated to be around $-1.2\sigma$ for the model with flat evolution (implying that the true energy is lower than measured) and almost no shift is favored for the models with SFR evolution. In general, the shift of the energy scale has a smaller impact on the deviance than the shift of the \xmax scale.
Here, a negative shift is preferred, which is around $-1\sigma$ for $m=0$ and $-1.5\sigma$ for $m=3.4$, in accordance with results for the reference models (see \cref{sec:ref}). This shifts the \xmax scale up by around $10-15\,\text{g/cm}^{2}$ (see \cref{sec:systematic}), which is in line with the findings from Ref.~\cite{j_vicha_for_the_pierre_auger_collaboration_testing_2023}. According to these results, the true \xmax values are lower than reconstructed, and the true composition is correspondingly heavier. The main effect on the model parameters is a softening of the spectral index to around $\gamma\approx-1$ for both evolutions. 
The softening is also visible in the modeled spectra on Earth in \cref{fig:spectra_syst}. 
%With the softer spectrum and systematic shifts, the agreement between model and data in the region around Centaurus A becomes even better, as shown in \cref{fig:spectra_syst} (\textit{lower row}).
The modeled signal fraction and blurring are almost completely independent of the systematic uncertainties on the energy and \xmax scales, so the contribution from the catalog sources is also similar to the one without systematics displayed in \cref{fig:spectra}. The modeled arrival directions look similar to the ones without systematics (\cref{fig:ADs}), so they are not shown again. Also, the different element contributions are almost unchanged, so still no elements lighter than the nitrogen group are expected at the highest energies.
The fluctuations between posterior draws have increased, as is visible in \cref{fig:spectra_syst}, which is because of the added freedom allowed by the two additional fit parameters for the experimental systematic uncertainties.

\begin{figure}[ht]
\centering
\includegraphics[width=0.32\textwidth]{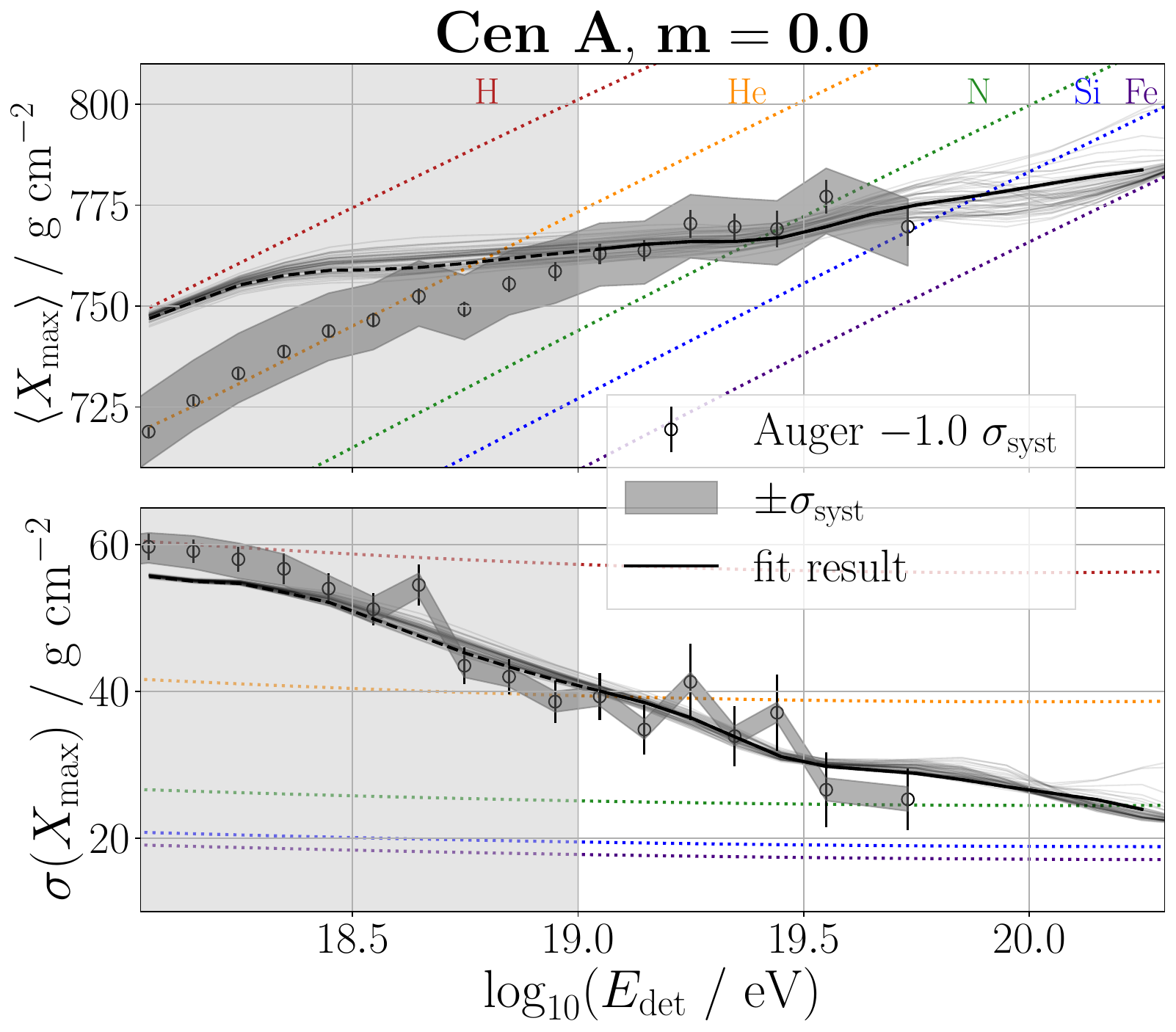}
\includegraphics[width=0.32\textwidth]{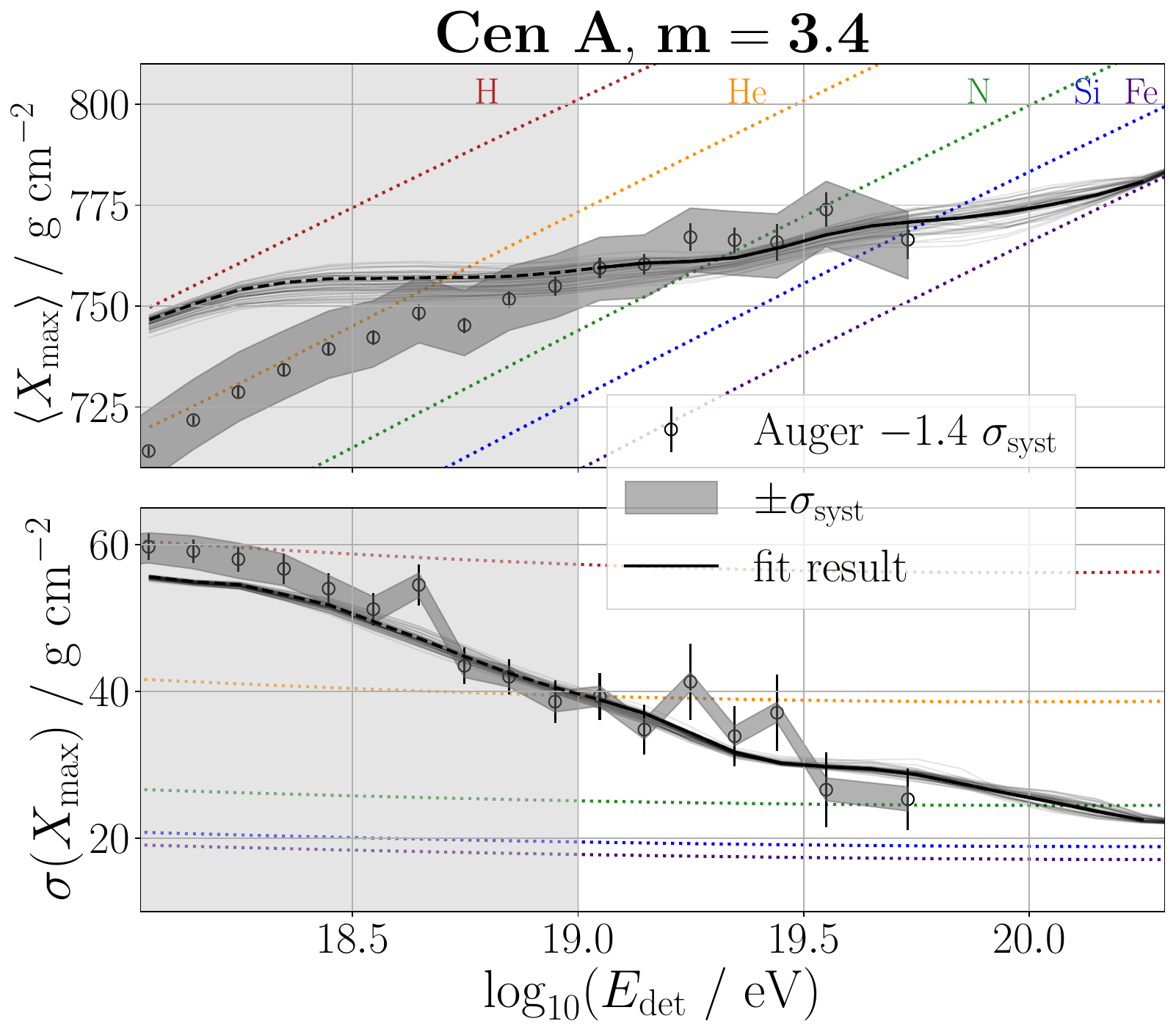}
\includegraphics[width=0.32\textwidth]{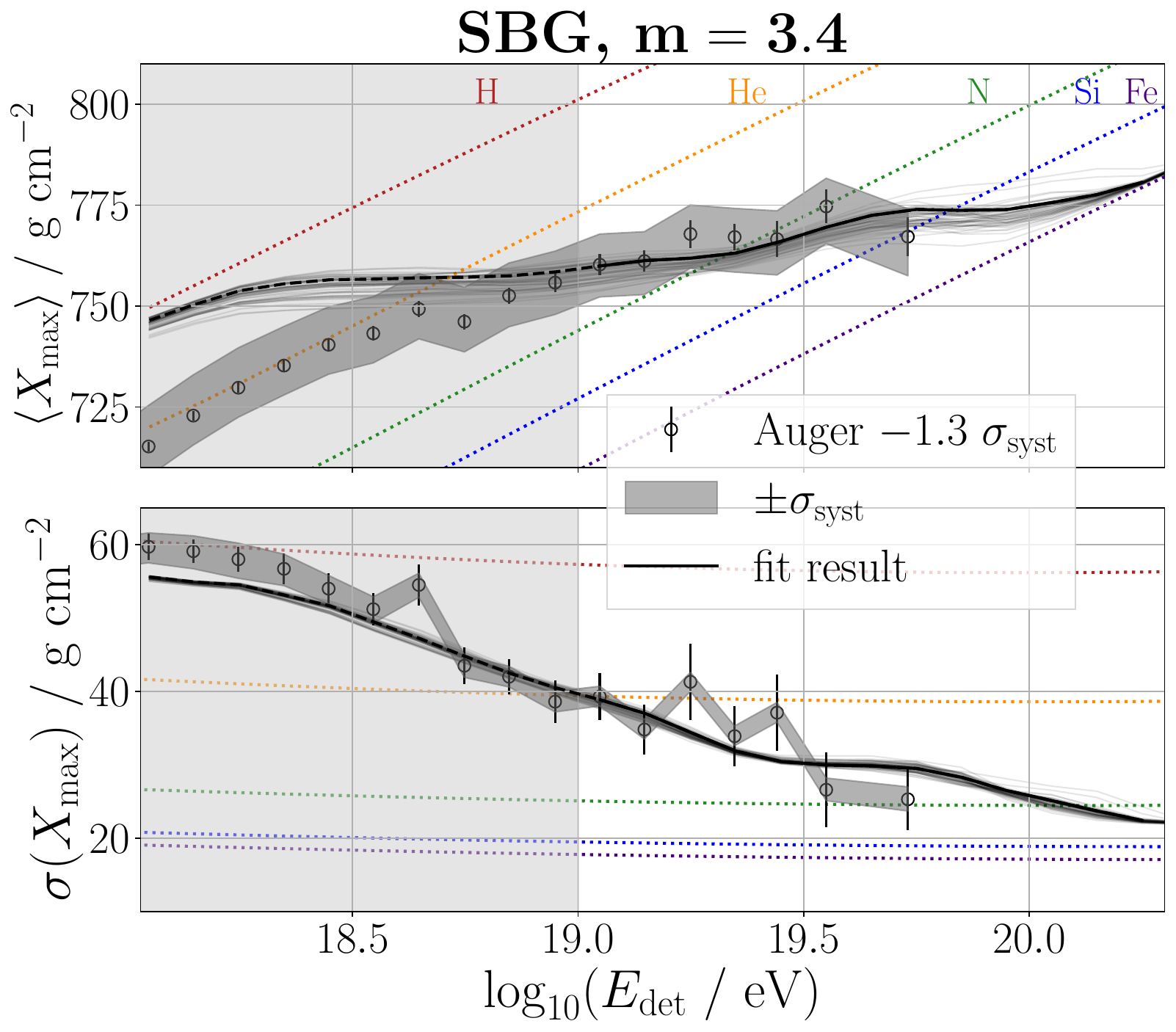}
\caption{Same as \cref{fig:xmax} but including experimental systematic uncertainties as nuisance parameters.}
\label{fig:xmax_syst}
\end{figure}

As displayed in \cref{fig:xmax_syst}, the measured mean shower maximum depth is now substantially better described than without the shift of the \xmax scale.

\renewcommand{\arraystretch}{1.2} % Default value: 1
\begin{table}[t]
\centering
\resizebox{\textwidth}{!}{%
\begin{tabular}{l | l l | l l | l l }
& \multicolumn{2}{c|}{\textbf{Cen A, }$\boldsymbol{m=0}$ (flat)} & \multicolumn{2}{c|}{\textbf{Cen A, }$\boldsymbol{m=3.4}$ (SFR)} & \multicolumn{2}{c}{\textbf{SBG, }$\boldsymbol{m=3.4}$ (SFR)}  \\ 
& posterior & MLE & posterior & MLE & posterior & MLE\\
\hline \hline
$\gamma$ & $-0.89_{-0.33}^{+0.37}$ & $-0.65$ & $-1.19_{-0.39}^{+0.45}$ & $-1.41$ & $-1.02_{-0.36}^{+0.43}$ & $-1.25$ \\
$\log_{10} (R_\mathrm{cut}$/V) & \phantom{+}$18.20_{-0.05}^{+0.04}$ & \phantom{+}18.23 & \phantom{+}$18.21_{-0.05}^{+0.04}$ & \phantom{+}18.20 & \phantom{+}$18.24_{-0.06}^{+0.04}$ & \phantom{+}18.22 \\
$f_0$ & \phantom{+}$0.07_{-0.05}^{+0.01}$ & \phantom{+}$0.029$ & \phantom{+}$0.07_{-0.05}^{+0.01}$ & \phantom{+}$0.031$ & \phantom{+}$0.19_{-0.11}^{+0.07}$ & \phantom{+}0.23 \\
$\delta_0 / ^\circ$ & \phantom{+}$30.5_{-20.2}^{+2.0}$ & \phantom{+}14.4 & \phantom{+}$27.4_{-17.0}^{+4.2}$ & \phantom{+}14.3 & \phantom{+}$18.8_{-3.6}^{+5.9}$ & \phantom{+}21.9 \\
$I_\mathrm{H}$ & $5.8_{-2.6}^{+2.9} \times 10^{-2}$ & $4.2\times 10^{-4}$ & $1.2_{-1.2}^{+0.2} \times 10^{-2}$ & $3.0\times 10^{-4}$ & $1.2_{-1.2}^{+0.1} \times 10^{-2}$ & $1.0\times 10^{-4}$ \\
$I_\mathrm{He}$ & $2.7_{-0.4}^{+0.4} \times 10^{-1}$ & $3.5\times 10^{-1}$ & $9.9_{-2.9}^{+3.8} \times 10^{-2}$ & $1.2\times 10^{-1}$ & $1.1_{-0.4}^{+0.3} \times 10^{-1}$ & $1.4\times 10^{-1}$ \\
$I_\mathrm{N}$ & $5.6_{-0.4}^{+0.4} \times 10^{-1}$ & $5.0\times 10^{-1}$ & $6.7_{-0.7}^{+0.7} \times 10^{-1}$ & $6.8\times 10^{-1}$ & $7.2_{-0.6}^{+0.6} \times 10^{-1}$ & $7.3\times 10^{-1}$ \\
$I_\mathrm{Si}$ & $9.0_{-3.4}^{+3.9} \times 10^{-2}$ & $1.4\times 10^{-1}$ & $1.5_{-0.6}^{+0.5} \times 10^{-1}$ & $1.6\times 10^{-1}$ & $1.2_{-0.5}^{+0.5} \times 10^{-1}$ & $9.8\times 10^{-2}$ \\
$I_\mathrm{Fe}$ & $2.3_{-1.2}^{+0.9} \times 10^{-2}$ & $1.8\times 10^{-2}$ & $5.1_{-1.8}^{+1.5} \times 10^{-2}$ & $4.4\times 10^{-2}$ & $4.7_{-1.7}^{+1.3} \times 10^{-2}$ & $3.8\times 10^{-2}$ \\
\hline
$\nu_{E} / \sigma$ & $-1.24_{-0.50}^{+0.68}$ & $-1.35$ & \phantom{+}$0.23_{-0.60}^{+0.42}$ & \phantom{+}$0.13$ & \phantom{+}$0.35_{-0.65}^{+0.44}$ & \phantom{+}$0.40$ \\
$\nu_{X\mathrm{max}} / \sigma$ & $-0.94_{-0.24}^{+0.29}$ & $-0.97$ & $-1.60_{-0.25}^{+0.30}$ & $-1.45$ & $-1.55_{-0.25}^{+0.26}$ & $-1.33$ \\
\hline
$\boldsymbol{\log b }$ & $-254.6  \pm 0.1$ &  & $-264.5 \pm 0.2$ &  & $-258.6 \pm 0.2$ &  \\
$\boldsymbol{D_\mathrm{syst}}$ &  & \phantom{+}2.8 &  & \phantom{+}2.1 &  & \phantom{+}1.9 \\
$\boldsymbol{D_E} \ (N_J=14)$ &  & \phantom{+}13.6 &  & \phantom{+}21.9 &  & \phantom{+}25.3 \\
$\boldsymbol{D_{X_\mathrm{max}}} \ (N_{{X}_\mathrm{max}}=74)$ &  & \phantom{+}107.4 &  & \phantom{+}113.6 &  & \phantom{+}112.7 \\
$\boldsymbol{D}$ &  & \phantom{+}123.8 &  & \phantom{+}137.7 &  & \phantom{+}139.9 \\
$\boldsymbol{\log}$ $\boldsymbol{\mathcal{L}_\mathrm{ADs}}$ &  & \phantom{+}$9.4$ &  & \phantom{+}$9.5$ &  & \phantom{+}$13.5$ \\
$\boldsymbol{\log}$ $\boldsymbol{\mathcal{L}}$ &  & $-228.51$ &  & $-235.3$ &  & $-232.4$
\end{tabular}}
\caption{Centaurus A and SBG models with experimental systematic uncertainties included as nuisance parameters}
\label{tab:results_syst}
\end{table}

\subsubsection{Evaluation of the model performances}
\label{sec:model_performances}

% large uncertainty on TS -> UHECR proceeding
% especially on below/above 0, but always peak around 40 EeV
% This is just best-fit, other parts of posterior may look differently (e.g. SBG E negative)

\begin{figure}[ht]
\centering
\includegraphics[width=1.0\textwidth]{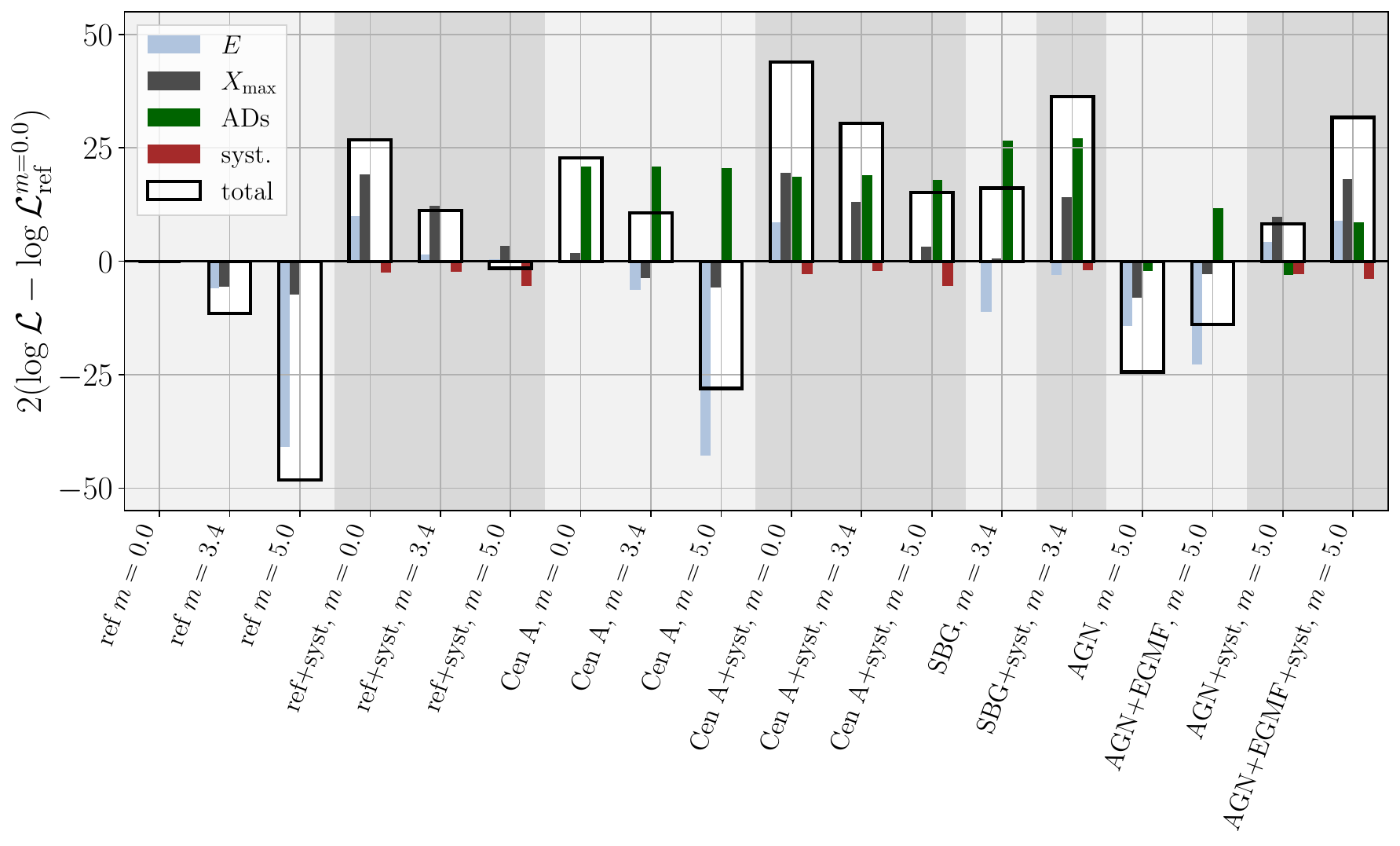}
\caption{Overview of the likelihood values for all tested models: for better comparability, for all models 2 times the likelihood ratio with the reference model with $m=0.0$ is given. The white bar with black edges gives the total likelihood, and the four smaller bars inside give the contributions of the observables and the systematics to the total likelihood.}
\label{fig:TS_all_overview}
\end{figure}

How well each of the tested models describes the measured data can be quantified for example by comparing the values of the likelihood function for the best-fit model parameters. An overview of all likelihood values  is given in \cref{fig:TS_all_overview}, including the contributions by the different observables. Here, the general trends that have been described above are visible, for example, that the models with strong source evolution are disfavored. This is mostly due to a poor description of the energy spectrum (blue). Also, when comparing the models with and without systematics, one can see that the likelihood ratio improves consistently when systematic shifts are allowed, mainly due to a better description of the shower maximum depth distributions (grey), but also the energy spectrum shows an improvement for the cases with strong evolution. The arrival direction likelihood (green) is almost independent of the source evolution and the systematics, and is always the largest for the SBG catalog.
%, it is consistently around 20 for the Centaurus A models, and around 27 for the SBG models.

For a more quantitative comparison, we use the test statistic calculated as 2 times the likelihood ratio between a model and the respective reference model with the same evolution and (no) systematics:
\begin{equation}
    \mathrm{TS}_\mathrm{tot} = \sum_{\mathrm{obs}=E,\xmax,\mathrm{ADs}} 2 (\log \mathcal{L}^{m=x} - \log\mathcal{L}_\mathrm{ref}^{m=x})^\mathrm{obs}.
\end{equation}
Hence, the test statistic describes the improvement of adding a specific catalog to a model compared to just homogeneous sources.
The values for the test statistic of each model are given in \cref{tab:TS}. As is apparent from the table, the arrival directions observable provides the largest contribution to the total test statistic. This is understandable, as the reference model already provides a proper fit of the energy spectrum and \xmax data~\cite{the_pierre_auger_collaboration_a_abdul_halim_constraining_2023}, so the subdominant contribution by the nearby source candidates only has a minor impact on these observables. For the arrival directions, however, the improvement from fully isotropic arrival directions in the reference model, to the anisotropic ones provided by the model including source candidates (Fig.~\ref{fig:ADs}), is substantial. Hence, the energy spectrum and \xmax distributions are necessary for constraining the source emission, while the arrival directions are most important for the differentiation of different source candidates.
This is as expected from a simulation study~\cite{t_bister_for_the_pierre_auger_collaboration_sensitivity_2023}. But, from that analysis, it also becomes clear that the exact values of TS should be treated with caution as they can vary considerably and depend on e.g. the distribution of the arrival directions in an energy bin.

\begin{table}[ht]
%\centering
\resizebox{1.0\textwidth}{!}{%
\begin{tabular}{l | c  c | c  c | c  c | c  c | c  c}
 & \multicolumn{2}{c|}{\textbf{Cen A}, \boldmath{$m=0.0$}} & \multicolumn{2}{c|}{\textbf{Cen A}, \boldmath{$m=3.4$}} & \multicolumn{2}{c|}{\textbf{SBG}, \boldmath{$m=3.4$}} & \multicolumn{2}{c|}{ \boldmath{$\gamma$}\textbf{AGN}, \boldmath{$m=5.0$}} & \multicolumn{2}{c}{ \boldmath{$\gamma$}\textbf{AGN+EGMF}**, \boldmath{$m=5.0$}}\\
 \hline
 & & + syst & & + syst & & + syst & & + syst & & + syst\\
\hline
\hline
TS$_\mathrm{tot}$ & 22.8 & 17.3 & 22.2 & 19.1 & 27.6 & 25.6 & 23.9* & 9.8* & 34.3* & 33.2*\\

TS$_E$ & $-0.1$ & $-1.4$ & $-0.4$ & $-1.1$ & $-5.2$ & $-4.5$ & 26.8 & 3.9 & 18.2 & 8.4\\
TS$_{\xmax}$ & 1.9 & 0.2 & 1.8 & $1.0$ & $6.2$ & $2.0$ & $-0.8$ & 6.4 & 4.4 & 14.7\\
TS$_\mathrm{ADs}$ & 20.9 & 18.7 & 20.8 & 19.0 & $26.6$ &  27.1 & $-2.1$ & $-3.0$ & 11.7 & 8.6\\
\end{tabular}}
\caption{Overview of test statistic values for the tested models. In the case with systematics, the contribution from \cref{eq:likelihood_syst} is taken into account additionally in TS$_\mathrm{tot}$. *\,Note that the test statistic of each model is always calculated with respect to the reference model with the same source evolution and (no) systematics. This implies that e.g. the test statistics of the $\gamma$-AGN model is large only due to the poor fit of the reference model with $m=5.0$ (compare to \cref{fig:TS_all_overview}). **\,Note also that the model with EGMF has an additional fit parameter $\beta_e$ compared to the other models (see \cref{sec:AGN_EGMF}).}
\label{tab:TS}
% Or use%\vspace*{5cm}  % with the correct table height
\end{table}

Additionally, it is important to note that the test statistic only quantifies the best-fit result. As indicated by the broad posterior distributions (see \cref{fig:posterior}), alternative combinations of model parameters can lead to similar values of the total likelihood function. This means that the SBG model can also describe the data almost equally well with a smaller signal fraction and harder spectral index, leading to a better description of the energy spectrum but simultaneously a worse description of the arrival directions. So, the values given for the test statistic of individual observables in \cref{tab:TS} depend largely on the best-fit parameter combination. 

Due to this, it is often useful to not only compare the test statistic of the best-fit solution, but also the Bayesian evidence as an additional test. As introduced in \cref{sec:fitting_techniques}, it can be used to compare the fit quality of a whole model, not just the best-fit parameter combination. The values for the Bayesian evidence $\log b$ for each model are stated in tables~\ref{tab:results} and \ref{tab:results_syst}. When comparing the values of the evidence, it is clear that they follow the same trend as the maximum likelihood value which means that the maximum-likelihood value / total test statistic is representative of the fit quality of the whole model. This allows us to focus on the maximum-likelihood values in the following.

\begin{figure}[th]
\centering
\includegraphics[width=0.49\textwidth]{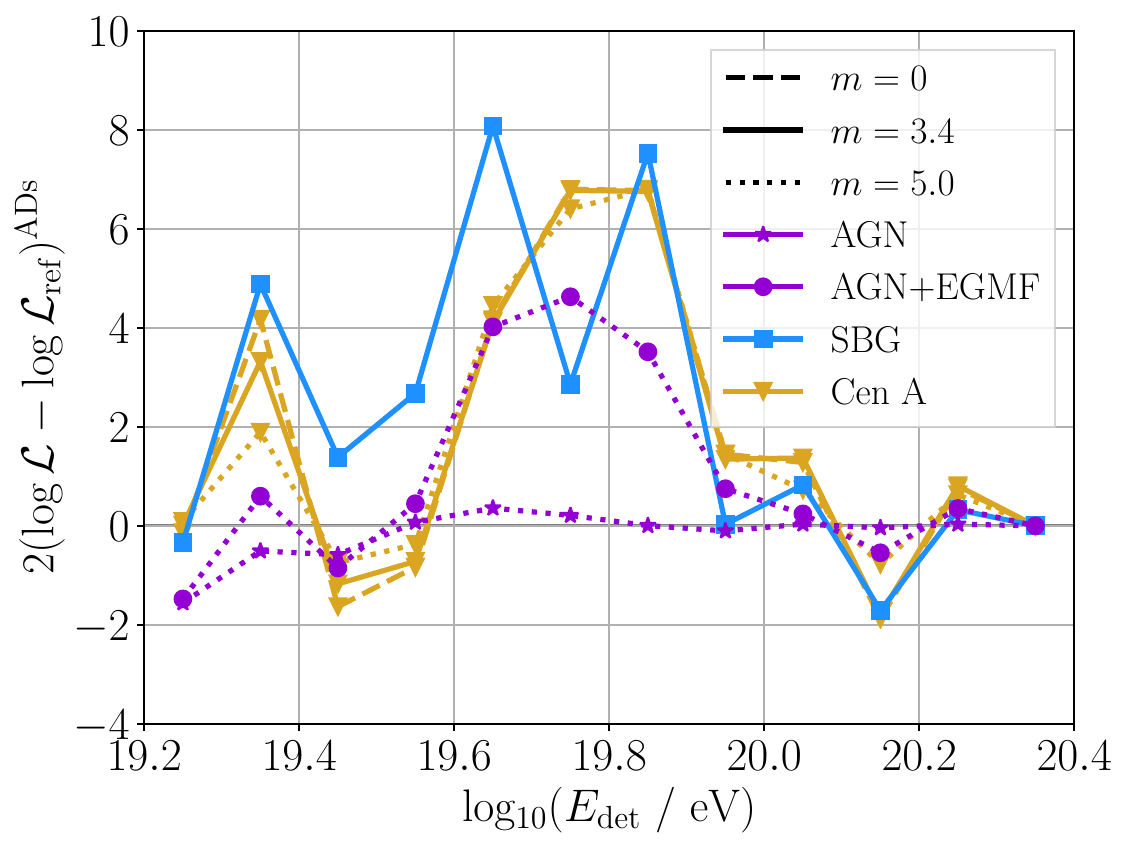}
\includegraphics[width=0.49\textwidth]{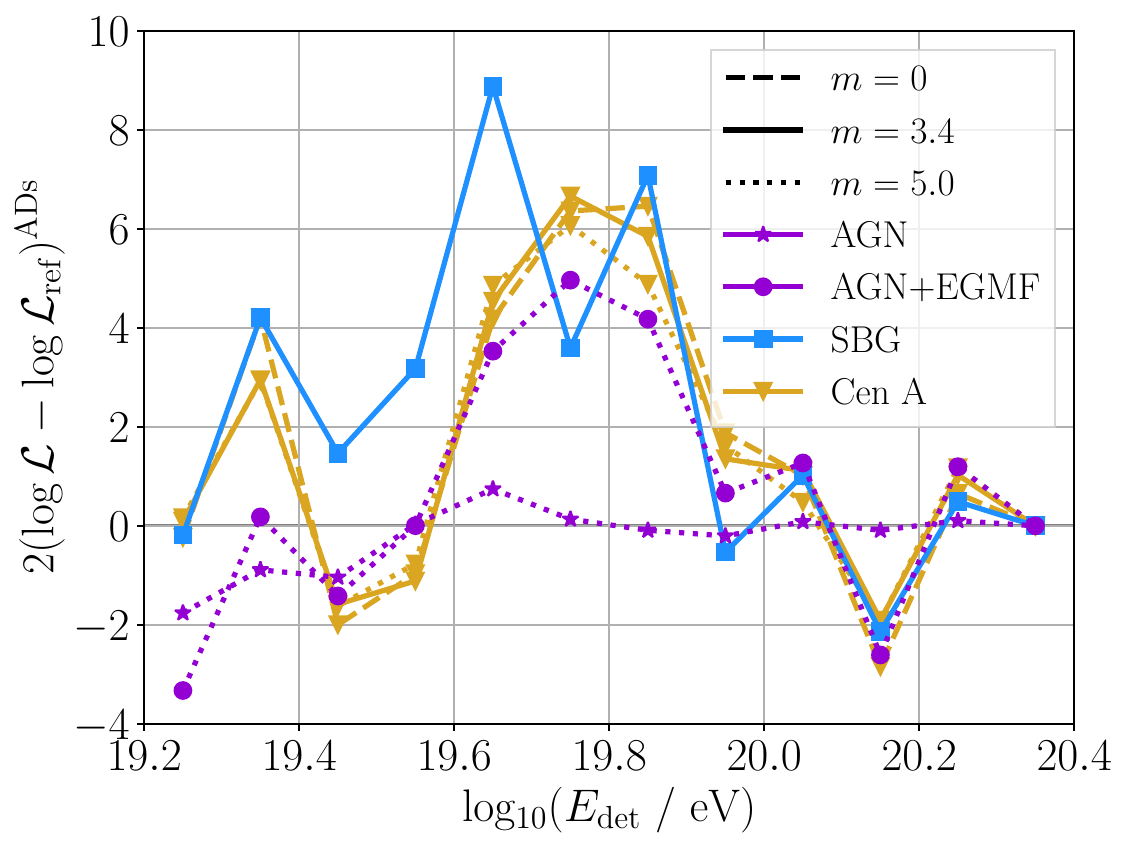}
\caption{Arrival directions likelihood ratio / test statistic as a function of the energy bin (marker sits on the bin center) for the tested source classes and evolutions. The test statistic is always calculated compared to the reference model with the same evolution and without (\textit{left}) / with (\textit{right}) systematics. The sum over all energy bins corresponds to the values for TS$_\mathrm{ADs}$ given in \cref{tab:TS}.}
\label{fig:TS_AD_models}
\end{figure}

The arrival directions test statistic contributed by each separate energy bin can be seen in \cref{fig:TS_AD_models}. Here, a general trend can be observed, independently of the systematic effects and of the source evolution. The arrival directions are well described in the energy bin $\edet=19.3$ and the bins around $\edet=19.7$ for both the SBG and the Centaurus A models. Peaks of the arrival direction test statistic at threshold energies of 40\,EeV$\simeq10^{19.6}~\mathrm{eV}$ and 60\,EeV$\simeq10^{19.8}~\mathrm{eV}$ have also been observed in Refs.~\cite{the_pierre_auger_collaboration_p_abreu_arrival_2022, the_pierre_auger_collaboration_a_aab_et_al_indication_2018}. But, only with the present analysis which models the energy dependency of each mass group's contribution, these can be compared to the modeled spectra of the strongest sources in the catalog (\cref{fig:spectra} and \cref{fig:spectra_syst}). Here, one can see that at $\edet=19.3$ the helium contribution has its peak, while at $\edet=19.6$ nitrogen is predominant, and at $\edet=19.8$ silicon begins to emerge.
At these energies, the rigidities correspond to $R_\mathrm{He}=\frac{10^{19.3}}{2}\mathrm{V}\simeq10~\mathrm{EV}$, $R_\mathrm{N}=\frac{10^{19.6}}{7}\mathrm{V}\simeq6~\mathrm{EV}$ and $R_\mathrm{Si}=\frac{10^{19.8}}{14}\mathrm{V}\simeq5~\mathrm{EV}$, respectively.
The similar rigidities of around 6\,eV for the nitrogen and silicon contributions lead to a relatively constant blurring over the whole energy range which could explain why the test statistic of~\cite{the_pierre_auger_collaboration_a_aab_et_al_indication_2018, the_pierre_auger_collaboration_p_abreu_arrival_2022} is so large (TS$_\text{ADs-only}^\mathrm{SBG}\simeq25.0$) even though the correlation analysis is performed together for all CRs above an energy threshold with one fixed blurring. It is important, however, to note that the energy dependency of the test statistic is subject to large statistical fluctuations~\cite{t_bister_for_the_pierre_auger_collaboration_sensitivity_2023}, so especially the less pronounced peak at $\edet=19.3$ could just be a statistical fluctuation.

The energy dependency of the arrival direction test statistic reveals another interesting finding, which is that the SBG model consistently describes the measured arrival directions better than the Centaurus A model in the energy range $\edet=19.4-19.6$. For these energies, the modeled arrival direction pdfs (\cref{fig:ADs}) are very similar for both models in the region around Centaurus A / NGC 4945, but differ in the southern region around NGC 253. From \cref{tab:TS} it can be concluded that $\sim20$ of the total arrival directions test statistic for the SBG model originates from the Centaurus A / NGC 4945 region. By removing NGC~253 from the catalog we explicitly checked that that source contributes around $\sim4$ to 5 to the arrival directions test statistic.

\subsection[The $\gamma$-AGN model]{The $\boldsymbol{\gamma}$-AGN model} \label{sec:AGNs}
% ADs shift to Cen A at 100 EeV
% systematics do not change arrival directions
% with systematics E and Xmax fit relatively well
When the $\gamma$-AGN model with source evolution $m=5.0$ is applied to the data, as expected from the reference model (\cref{sec:ref}) a very hard spectral index of $\gamma\approx-3.5$ is found in combination with a small maximum rigidity of $\rcut\approx18.1$. The best-fit signal fraction is around $f_0\approx15\%$. The $\gamma$-AGN catalog is dominated by a single source between 10\,EeV and 100\,EeV, the faraway blazar Markarian 421, which leads to a constant catalog contribution in that energy range of around 15\%. 

When comparing the likelihood of the $\gamma$-AGN model with the respective reference model with $m=5.0$ (\cref{fig:TS_all_overview} and \cref{tab:TS}), it is clear that the inclusion of the catalog sources mainly leads to an improvement of the energy spectrum description. This is due to the reduced overshooting of the spectrum at low energies by secondary products from faraway background sources, which is reduced by the contribution of the closer catalog sources. Overall, the likelihood of the $\gamma$-AGN model is small, and it does not reach values larger than the reference models with less evolution. For that reason, the best-fit parameters will not be discussed in more depth.

\begin{figure}[ht]
\centering
\includegraphics[width=0.36\textwidth]{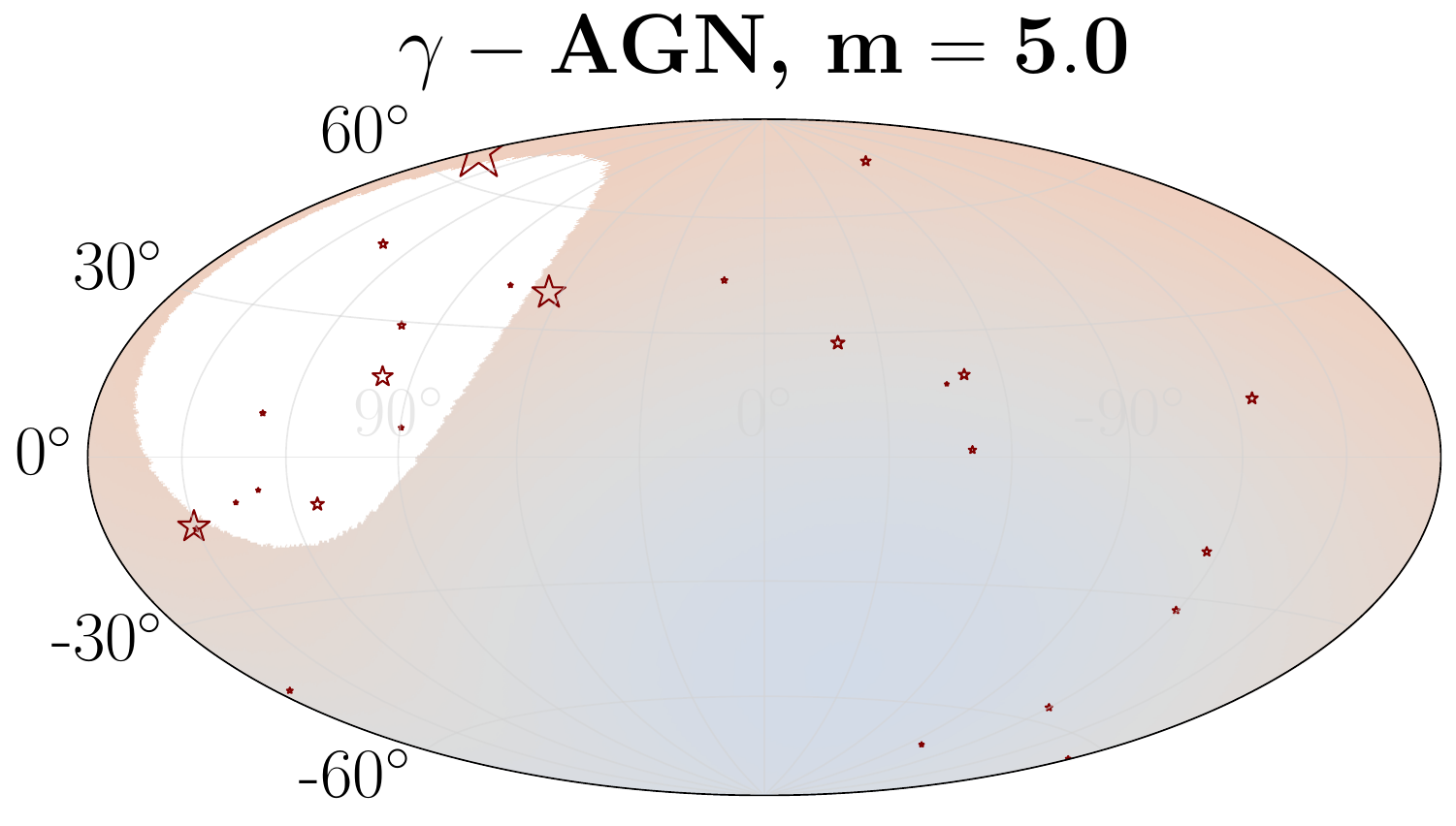}
\includegraphics[width=0.36\textwidth]{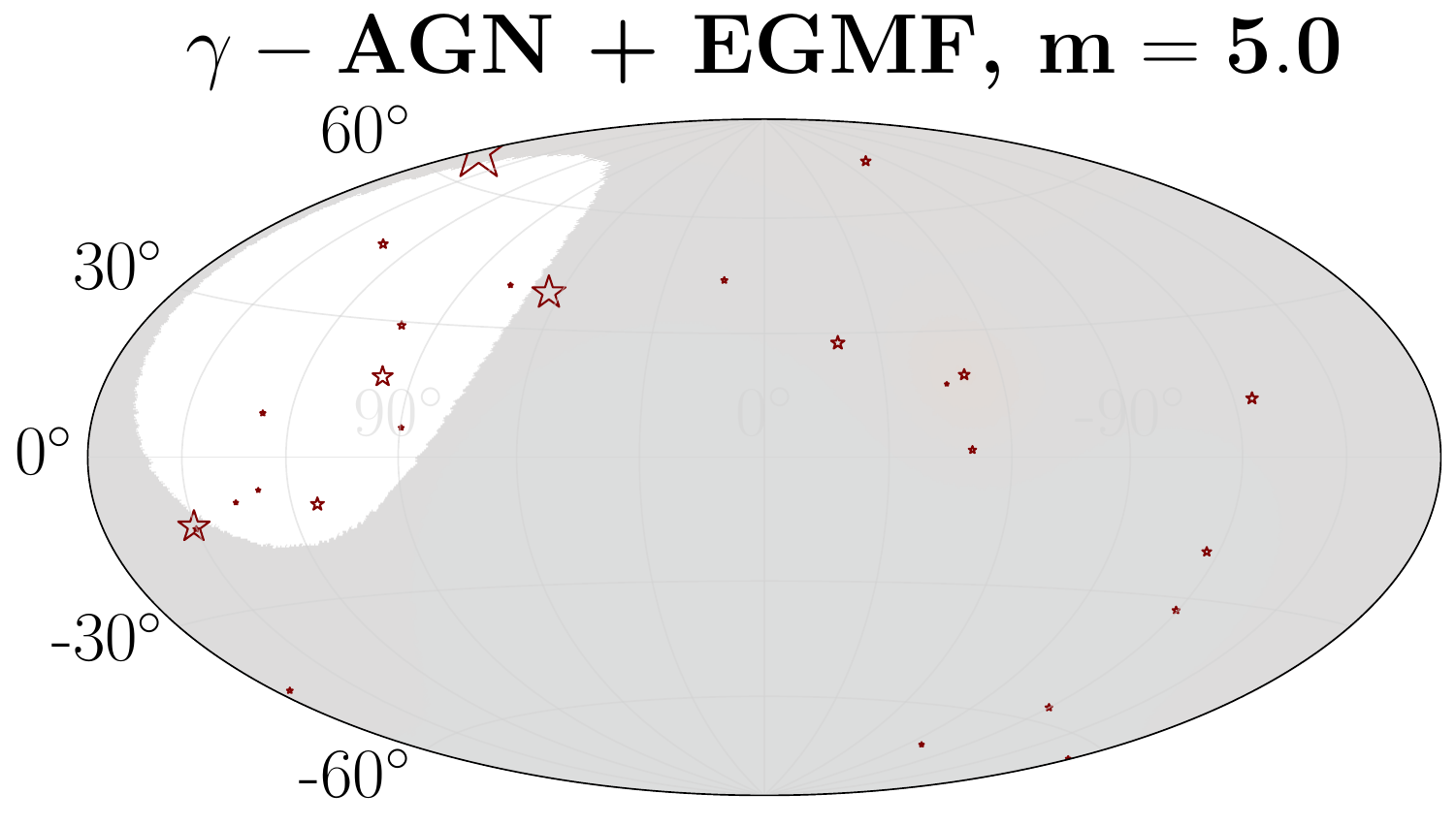}

\includegraphics[width=0.36\textwidth]{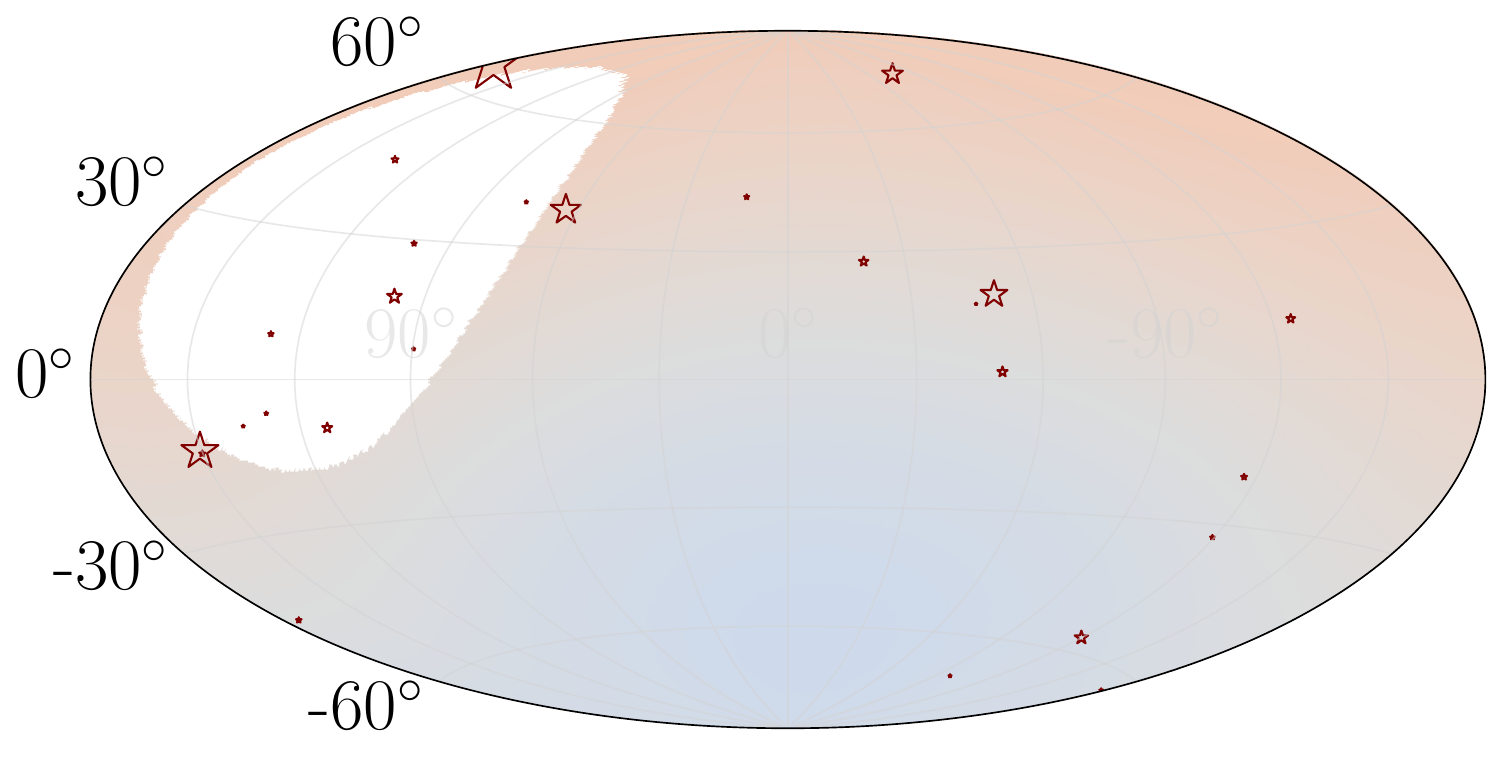}
\includegraphics[width=0.36\textwidth]{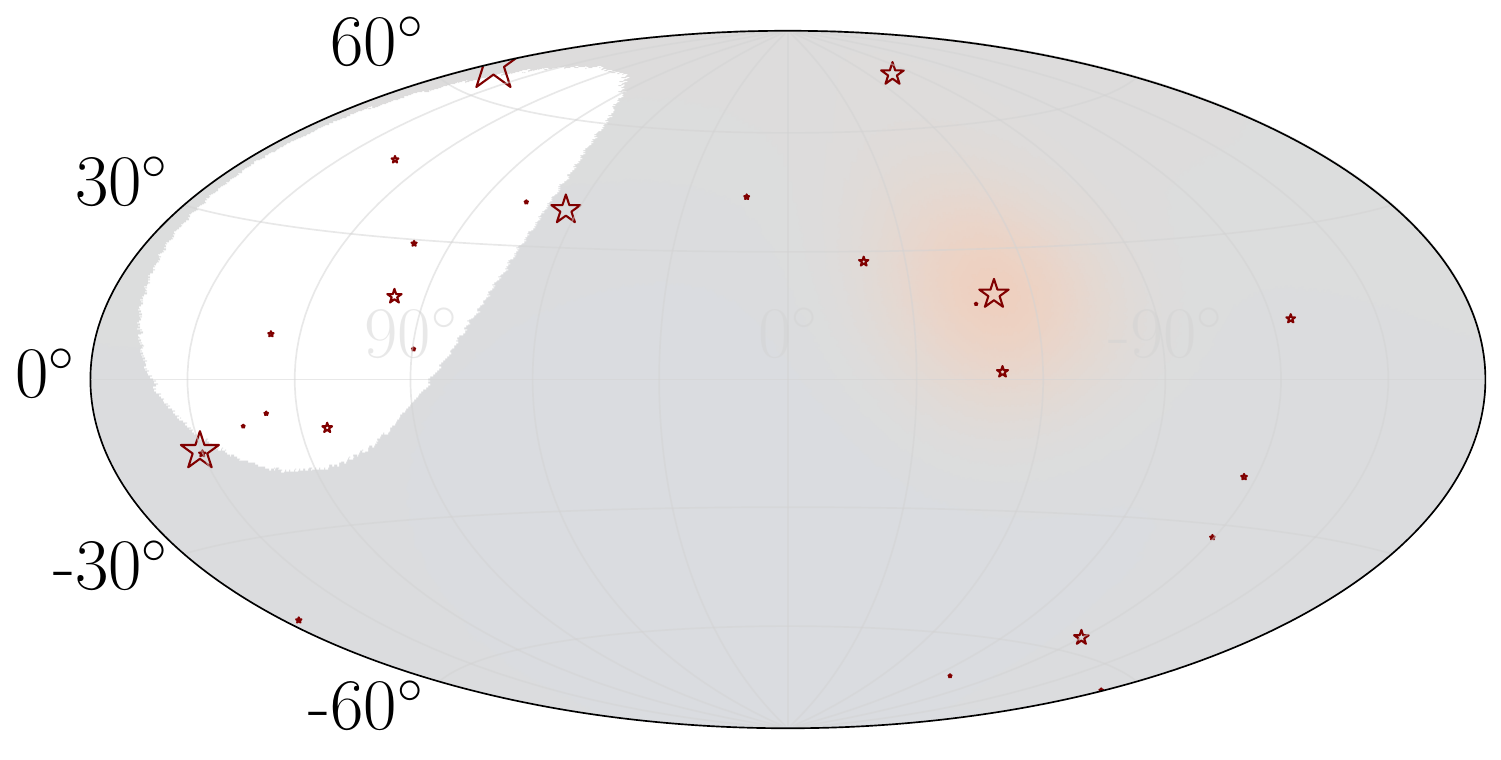}

\includegraphics[width=0.36\textwidth]{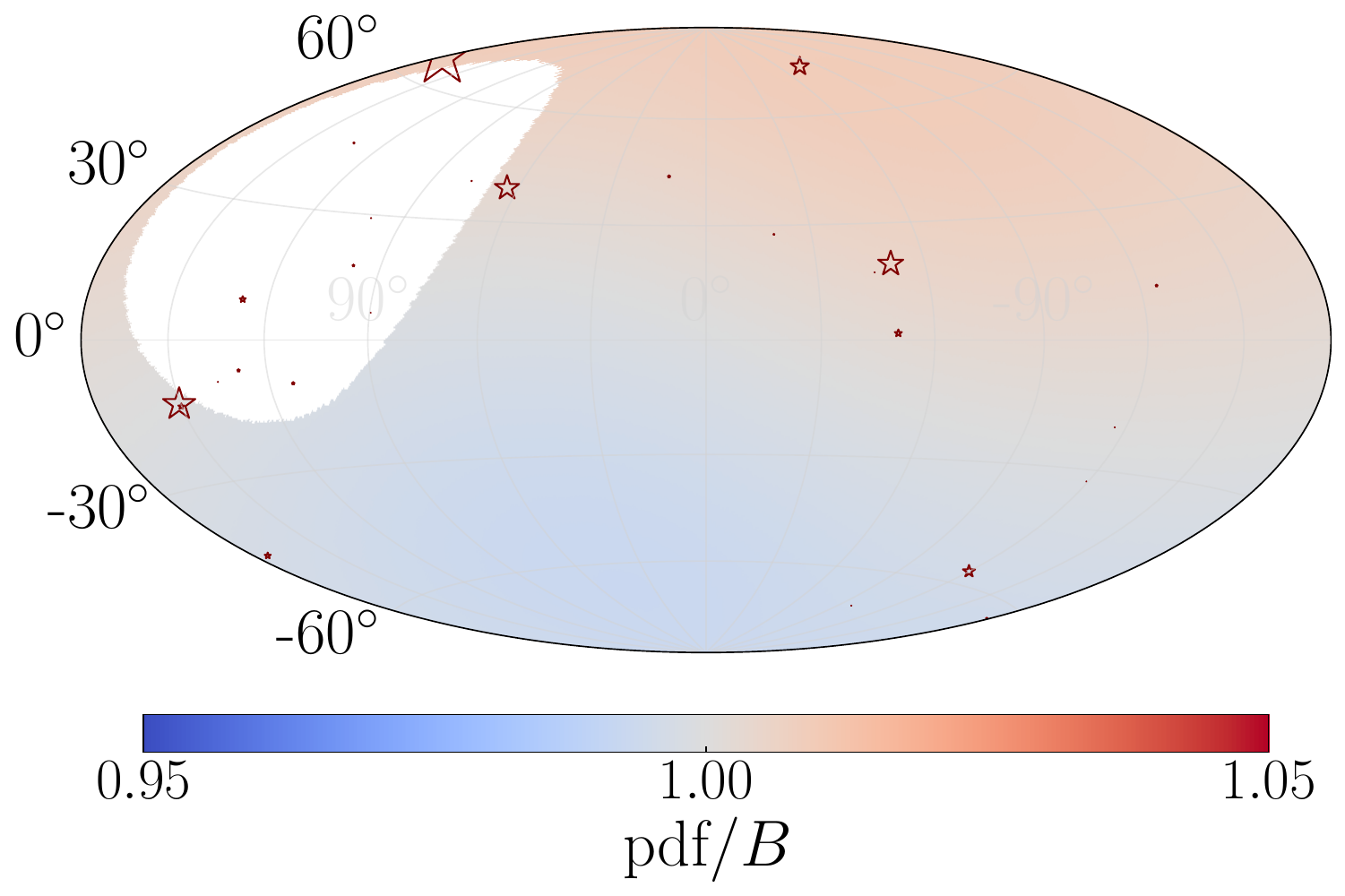}
\includegraphics[width=0.36\textwidth]{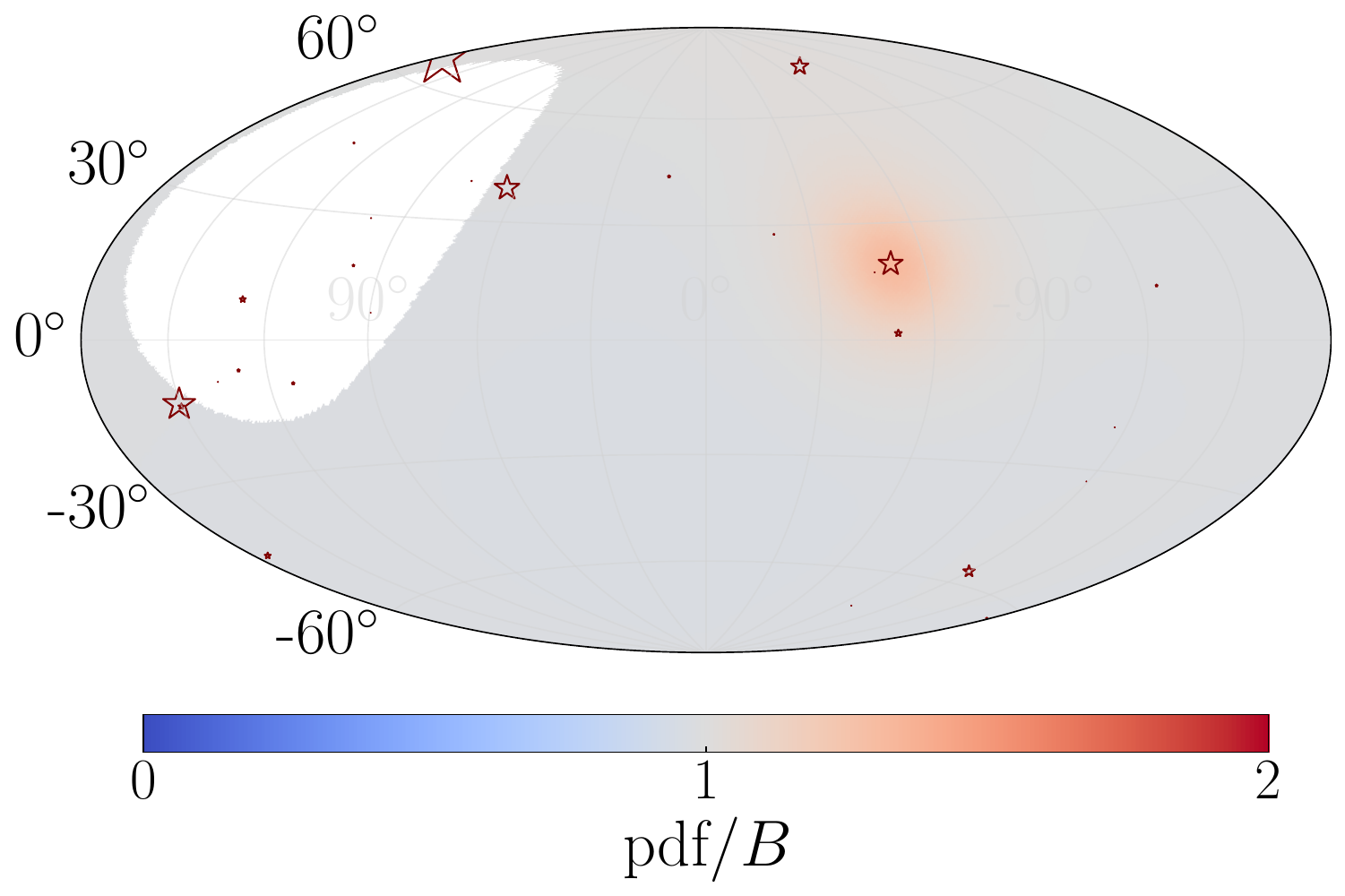}
\caption{Modeled arrival directions for the $\gamma$-AGN model, \textit{left} for the model without and \textit{right} with distance-dependent blurring. Again, the energy bins $\edet=19.3$ (\textit{upper row}), \edet=19.6 (\textit{middle row}) and $\edet=19.9$ (\textit{lower row}) are shown as examples. Note the different color bar normalizations of the two models.}
\label{fig:ADs_AGN}
\end{figure}

Also, it is visible that the arrival directions test statistic is negative. The best-fit arrival directions are depicted in \cref{fig:ADs_AGN} (\textit{upper row}). Here, the dominance of Markarian 421 is clearly visible. Only at the highest energies around 100\,EeV where the statistics of the data set is very small, Centaurus A starts contributing. The best-fit blurring reaches the maximum value allowed in the sampling space (see \cref{sec:priors}). This means, if even larger values of the blurring were accepted, the arrival directions would be modeled as completely isotropic, leading to TS$_\mathrm{ADs}^{\gamma-\mathrm{AGN}}=0$. When looking at the energy dependency of the arrival directions test statistic displayed in \cref{fig:TS_AD_models}, one can see that the negative contribution comes from the smallest energy bins where only Markarian 421 contributes. Note that the poor description of the arrival directions with the $\gamma$-AGN model does not depend on the source evolution of the background sources.

\paragraph{Influence of a distance-dependent blurring:} \label{sec:AGN_EGMF}
We tested if a distance-dependent blurring as expected from an extragalactic magnetic field could improve the agreement between the $\gamma$-AGN model and the measured data\footnote{For the SBG model, we also tested the influence of a distance-dependent blurring. However, as almost all contributing SBGs are located at similar distances of $\sim4$\,Mpc, the distance-dependent and distance-independent blurrings are fully correlated. The likelihood does not increase by more than $\sim0.5$ which is compensated by the additional degree of freedom, so this model is not discussed further here.}. For that, we include an additional fit parameter $\beta_e$ in the arrival directions modeling, and replace \cref{eq:delta_tot} by the following
\begin{equation} \label{eq:delta_tot_egmf}
    \delta(E^e_\mathrm{det}, A^k_\mathrm{det}, C^s) = \frac{\sqrt{\delta_0^2 + \beta_e^2 \ (d^s / \mathrm{Mpc})}}{R_\mathrm{det}(E^e_\mathrm{det}, A^k_\mathrm{det}) / 10 \ \mathrm{EV}}.
\end{equation}
The prior for the distance-dependent blurring is flat $\beta_e\in[0^\circ, 11.5^\circ]$. This means, a 10\,EV particle can at most obtain a distance-dependent blurring of $\sim20^\circ$ for a source distance of 3\,Mpc. This maximum value was chosen so that the magnetic field parameters do not become much larger than theoretical limits~\cite{durrer_cosmological_2013}. 

The found model parameters for the $\gamma$-AGN model with distance-dependent blurring do not differ much from the ones without it. The posterior distribution of $\beta_e$ is broad between $\beta_e=5^\circ$ and the maximum allowed $\beta_e=11.5^\circ$, while the distance-dependent blurring has decreased to $\delta_0\simeq20^\circ$. This leads to the arrival directions displayed in \cref{fig:ADs_AGN}. In contrast with the case without distance-dependent blurring, now the contribution by faraway Markarian 421 is completely blurred out, and only Centaurus A is visible above $\edet\simeq19.5$. These arrival directions now describe the data better than the isotropic reference model, as visible in \cref{fig:TS_all_overview}. Nevertheless, with TS$_\mathrm{ADs}^{\gamma\mathrm{-AGN+EGMF}}\simeq10$, it still does not describe the measured arrival directions as well as the models with only Centaurus A as a single source do. We tested explicitly if this model can reproduce the results from Refs.~\cite{the_pierre_auger_collaboration_a_aab_et_al_indication_2018, the_pierre_auger_collaboration_p_abreu_arrival_2022} and found that it cannot reach large enough arrival direction test statistics.

For that reason, we can conclude that the $\gamma$-AGN model where the flux contribution scales linearly with the $\gamma$-ray intensity, and which thus favors blazars like Markarian 421, is not supported by the data of the Pierre Auger Observatory. The measured arrival directions cannot be described sufficiently well even when including an extragalactic magnetic field with a strength in the range of theoretical limits. Additionally, as discussed above, the strong source evolution associated with high-luminosity AGNs leads to an overproduction of secondaries during propagation, which causes models with strong evolution to poorly describe the measured energy spectrum.

\section{Discussion and conclusion} \label{sec:conclusion}
The arrival directions of UHECRs exhibit correlations with catalogs of powerful extragalactic objects~\cite{the_pierre_auger_collaboration_p_abreu_arrival_2022, the_pierre_auger_collaboration_a_aab_et_al_indication_2018}. In this work, these correlations with a catalog of starburst galaxies and one of jetted active galactic nuclei have been investigated further using, for the first time, a combined fit of energy spectrum, shower maximum depth distributions, and arrival directions. The astrophysical model that is used in the fit includes propagation effects and an energy-dependent modeling of the contribution of each source depending on its distance and injection spectrum. Also, a rigidity-dependent magnetic field blurring is included, which is calculated individually for each source depending on the element contributions of the respective source on Earth. At the current stage, no coherent magnetic field deflections are employed in the model, so all obtained results are under the assumption of mostly turbulent deflections by the Galactic and extragalactic magnetic fields, at least in the directions of the strongest sources. In the future, coherent deflections could also be included in the astrophysical model, hoping also that a better convergence between different Galactic magnetic field models will be reached.
%It is tested whether or not the measured arrival directions and shower depth distributions on Earth as functions of energy and the energy spectrum can be described by the model. 

The model has several free fit parameters, in particular those describing the injected spectrum and composition at the sources. Compared to the reference models with only homogeneous background sources (as used in Refs.~\cite{the_pierre_auger_collaboration_a_aab_combined_2017, the_pierre_auger_collaboration_a_abdul_halim_constraining_2023}), two additional fit parameters are introduced, the signal fraction of the nearby sources at fixed energy, and the size of the magnetic field blurring that scales with rigidity. By modeling the arrival directions as a sum of weighted von~Mises--Fisher distributions for each source and each mass number group in each energy bin, the model naturally produces overdensities which decrease in size for increasing rigidity. Also, the flux modulations become stronger as the contribution from the nearby catalog sources rises with the energy. This enables the modeling of the energy evolution of the arrival directions, which can then be compared to the data in energy bins, and not just above an energy threshold as in previous analyses~\cite{the_pierre_auger_collaboration_a_abdul_halim_constraining_2023, the_pierre_auger_collaboration_a_aab_combined_2017}.

The largest test statistic of TS$_\mathrm{tot}^\mathrm{SBG}=27.6$ (indicating how much the likelihood improves by adding catalog sources, compared to the reference model with only homogeneous background) is found for the starburst galaxy model 
%The best agreement with the data for models with a background source evolution following the star-formation rate is found for the SBG model
with a hard spectral index, a signal fraction of $\sim20\%$ at 40\,EeV and a blurring\footnote{Note that in the non-resonant scattering regime, the expected root-mean-square blurring from turbulent magnetic fields is $\delta_\mathrm{rms} \simeq 20^\circ \frac{B}{2.5\,\mathrm{nG}} \frac{10\,\mathrm{EV}}{E/Z} \sqrt{\frac{d_\mathrm{source}}{4\,\mathrm{Mpc}}} \sqrt{\frac{l_\mathrm{c}}{\mathrm{Mpc}}}$~\cite{mollerach_progress_2018}, allowing for conclusions on the corresponding coherence length $l_\mathrm{c}$ and magnetic field strength $B$ for the observed blurring of $20^\circ$ at $R=10$ EV.} of $\sim20^\circ$ for a particle with a rigidity of 10\,EV.
% For this model, a test statistic of TS$_\mathrm{tot}^\mathrm{SBG}=27.6$ is reached. 
Compared to the analysis using only the (energy-independent) arrival directions~\cite{the_pierre_auger_collaboration_p_abreu_arrival_2022, the_pierre_auger_collaboration_a_aab_et_al_indication_2018} with TS$_\text{ADs-only}^\mathrm{SBG}=25.0$, this corresponds to an increase, which indicates a good description of the energy evolution of the arrival directions by the SBG model.

Even when considering the most important systematic effects on the energy and \xmax scales, the total test statistic of the combined fit is still TS$_\mathrm{tot}^\mathrm{SBG+syst}=25.6$. Additionally, different to Refs.~\cite{the_pierre_auger_collaboration_p_abreu_arrival_2022, the_pierre_auger_collaboration_a_aab_et_al_indication_2018} no scan of the energy threshold is necessary as the whole energy dependency is included in the astrophysical model for the combined fit. In Ref.~\cite{t_bister_for_the_pierre_auger_collaboration_combined_2021} the conversion of the total test statistic to a significance was investigated. We found that the significance can approximately be calculated using a $\chi^2$ distribution with two degrees of freedom for the two additional fit parameters ($f_0$, $\delta_0$) compared to the respective reference model. With this, the 1-sided significance of the SBG model amounts to $\sim4.5\sigma$ (or $\sim4.7\sigma$ when not including the effects of experimental systematic uncertainties).
The total test statistic is dominated by a region in which two interesting source candidates reside, NGC 4945 and Centaurus A. In a separate investigation of a model where we picked just Centaurus A on top of a homogeneous background, a test statistic of around TS$_\mathrm{tot}^\mathrm{Cen A}\simeq20$ is found, which is almost independent of the background evolution and systematic uncertainties. The contribution of Centaurus A to the total flux is estimated to be around $3\%$ at 40\,EeV (although this fraction could be as large as $25\%$ for the case of flat evolution), with a dominant nitrogen component reaching Earth above this energy. The signal fraction and blurring for the model with Centaurus A as a single source are in agreement with the ones obtained for NGC 4945 in the case of the SBG model.

According to the results of this work, the $\gamma$-AGN model, which exhibits a test statistic of TS$_\text{ADs-only}^{\gamma\mathrm{-AGN}}=17.9$ in the arrival direction correlation analysis~\cite{the_pierre_auger_collaboration_p_abreu_arrival_2022}, does not agree with the energy-dependent arrival directions. This is mostly due to the strong contribution of the faraway blazar Markarian 421 due to its large UHECR flux weight, calculated from the $\gamma$-ray flux in the model. From this, it can be concluded that the $\gamma$-ray emission, which tends to favor the narrowly beamed jetted blazars pointing towards Earth, may not be an optimal tracer of UHECR emission for which the beaming is expected to be much broader due to deflections by cosmic magnetic fields~\cite{murase_blazars_2012}.
The $\gamma$-AGN model also does not describe the energy spectrum adequately, mostly due to the overproduction of secondaries at energies of 5\,EeV to 10\,EeV owing to the strong source evolution considered, something which should also affect other potential scenarios associated with high-luminosity AGNs.
Nevertheless, as described above, a model based solely on the nearby AGN Centaurus A, which is also part of the jetted AGN catalog, does reproduce the data well.

Future directions for the present analysis could be the inclusion of deflections by the regular part of the magnetic field of the Galaxy as described above. Also, other source catalogs with different flux weighting, as well as an extension of the analysis to lower energies, where other populations of sources may also contribute, could be investigated. Additionally, the analysis will profit from the on-going upgrade of the Pierre Auger Observatory, AugerPrime~\cite{the_pierre_auger_collaboration_a_aab_pierre_2016}, which will enhance the sensitivity of the surface detector to the masses of individual cosmic rays, allowing for better predictions of their rigidities and possible magnetic field deflections.

\appendix
\section{Results for the reference case of only homogeneously distributed sources}
\label{sec:ref}
In \cref{tab:results_ref} and \cref{tab:results_ref_syst}, the fit parameters for the reference models of only homogeneous background sources are given. These serve as a basis and comparison for the results of the models including catalog sources discussed in \cref{sec:results}. 
The fit parameters are in agreement with the ones found in other recent works of the Pierre Auger Collaboration~\cite{the_pierre_auger_collaboration_a_abdul_halim_constraining_2023}. Note however, that Ref.~\cite{the_pierre_auger_collaboration_a_abdul_halim_constraining_2023} uses SimProp instead of CRPropa3 and includes a local overdensity as well as a contribution by an additional lower-energy component, so no exact agreement of the fit parameters is expected.

\renewcommand{\arraystretch}{1.2} % Default value: 1
\begin{table}[ht]
\centering
\resizebox{\textwidth}{!}{%
\begin{tabular}{l | l l | l l | l l }
& \multicolumn{2}{c|}{\textbf{Reference, }$\boldsymbol{m=0}$ (flat)} & \multicolumn{2}{c|}{\textbf{Reference, }$\boldsymbol{m=3.4}$ (SFR)} & \multicolumn{2}{c}{\textbf{Reference, }$\boldsymbol{m=5.0}$ (AGN-like)}  \\ 
& posterior & MLE & posterior & MLE & posterior & MLE\\
\hline \hline
$\gamma$ & $-2.18_{-0.33}^{+0.31}$ & $-2.25$ & $-3.23_{-0.21}^{+0.24}$ & $-3.09$ & $-3.92_{-0.08}^{+0.01}$ & $-4.0$ \\
$\log_{10} (R_\mathrm{cut}$/V) & \phantom{+}$18.18_{-0.03}^{+0.03}$ & \phantom{+}18.17 & \phantom{+}$18.09_{-0.02}^{+0.02}$ & \phantom{+}18.10 & \phantom{+}$18.04_{-0.01}^{+0.01}$ & \phantom{+}18.04 \\
$I_\mathrm{H}$ & $6.5_{-1.6}^{+2.0} \times 10^{-2}$ & $7.0\times 10^{-2}$ & $9.0_{-9.0}^{+2.2} \times 10^{-3}$ & $<10^{-10}$ & $1.8_{-1.8}^{+0.3} \times 10^{-3}$ & $<10^{-10}$ \\
$I_\mathrm{He}$ & $2.0_{-0.3}^{+0.3} \times 10^{-1}$ & $2.0\times 10^{-1}$ & $1.3_{-0.2}^{+0.2} \times 10^{-1}$ & $1.4\times 10^{-1}$ & $4.5_{-1.8}^{+1.8} \times 10^{-2}$ & $<10^{-10}$ \\
$I_\mathrm{N}$ & $6.0_{-0.3}^{+0.4} \times 10^{-1}$ & $6.0\times 10^{-1}$ & $7.2_{-0.3}^{+0.3} \times 10^{-1}$ & $7.1 \times 10^{-1}$ & $8.1_{-0.3}^{+0.3} \times 10^{-1}$ & $8.8 \times 10^{-1}$ \\
$I_\mathrm{Si}$ & $1.2_{-0.3}^{+0.2} \times 10^{-1}$ & $1.2\times 10^{-1}$ & $1.2_{-0.2}^{+0.3} \times 10^{-1}$ & $1.3 \times 10^{-1}$ & $1.0_{-0.3}^{+0.3} \times 10^{-1}$ & $8.2 \times 10^{-2}$ \\
$I_\mathrm{Fe}$ & $2.1_{-0.9}^{+0.6} \times 10^{-2}$ & $1.9\times 10^{-2}$ & $2.6_{-0.9}^{+0.7} \times 10^{-2}$ & $2.3 \times 10^{-2}$ & $3.8_{-1.1}^{+1.1} \times 10^{-2}$ & $3.6 \times 10^{-2}$ \\
\hline
$\boldsymbol{\log b}$ & $-270.6  \pm 0.2$ &  & $-277.7  \pm 0.3$ &  & $-297.0 \pm 1.3$ &  \\
$\boldsymbol{D_E} \ (N_J=14)$ &  & \phantom{+}22.2 &  & \phantom{+}28.1 &  & \phantom{+}63.2 \\
$\boldsymbol{D_{X_\mathrm{max}}} \ (N_{{X}_\mathrm{max}}=74)$ &  & \phantom{+}126.8 &  & \phantom{+}132.4 &  & \phantom{+}134.1 \\
$\boldsymbol{D}$ &  & \phantom{+}149.0 &  & \phantom{+}160.5 &  & \phantom{+}197.3 \\
\textbf{log} $\boldsymbol{\mathcal{L}_\mathrm{ADs}}$ &  & \phantom{+}$0$ &  & \phantom{+}$0$ &  & \phantom{+}$0$ \\
\textbf{log} $\boldsymbol{\mathcal{L}}$ &  & $-250.5$ &  & $-256.2$ &  & $-274.6$
\end{tabular}}
\caption{Results for the reference models of only homogeneously distributed sources.}
\label{tab:results_ref}
\end{table}

\renewcommand{\arraystretch}{1.2} % Default value: 1
\begin{table}[ht]
\centering
\resizebox{\textwidth}{!}{%
\begin{tabular}{l | l l | l l | l l }
& \multicolumn{2}{c|}{\textbf{Reference, }$\boldsymbol{m=0}$ (flat)} & \multicolumn{2}{c|}{\textbf{Reference, }$\boldsymbol{m=3.4}$ (SFR)} & \multicolumn{2}{c}{\textbf{Reference, }$\boldsymbol{m=5.0}$ (AGN-like)}  \\ 
& posterior & MLE & posterior & MLE & posterior & MLE\\
\hline \hline
$\gamma$ & $-1.01_{-0.33}^{+0.41}$ & $-1.13$ & $-1.34_{-0.50}^{+0.63}$ & $-1.39$ & $-1.19_{-0.34}^{+0.41}$ & $-1.40$ \\
$\log_{10} (R_\mathrm{cut}$/V) & \phantom{+}$18.19_{-0.05}^{+0.03}$ & \phantom{+}18.19 & \phantom{+}$18.20_{-0.06}^{+0.05}$ & \phantom{+}18.19 & \phantom{+}$18.25_{-0.05}^{+0.04}$ & \phantom{+}18.25 \\
$I_\mathrm{H}$ & $4.8_{-3.2}^{+2.3} \times 10^{-2}$ & $5.1\times 10^{-2}$ & $1.1_{-1.1}^{+0.1} \times 10^{-2}$ & $<10^{-10}$ & $1.1_{-1.1}^{+0.2} \times 10^{-2}$ & $3.5 \times 10^{-8}$ \\
$I_\mathrm{He}$ & $2.9_{-0.4}^{+0.4} \times 10^{-1}$ & $2.9\times 10^{-1}$ & $1.1_{-0.3}^{+0.4} \times 10^{-1}$ & $1.3 \times 10^{-1}$ & $3.5_{-3.5}^{+0.8} \times 10^{-2}$ & $5.7 \times 10^{-4}$ \\
$I_\mathrm{N}$ & $5.3_{-0.4}^{+0.4} \times 10^{-1}$ & $5.2\times 10^{-1}$ & $6.4_{-0.7}^{+0.7} \times 10^{-1}$ & $6.4 \times 10^{-1}$ & $5.9_{-0.5}^{+0.7} \times 10^{-1}$ & $6.8 \times 10^{-1}$ \\
$I_\mathrm{Si}$ & $1.2_{-0.4}^{+0.4} \times 10^{-1}$ & $1.2\times 10^{-1}$ & $1.8_{-0.5}^{+0.6} \times 10^{-1}$ & $1.8 \times 10^{-1}$ & $2.8_{-0.5}^{+0.5} \times 10^{-1}$ & $2.4 \times 10^{-1}$ \\
$I_\mathrm{Fe}$ & $2.2_{-1.1}^{+0.9} \times 10^{-2}$ & $1.7\times 10^{-2}$ & $5.7_{-2.2}^{+1.4} \times 10^{-2}$ & $4.8 \times 10^{-2}$ & $9.3_{-2.2}^{+1.9} \times 10^{-2}$ & $8.3 \times 10^{-2}$ \\
\hline
$\nu_E / \sigma$ & $-1.43_{-0.67}^{+0.57}$ & $-1.37$ & \phantom{+}$0.19_{-0.60}^{+0.39}$ & \phantom{+}$0.18$ & $\phantom{+}1.21_{-0.30}^{+0.29}$ & \phantom{+}$1.17$ \\
$\nu_\mathrm{Xmax} / \sigma$ & $-0.88_{-0.28}^{+0.33}$ & $-0.78$ & $-1.62_{-0.36}^{+0.34}$ & $-1.53$ & $-2.23_{-0.24}^{+0.27}$ & $-2.00$ \\
\hline
% $\boldsymbol{\log b}$ & $-132373.5 \pm 0.1$ &  & $-132382.6 \pm 0.1$ &  & $-132395.9 \pm 1.6$ &  \\
% $\boldsymbol{D_\mathrm{syst}}$ &  & \phantom{+00000}2.5 &  & \phantom{+00000}2.4 &  & \phantom{+00000}5.4 \\
% $\boldsymbol{D_E}$ &  & \phantom{+0000}12.2 &  & \phantom{+0000}20.8 &  & \phantom{+0000}21.8 \\
% $\boldsymbol{D_{X\mathrm{max}}}$ &  & \phantom{+000}107.6 &  & \phantom{+000}114.6 &  & \phantom{+000}123.4 \\
% $\boldsymbol{D}$ &  & \phantom{+000}122.3 &  & \phantom{+000}137.8 &  & \phantom{+000}150.6 \\
% \textbf{log} $\boldsymbol{\mathcal{L}_\mathrm{ADs}}$ &   & $-132115.6$ &  & $-132115.6$ &  & $-132115.6$ \\
% \textbf{log} $\boldsymbol{\mathcal{L}}$  &  & $-132352.8$ &  & $-132360.5$ &  & $-132366.9$
$\boldsymbol{\log b}$ & $-257.9 \pm 0.1$ &  & $-267.0 \pm 0.1$ &  & $-280.3 \pm 1.6$ &  \\
$\boldsymbol{D_\mathrm{syst}}$ &  & \phantom{+}2.5 &  & \phantom{+}2.4 &  & \phantom{+}5.4 \\
$\boldsymbol{D_E} \ (N_J=14)$ &  & \phantom{+}12.2 &  & \phantom{+}20.8 &  & \phantom{+}21.8 \\
$\boldsymbol{D_{X_\mathrm{max}}} \ (N_{{X}_\mathrm{max}}=74)$ &  & \phantom{+}107.6 &  & \phantom{+}114.6 &  & \phantom{+}123.4 \\
$\boldsymbol{D}$ &  & \phantom{+}122.3 &  & \phantom{+}137.8 &  & \phantom{+}150.6 \\
\textbf{log} $\boldsymbol{\mathcal{L}_\mathrm{ADs}}$ &   & \phantom{+}$0$ &  & \phantom{+}$0$ &  & \phantom{+}$0$ \\
\textbf{log} $\boldsymbol{\mathcal{L}}$  &  & $-237.2$ &  & $-244.9$ &  & $-251.3$
\end{tabular}}
\caption{Results for the reference models of only homogeneously distributed sources including experimental systematic uncertainties as nuisance parameters.}
\label{tab:results_ref_syst}
\end{table}

\pagebreak
\input{acknowledgments}


% \begin{thebibliography}{99}

% \bibitem{a}
% Author, \emph{Title}, \emph{J.\ Abbrev.} {\bf vol} (year) pg.

% \bibitem{b}
% Author, \emph{Title},
% arxiv:1234.5678.

% \bibitem{c}
% Author, \emph{Title},
% Publisher (year).

% % For bibtex please use the JHEP.bst bibilography style.

% \end{thebibliography}
\bibliographystyle{JHEP}
\bibliography{references}

\end{document}

%% file: jcappub_authorlist.tex
\author[13]{A.~Abdul Halim,}
\author[72]{P.~Abreu,}
\author[54,52]{M.~Aglietta,}
\author[1]{I.~Allekotte,}
\author[70]{K.~Almeida Cheminant,}
\author[7,12]{A.~Almela,}
\author[45,46]{R.~Aloisio,}
\author[79]{J.~Alvarez-Mu\~niz,}
\author[79]{J.~Ammerman Yebra,}
\author[54,52]{G.A.~Anastasi,}
% ...

\affiliation[1]{Centro At\'omico Bariloche and Instituto Balseiro (CNEA-UNCuyo-CONICET), San Carlos de Bariloche, Argentina}
% ...
\affiliation[7]{Instituto de Tecnolog\'\i{}as en Detecci\'on y Astropart\'\i{}culas (CNEA, CONICET, UNSAM), Buenos Aires, Argentina}
% ...
\affiliation[12]{Universidad Tecnol\'ogica Nacional -- Facultad Regional Buenos Aires, Buenos Aires, Argentina}
\affiliation[13]{University of Adelaide, Adelaide, S.A., Australia}
% ...
\affiliation[45]{Gran Sasso Science Institute, L'Aquila, Italy}
\affiliation[46]{INFN Laboratori Nazionali del Gran Sasso, Assergi (L'Aquila), Italy}
% ...
\affiliation[52]{INFN, Sezione di Torino, Torino, Italy}
% ...
\affiliation[54]{Osservatorio Astrofisico di Torino (INAF), Torino, Italy}
% ...
\affiliation[70]{Institute of Nuclear Physics PAN, Krakow, Poland}
% ...
\affiliation[72]{Laborat\'orio de Instrumenta\c{c}\~ao e F\'\i{}sica Experimental de Part\'\i{}culas -- LIP and Instituto Superior T\'ecnico -- IST, Universidade de Lisboa -- UL, Lisboa, Portugal}
% ...
\affiliation[79]{Instituto Galego de F\'\i{}sica de Altas Enerx\'\i{}as (IGFAE), Universidade de Santiago de Compostela, Santiago de Compostela, Spain}
% ...
\affiliation[c]{School of Physics and Astronomy, University of Leeds, Leeds, United Kingdom}
% ...

%% file: acknowledgments.tex
% created on 2022-07-26
\section*{Acknowledgments}

\begin{sloppypar}
The successful installation, commissioning, and operation of the Pierre
Auger Observatory would not have been possible without the strong
commitment and effort from the technical and administrative staff in
Malarg\"ue. We are very grateful to the following agencies and
organizations for financial support:
\end{sloppypar}

\begin{sloppypar}
Argentina -- Comisi\'on Nacional de Energ\'\i{}a At\'omica; Agencia Nacional de
Promoci\'on Cient\'\i{}fica y Tecnol\'ogica (ANPCyT); Consejo Nacional de
Investigaciones Cient\'\i{}ficas y T\'ecnicas (CONICET); Gobierno de la
Provincia de Mendoza; Municipalidad de Malarg\"ue; NDM Holdings and Valle
Las Le\~nas; in gratitude for their continuing cooperation over land
access; Australia -- the Australian Research Council; Belgium -- Fonds
de la Recherche Scientifique (FNRS); Research Foundation Flanders (FWO);
Brazil -- Conselho Nacional de Desenvolvimento Cient\'\i{}fico e Tecnol\'ogico
(CNPq); Financiadora de Estudos e Projetos (FINEP); Funda\c{c}\~ao de Amparo \`a
Pesquisa do Estado de Rio de Janeiro (FAPERJ); S\~ao Paulo Research
Foundation (FAPESP) Grants No.~2019/10151-2, No.~2010/07359-6 and
No.~1999/05404-3; Minist\'erio da Ci\^encia, Tecnologia, Inova\c{c}\~oes e
Comunica\c{c}\~oes (MCTIC); Czech Republic -- Grant No.~MSMT CR LTT18004,
LM2015038, LM2018102, CZ.02.1.01/0.0/0.0/16{\textunderscore}013/0001402,
CZ.02.1.01/0.0/0.0/18{\textunderscore}046/0016010 and
CZ.02.1.01/0.0/0.0/17{\textunderscore}049/0008422; France -- Centre de Calcul
IN2P3/CNRS; Centre National de la Recherche Scientifique (CNRS); Conseil
R\'egional Ile-de-France; D\'epartement Physique Nucl\'eaire et Corpusculaire
(PNC-IN2P3/CNRS); D\'epartement Sciences de l'Univers (SDU-INSU/CNRS);
Institut Lagrange de Paris (ILP) Grant No.~LABEX ANR-10-LABX-63 within
the Investissements d'Avenir Programme Grant No.~ANR-11-IDEX-0004-02;
Germany -- Bundesministerium f\"ur Bildung und Forschung (BMBF); Deutsche
Forschungsgemeinschaft (DFG); Finanzministerium Baden-W\"urttemberg;
Helmholtz Alliance for Astroparticle Physics (HAP);
Helmholtz-Gemeinschaft Deutscher Forschungszentren (HGF); Ministerium
f\"ur Kultur und Wissenschaft des Landes Nordrhein-Westfalen; Ministerium
f\"ur Wissenschaft, Forschung und Kunst des Landes Baden-W\"urttemberg;
Italy -- Istituto Nazionale di Fisica Nucleare (INFN); Istituto
Nazionale di Astrofisica (INAF); Ministero dell'Istruzione,
dell'Universit\'a e della Ricerca (MIUR); CETEMPS Center of Excellence;
Ministero degli Affari Esteri (MAE); M\'exico -- Consejo Nacional de
Ciencia y Tecnolog\'\i{}a (CONACYT) No.~167733; Universidad Nacional Aut\'onoma
de M\'exico (UNAM); PAPIIT DGAPA-UNAM; The Netherlands -- Ministry of
Education, Culture and Science; Netherlands Organisation for Scientific
Research (NWO); Dutch national e-infrastructure with the support of SURF
Cooperative; Poland -- Ministry of Education and Science, grant
No.~DIR/WK/2018/11; National Science Centre, Grants
No.~2016/22/M/ST9/00198, 2016/23/B/ST9/01635, and 2020/39/B/ST9/01398;
Portugal -- Portuguese national funds and FEDER funds within Programa
Operacional Factores de Competitividade through Funda\c{c}\~ao para a Ci\^encia
e a Tecnologia (COMPETE); Romania -- Ministry of Research, Innovation
and Digitization, CNCS/CCCDI UEFISCDI, grant no. PN19150201/16N/2019 and
PN1906010 within the National Nucleus Program, and projects number
TE128, PN-III-P1-1.1-TE-2021-0924/TE57/2022 and PED289, within PNCDI
III; Slovenia -- Slovenian Research Agency, grants P1-0031, P1-0385,
I0-0033, N1-0111; Spain -- Ministerio de Econom\'\i{}a, Industria y
Competitividad (FPA2017-85114-P and PID2019-104676GB-C32), Xunta de
Galicia (ED431C 2017/07), Junta de Andaluc\'\i{}a (SOMM17/6104/UGR,
P18-FR-4314) Feder Funds, RENATA Red Nacional Tem\'atica de
Astropart\'\i{}culas (FPA2015-68783-REDT) and Mar\'\i{}a de Maeztu Unit of
Excellence (MDM-2016-0692); USA -- Department of Energy, Contracts
No.~DE-AC02-07CH11359, No.~DE-FR02-04ER41300, No.~DE-FG02-99ER41107 and
No.~DE-SC0011689; National Science Foundation, Grant No.~0450696; The
Grainger Foundation; Marie Curie-IRSES/EPLANET; European Particle
Physics Latin American Network; and UNESCO.
\end{sloppypar}